\pdfoutput=1
\documentclass[onecolumn]{aastex63}

\usepackage{framed,multirow}
\usepackage{times}

\usepackage{latexsym, amssymb, enumerate, amsmath}
\usepackage{defs}
\usepackage{fonts}
\usepackage{definitions}
\usepackage{epsfig}
\usepackage{epstopdf}
\usepackage{booktabs}
\usepackage{lineno}
\usepackage{xfrac}
\usepackage[normalem]{ulem}
\usepackage{mathrsfs}  
\usepackage{hyperref}
\newtheorem{remark}{Remark}

\usepackage[caption=false]{subfig} 
\usepackage{enumitem}
\usepackage{algorithm}
\usepackage{algpseudocode}

\usepackage{float}

\usepackage{xargs}                      % Use more than one optional parameter in a new commands
\usepackage{upgreek}

\received{---}
\revised{---}
\accepted{---}
\submitjournal{ApJS}

\shorttitle{Solvers for neutrino-matter coupling}
\shortauthors{Laiu et al.}

\begin{document}
\title{A DG-IMEX method for two-moment neutrino transport: Nonlinear solvers for neutrino-matter coupling%
	\footnote{This manuscript has been authored, in part, by UT-Battelle, LLC, under contract DE-AC05-00OR22725 with the US Department of Energy (DOE). The US government retains and the publisher, by accepting the article for publication, acknowledges that the US government retains a nonexclusive, paid-up, irrevocable, worldwide license to publish or reproduce the published form of this manuscript, or allow others to do so, for US government purposes. DOE will provide public access to these results of federally sponsored research in accordance with the DOE Public Access Plan (\texttt{http://energy.gov/downloads/doe-public-access-plan}).}	
	}

\correspondingauthor{M. Paul Laiu}
\email{laiump@ornl.gov}

\author[0000-0002-2215-6968]{M. Paul Laiu}
\affiliation{Multiscale Methods Group, Computer Science and Mathematics Division, Oak Ridge National Laboratory, Oak Ridge, TN 37831, USA}

\author[0000-0003-1251-9507]{Eirik Endeve}
\affiliation{Multiscale Methods Group, Computer Science and Mathematics Division, Oak Ridge National Laboratory, Oak Ridge, TN 37831, USA}
\affiliation{Department of Physics and Astronomy, University of Tennessee Knoxville, Knoxville, TN 37996, USA}

\author[0000-0001-7951-049X]{Ran Chu}
\affiliation{Department of Physics and Astronomy, University of Tennessee Knoxville, Knoxville, TN 37996, USA}

\author[0000-0003-3023-7140]{J. Austin Harris}
\affiliation{National Center for Computational Sciences, Oak Ridge National Laboratory, Oak Ridge, TN 37831, USA}

\author[0000-0002-5358-5415]{O. E. Bronson Messer}
\affiliation{National Center for Computational Sciences, Oak Ridge National Laboratory, Oak Ridge, TN 37831, USA}
\affiliation{Physics Division, Oak Ridge National Laboratory, Oak Ridge, TN 37831, USA}
\affiliation{Department of Physics and Astronomy, University of Tennessee Knoxville, Knoxville, TN 37996, USA}

\begin{abstract}
Neutrino-matter interactions play an important role in core-collapse supernova (CCSN) explosions as they contribute to both lepton number and/or four-momentum exchange between neutrinos and matter, and thus act as the agent for neutrino-driven explosions. 
Due to the multiscale nature of neutrino transport in CCSN simulations, an implicit treatment of neutrino-matter interactions is desired, which requires solutions of coupled nonlinear systems in each step of the time integration scheme.  
In this paper we design and compare nonlinear iterative solvers for implicit systems with energy coupling neutrino-matter interactions commonly used in CCSN simulations.  
Specifically, we consider electron neutrinos and antineutrinos, which interact with static matter configurations through the Bruenn~85 opacity set.  
The implicit systems arise from the discretization of a non-relativistic two-moment model for neutrino transport, which employs the discontinuous Galerkin (DG) method for phase-space discretization and an implicit-explicit (IMEX) time integration scheme.  
In the context of this DG-IMEX scheme, we propose two approaches to formulate the nonlinear systems --- a coupled approach and a nested approach.  
For each approach, the resulting systems are solved with Anderson-accelerated fixed-point iteration and Newton's method.  
The performance of these four iterative solvers has been compared on relaxation problems with various degree of collisionality, as well as proto-neutron star deleptonization problems with several matter profiles adopted from spherically symmetric CCSN simulations.
Numerical results suggest that the nested Anderson-accelerated fixed-point solver is more efficient than other tested solvers for solving implicit nonlinear systems with energy coupling neutrino-matter interactions.
\end{abstract}
%\linenumbers

%% main text

\section{Introduction}
\label{sec:intro}

\begin{figure}[h]
	\centering
	{\includegraphics[width=0.6\linewidth]{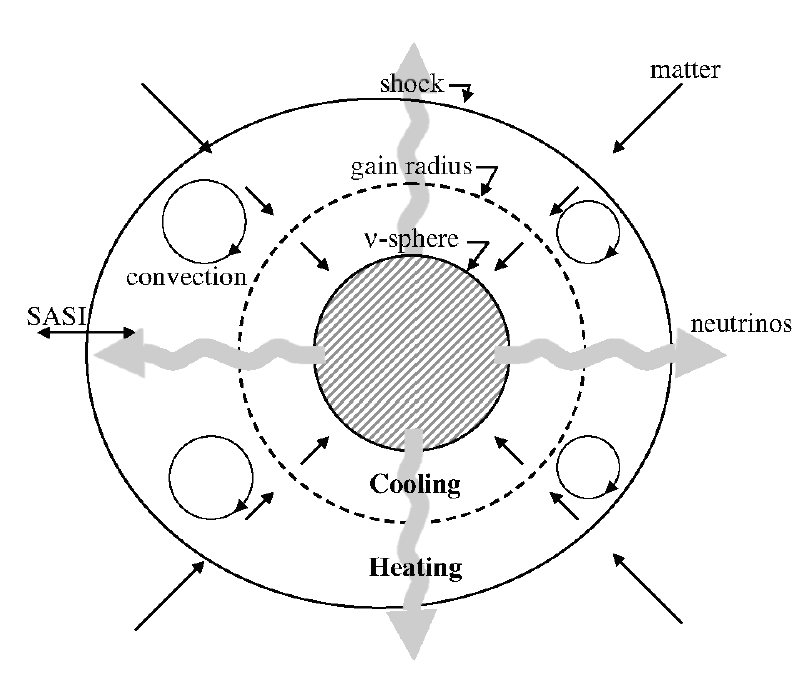}}
	\caption{Schematic diagram of the neutrino reheating phase during a CCSN explosion.  (This figure is originally from \cite{Bruenn:2010}, and is reproduced here with permission from AIP~Publishing.)  Neutrinos emanate from the neutrinosphere --- a proxy for the surface of the proto-neutron star (gray shaded region) --- and the cooling layer above it.  Neutrino heating mediated by neutrino-matter interactions in the gain layer between the gain radius and the shock provides energy to reinvigorate the stalled supernova shock. During shock revival, the flow below the shock may be modulated by neutrino-driven convection and the standing accretion shock instability (SASI).}
	\label{fig:diagramCCSN}
\end{figure}

Core-collapse supernovae (CCSNe), the explosive deaths of massive stars, are to a large extent neutrino driven.  
About 99\% of the gravitational potential energy released in a core-collapse event ($\sim10^{53}$~erg) is radiated away by neutrinos, which also act as a driver for the expulsion of matter.  
Near the end of a massive star's life (i.e. a star with mass exceeding about ten solar masses) its iron core, which does not produce energy by nuclear burning, is held up against gravity by the pressure from degenerate electrons.  
However, the mass of the iron core continues to increase due to silicon burning on its surface.  
Once the mass of the iron core reaches about $1.4$ solar masses (the Chandrasekhar limit), the electron degeneracy pressure becomes insufficient in balancing gravity, and the core collapses in on itself.  
The collapse proceeds until the central rest mass density exceeds nuclear matter densities ($>10^{14}$~g~cm$^{-3}$), when the matter equation of state (EoS) stiffens to halt the collapse and a shock wave is launched into the collapsing outer core.  
The outward-propagating shock wave loses energy by dissociating iron nuclei into free nucleons.  
Furthermore, electron capture on nucleons in the hot matter behind the shock produces copious amounts of neutrinos and antineutrinos, which can escape the system once the shock reaches sufficiently low densities (about $10^{12}$~g~cm$^{-3}$).  
The combination of neutrino emission and iron dissociation weakens the shock, which eventually stalls at a radius of about $100-200$~km from the center of the star.  
At this point, the region below the shock can be divided into the \emph{cooling layer} and the \emph{gain layer}, separated by the gain surface, where neutrino heating and cooling balance.  
In the cooling layer, extending from the surface of the proto-neutron star to the gain surface, there is net energy loss by neutrino emission.  
At lower densities, in the heating layer extending from the gain surface to the shock surface, there is net heating by neutrinos emanating from below; see diagram in Figure~\ref{fig:diagramCCSN}.  
In the neutrino reheating explosion mechanism \citep{betheWilson_1985}, energy deposition by neutrinos in the heating layer revives the supernova shock wave to disrupt the massive star in a CCSN explosion.  
This basic description is supported by recent numerical simulations (e.g., \citet{lentz_etal_2015,melson_etal_2015,burrows_etal_2020}), but the details are more complicated: The CCSN explosion emerges from a complex interplay between between neutrino transport and hydrodynamic (or magnetohydrodynamic) processes, playing out within a curved spacetime (see, e.g., \cite{janka_2012,burrows_2013,hix_etal_2014,muller_2016} for reviews).  

Multiple interaction processes contribute to lepton number and/or four-momentum exchange between neutrinos and matter in the CCSN explosion.  
Neutrino emission caused by electron capture on nucleons and the inverse process of neutrino capture (absorption) dominate and contribute to both lepton number and four-momentum exchange.  
During stellar core collapse, electron capture on nuclei is critical to the collapse dynamics; e.g., \cite{hix_etal_2003}.  
Neutrino scattering on electrons and nucleons (and nuclei during collapse) are also major contributors to the neutrino opacity.  
In the seminal work of \cite{bruenn_1985}, neutrino-nucleon scattering is treated as isoenergetic (the neutrino changes direction but not energy in its encounter with the nucleon), while full four-momentum exchange is considered for neutrino-electron scattering (NES).  
In later works (e.g., \cite{thompson_etal_2000,lentz_etal_2012b,muller_etal_2012a,burrows_etal_2018}), it has been demonstrated that neutrino-nucleon scattering contributes to neutrino-matter thermalization in a nontrivial way.  
(Without modern nuclear electron capture rates, NES also helps shape conditions in the core prior to the formation of the supernova shock \citep{mezzacappaBruenn_1993c,lentz_etal_2012b}.)  
Neutrino pair creation and annihilation, via electron-positron pairs \citep{bruenn_1985} or nucleon-nucleon bremsstrahlung \citep{hannestadRaffelt_1998}, is yet another significant process in CCSN neutrino transport.  
In particular, nucleon-nucleon bremsstrahlung is a dominant source of $\mu$ and $\tau$ neutrinos.  
Several studies (e.g., \cite{lentz_etal_2012b,just_etal_2018,burrows_etal_2018}) have investigated the impact of various neutrino opacities on the CCSN explosion mechanism.  
Although it is difficult to pin down an exact set of necessary interactions, there is a consensus view that processes that couple globally in neutrino momentum-space and/or across neutrino species (e.g., NES and pair processes) must be included for a realistic description.  

It is computationally expensive to include neutrino transport with satisfactory realism in simulations of CCSNe.  
First of all, neutrinos interact relatively weakly with matter in the gain region, which demands a kinetic description.  
In addition, the hydrodynamics must be modeled without imposed spatial symmetries and with adequate resolution to capture processes that shape the explosion (see e.g., \cite{muller_2020}).  
The additional requirement of including momentum-space coupling neutrino-matter interactions makes realistic, large scale simulations a computational challenge.  
For example, if momentum space is discretized with $N_{\vect{p}}$ points, a simple evaluation of the collision operator at one spatial location requires $\mathcal{O}(N_{\vect{p}}^{2})$ operations, as opposed to $\mathcal{O}(N_{\vect{p}})$ operations for the emission, absorption and isoenergetic scattering operators.  
In current three-dimensional supernova models, the dimensionality of the neutrino transport problem is reduced by adopting one- or two-moment approaches that retain the neutrino energy dimension of momentum-space \citep{just_etal_2015,skinner_etal_2019,bruenn_etal_2020}, but this does not completely alleviate the computational challenge of including the critical energy coupling interactions.  

Because the neutrino mean free path can be much shorter than the spatial resolution afforded in parts of the computational domain, an implicit treatment of neutrino-matter interactions is desired.  
In spherically symmetric models \citep{ramppJanka_2002,liebendorfer_etal_2004} and multi-dimensional models employing the so-called ray-by-ray approximation \citep{bruenn_etal_2020}, the full set of transport equations is commonly solved with implicit time integration.  
So far, fully implicit time integration has not been the method of choice for truly multi-dimensional neutrino transport (but see \cite{sumiyoshiYamada_2012} for an exception).  
Instead, implicit-explicit (IMEX) methods \citep{ascher_etal_97,pareschiRusso_2005} have received more attention (see, e.g., \cite{oConnor_2015,just_etal_2015,kuroda_etal_2016,skinner_etal_2019,CHU201962}).  
In the IMEX approach, collisions are treated with implicit methods, while phase-space advection is treated with explicit methods (implicit integration for the momentum-space advection terms is also used \citep{kuroda_etal_2016}).  
Explicit integration for phase-space advection is advantageous because it avoids solving a distributed, sparse system of nonlinear equations, and, since neutrino-matter interactions are completely local in space, the implicit part is embarrassingly parallel.  
Moreover, the characteristic wave speeds associated with the transport and hydrodynamics equations are not too dissimilar in relativistic systems, and it is not clear that the expected additional cost of solving the full transport equation with implicit time integration and a larger time step will pay off.  

Energy coupling neutrino-matter interactions still dominate the computational cost of IMEX-based neutrino transport schemes employing a spectral two-moment approach, where the neutrino energy domain is discretized with $N_{\varepsilon}$ points.  
The cost per spatial point is expected to scale roughly as $\mathcal{O}(N_{\varepsilon}^{p})$, where the power $p$ is between two and three.  
The lower bound ($p=2$) is motivated by the expected cost of simply evaluating the energy coupling operators, while the $p=3$ scaling is due to the cost of inverting the Jacobian matrix resulting from an implicit solution algorithm based on Newton's method.  
In addition, the neutrino-matter interaction rates depend nonlinearly on the local thermodynamic properties of the matter, so that in a fully implicit approach \citep{kuroda_etal_2016}, the local computational cost scales linearly with the number of iterations needed to reach convergence.  

Several studies have investigated approximations that are motivated by potential gains in computational expediency.  
To alleviate the cost of including energy coupling neutrino-matter interactions, \cite{just_etal_2015} investigated the effect of evaluating the collision terms for energy coupling interactions with matter conditions taken from the known state at high densities (above $10^{11}$~g~cm$^{-3}$), and using explicit integration at lower densities, and found results that were practically identical to a run with a more implicit treatment.  
Another approach to circumvent stiffness induced by neutrino-matter interactions is to artificially reduce the rates at high densities and use explicit time integration throughout the computational domain \citep{thompson_etal_2003,oConnor_2015,burrows_etal_2018,just_etal_2018}.  
We also note a simplifying approach to pair processes involving approximating pair annihilation partners by local equilibrium distributions, which essentially renders this interaction local in momentum-space \citep{oConnor_2015}.  
While some approximations have shown to work well in a limited set of comparisons, they do introduce uncertainties and limit the applicability of the algorithm.  
Partially for these reasons, we do not follow the approach of altering the kernel or collision operator based on physical insight into the problem.  
Instead we seek to develop fully implicit solvers for energy coupling neutrino-matter interactions for CCSN simulations and related applications.

This paper details the base neutrino transport algorithms implemented in the toolkit for high-order neutrino-radiation hydrodynamics (\thornado\footnote{\url{https://github.com/endeve/thornado}, \url{http://dx.doi.org/10.13139/OLCF/1735948}}), which is being developed for simulations of CCSNe and related problems using the discontinuous Galerkin (DG) method for phase-discretization.  
Specifically, the phase-space discretization in \thornado\ is based on the \emph{nodal} DG method (see, e.g., \cite{hesthavenWarburton_2008}).  
The original DG method was developed by \cite{reedHill_1973} for solving neutron transport problems.  
Since then, it has been extended to the Runge-Kutta discontinuous Galerkin framework for solving more general hyperbolic partial differential equations (see, e.g., \cite{cockburn1997runge,cockburn_etal_1990,cockburn1988tvb,cockburn1989tvb,cockburn1991runge,cockburnShu_2001} for early developments).  
For more recent developments of DG methods, see \cite{shu_2016} and references therein.  
DG methods are particularly attractive for transport problems since they recover the correct asymptotic behavior in the so-called diffusion limit \citep{larsenMorel_1989,adams_2001}.  
They can also be easily applied to problems with curvilinear coordinates --- necessary when solving general relativistic problems \citep{teukolsky_2016}.  
However, the DG method has so far not been applied to neutrino transport in CCSN models (but see \cite{radice_etal_2013,endeve_etal_2015,CHU201962} for applications in simplified settings).  
The DG method approximates solutions with piecewise local polynomials, and tracks the evolution of coefficients associated with the polynomial expansion, thus, they are often referred to as modal DG methods.  
The nodal DG method uses a particular interpolating polynomial to construct the approximation, which allows it to track the evolution of nodal values at the interpolation points.  
This special polynomial approximation results in a simple projection operator from the target function to the polynomial space, which enables straightforward parallel implementation of the nodal DG method \citep{klockner_etal_2009}. 
The nodal DG method has been used in various applications, including solving kinetic equations (e.g., \cite{xiong_etal_2015,juno_2018}).  

In this paper we design and evaluate nonlinear solvers for neutrino-matter interactions in a two-moment model for neutrino transport within the IMEX framework.  
This DG-IMEX method is essentially the same as that described by \cite{CHU201962}, but is extended to include curvilinear spatial coordinates, multiple neutrino species, more realistic interactions, and coupling to a material background governed by a nuclear EoS.  
Specifically, we consider electron neutrinos and antineutrinos, and develop nonlinear solvers for the opacity set of \cite{bruenn_1985}, which includes emission and absorption due to electron and neutrino capture on nucleons and nuclei, isoenergetic neutrino scattering off nucleons and nuclei, inelastic NES, and neutrino-antineutrino pair production/annihilation from electron-positron pairs.  
We use the SFHo EoS \citep{steiner_etal_2013} in the numerical experiments.  
The microphysics (neutrino opacities and EoS) has been tabulated by the \weaklib\ library\footnote{\url{https://github.com/starkiller-astro/weaklib}, \url{http://dx.doi.org/10.13139/OLCF/1735948}}, which also provides routines for access and manipulation (e.g., interpolation and differentiation) of tabulated microphysics data.  

Several nonlinear solver strategies are considered for the neutrino-matter coupling problem.  
As a baseline for comparison we consider Newton's method, which can offer rapid convergence to the solution if the initial guess is sufficiently close and the objective function sufficiently regular.  
However, the necessity of using approximate derivatives due to tabulated opacity kernels to form the Jacobian matrix can hamper the convergence speed.  
In addition, the construction of the Jacobian from tabulated data and the solution of a dense linear system for each iteration is computationally expensive.  
We consider fixed-point iteration as an alternative to Newton's method.  
With fixed-point iteration, Jacobian matrix constructions and dense linear system solutions are not necessary, but the rate of convergence can be slow.  
We employ Anderson acceleration to improve the convergence rate of the fixed-point method.  
This acceleration technique was first proposed by \cite{Anderson-1965} for solving integral equations, and has been used to accelerate fixed-point solutions in several applications, including solving radiation-diffusion equations \citep{AN20171}, flow problems \citep{LOTT201292}, nuclear reactor simulations \citep{HAMILTON2016241}, as well as a variety of nonlinear problems \citep{Walker-Ni-2011}.  
Anderson acceleration speeds up the convergence of standard fixed-point iterations by taking an extrapolation step based on the recent iterates that aims to minimize the residual of the new iterate.  
The convergence properties of Anderson acceleration were analyzed in \cite{Toth-Kelley-2015,kelley2018numerical,evans2020proof}, which include (i) global convergence on linear problems under the standard contraction assumption, (ii) local convergence on nonlinear problems under assumptions similar to the standard ones for local convergence of Newton's method, and (iii) improved local convergence rate when applied on linearly converging fixed-point iterations.  

Although the Anderson-accelerated fixed-point algorithm is generally faster than Newton's method in the cases investigated in this paper, we find that the computational cost associated with reevaluating the neutrino opacities in each iteration remains relatively high.  
This observation motivates a nested approach, where the matter quantities are updated in an outer iteration loop, outside an inner iteration loop where the radiation field is iterated to convergence while the matter state is fixed.  
We consider two nested iteration schemes, both based on Anderson-accelerated fixed-point iteration in the outer loop.  
In the first case the inner loop solve is based on Newton's method.  
In this nested approach, the Jacobian associated with the inner solve can be computed analytically since the nonlinear functional is considered independent of the matter state.  
However, the dense linear system to be solved is of similar size as for the fully coupled Newton method.  
In the second case, the inner loop solve is based on Anderson-accelerated fixed-point iteration.  
We find that the nested iteration schemes require fewer opacity evaluations and are more efficient than the fully coupled schemes.  
We also find that the nested scheme with inner Newton iterations requires fewer iterations to converge for the highest mass densities than the nested scheme with inner fixed-point iterations.  
However, the cost per fixed-point iteration is lower than the cost of solving the dense linear system associated with each iteration in Newton's method.  
As a result, the nested iteration scheme with fixed-point iteration in both the inner and outer loops is the most efficient of the solvers considered.  

We note that the model considered in this paper lacks the physical fidelity of current CCSN models in some key aspects.  
First, we consider a static fluid and adopt a non-relativistic model.  
However, it is well-established that both special and general relativistic effects must be included in realistic models \citep{bruenn_etal_2001,lentz_etal_2012a,muller_etal_2012a}.  
Second, we only consider electron neutrinos and antineutrinos, and do not consider the nucleon-nucleon bremsstrahlung opacity \citep{hannestadRaffelt_1998}.  
This process is a dominant source for production of muon and tau neutrinos and antineutrinos \citep{thompson_etal_2000}, which contribute significantly to the total neutrino luminosity from CCSNe.  
Third, neutrino-nucleon scattering is treated as isoenergetic.  
Still, since the work of \cite{bruenn_1985}, it has been demonstrated that, despite a relatively small energy exchange per neutrino-nucleon interaction, the relatively large cross-section (when compared with NES) implies that this scattering process should be treated as inelastic \citep{reddy_etal_1998,muller_etal_2012a,burrows_etal_2018}.  
However, as the development of \thornado\ matures, we intend to account for this physics, and document on the performance in future publications.  
To the best of our knowledge, the present paper documents the most advanced application of the DG method to neutrino transport.  

This paper is organized as follows: in Section~\ref{sec:kinetic_eqn} we present the mathematical model we adopt for neutrino transport; in Section~\ref{sec:DG_IMEX} we introduce the DG-IMEX scheme implemented in \thornado; the nonlinear solvers are detailed in Section~\ref{sec:techniques}; and in Section~\ref{sec:num_results} we present numerical results, where we compare the performance of the solvers on (1) relaxation to equilibrium and (2) proto-neutron star deleptonization.  
We summarize our findings and draw conclusions in Section~\ref{sec:conclusion}.  
\section{Neutrino/Antineutrino transport equations}
\label{sec:kinetic_eqn}

\subsection{Boltzmann equation and neutrino-matter interactions}
\label{subec:Boltzmann}

In nuclear astrophysics applications, neutrino transport can be modeled by the Boltzmann equation.  
In this paper, we consider a non-relativistic Boltzmann equation
\begin{equation}
  \frac{1}{\SOL}\p_t{\fNe} + \cT(\fNe) = \cC(\fNe)\:,
\end{equation}
which governs the particle distribution function $\fNe(\vect{x},\vect{p},t)$ that describes the density of particles at position $\vect{x}=(x^1,x^2,x^3)\in\bbR^3$
with momentum $\vect{p}=(p^1,p^2,p^3)\in\bbR^3$
at time $t\in\bbR^+$, where $\SOL$ is the speed of light.
Adopting curvilinear phase-space coordinates, the advection operator $\cT$ takes the form (see, e.g., \cite{endeve_etal_2015})
\begin{equation}\label{eq:transportOperator}
  \cT(\fNe):=\frac{1}{\sqrt{\gamma}}\sum_{i=1}^{3}\p_{x^i} \big(\sqrt{\gamma}\,{H_{\vect{x}}^{i}}\fNe\big) 
  + \frac{1}{\sqrt{\lambda}}\sum_{i=1}^{3}\p_{p^i} \big(\sqrt{\lambda}\,{H_{\vect{p}}^{i}}\fNe\big)\:, 
\end{equation}
where $H_{\vect{x}}^{i} \fNe$ and $H_{\vect{p}}^{i} \fNe$ denote the position space flux and the momentum space flux in the corresponding $i$th direction, respectively, and $\gamma$, $\lambda$ are the determinants of the position space and momentum space metric tensors, respectively.  
While the form of the advection operator in Eq.~\eqref{eq:transportOperator} holds for more general (non-orthogonal) spacetime and momentum space bases \citep[see, e.g.,][]{cardall_etal_2013}, we restrict ourselves to orthogonal bases in this paper. 
The position space flux considered in this paper takes the form $H_{\vect{x}}^{i} \fNe = (\sfrac{p^i}{|\vect{p}|}) \fNe$, which is proportional to the particle propagation direction.  
We will restrict ourselves to spherical polar momentum coordinates, and in this case, $H_{\vect{p}}^{i} \fNe$ can be obtained from Eqs.~(A15) and (A16) in \cite{endeve_etal_2015}.  
The collision operator $\cC$ models interactions between particles and a material background, and includes emission, absorption, elastic scattering on nucleons and nuclei, neutrino-electron scattering, and thermal pair processes from electron-positron creation and annihilation.  
This is the neutrino opacity set described in \cite{bruenn_1985} (cf. Table~1 therein).  

In this work, we consider the transport of electron neutrinos ($\nu_e$) and antineutrinos ($\bar{\nu}_e$), which results in the coupled equations
\begin{subequations}\label{eq:coupled_transport}
\begin{align}
\label{eq:neutrino_transport}
\frac{1}{\SOL}\p_t{\fNe} + \cT(\fNe) &= \cC(\fNe,\fNeb)\:,\\
\label{eq:antineutrino_transport}
\frac{1}{\SOL}\p_t{\fNeb} + \cT(\fNeb) &= \bar{\cC}(\fNeb,\fNe)\:,
\end{align}
\end{subequations}
where the neutrino and antineutrino distribution functions are denoted with $\fNe$ and $\fNeb$, respectively, and the collision terms $\cC$ and $\bar{\cC}$ both depend on $\fNe$ and $\fNeb$, since they include the thermal pair production and annihilation processes of neutrino-antineutrino pairs.  
Before giving a detail formulation of the collision terms, we first change to spherical polar momentum coordinates $(\epsilonNu,\omegaNu)$ and decompose the neutrino three-momentum as $\vect{p}=\epsilonNu\,\vect{\ell}(\omegaNu)$, where $\epsilonNu\in\bbR^{+}$ denotes neutrino energy and the unit vector $\vect{\ell}(\omegaNu)\in\bbR^3$ only depends on the angular direction $\omegaNu\in\bbS^2$ relative to a local orthonormal basis.  
The momentum space volume element is then decomposed into $d\vect{p}= d V_{\epsilonNu}\,d\omegaNu$, where $d V_{\epsilonNu}:=\epsilonNu^2 d\epsilonNu$ is the spherical shell energy volume element and $d\omegaNu$ is the momentum space angular element.  
With this notation, the particle distribution $\fNe(\vect{x}, \vect{p}, t)$ can be written as $\fNe(\vect{x}, \omegaNu, \epsilonNu, t)$.
At each $\vect{x}$ and $t$, the neutrino collision operator then takes the form \citep{bruenn_1985}
\begin{align}\label{eq:collisionTerms}
\cC(\fNe, \fNeb)
& = \big(1- \fNe(\omegaNu,\epsilonNu)\big)\,\hat{\eta}(\epsilonNu) - \hat{\chi}(\epsilonNu)\, \fNe(\omegaNu,\epsilonNu)\nonumber\\
&+ \frac{\epsilonNu^2}{\SOL(\Planck\SOL)^3}\int_{\bbS^2} R^{\IS}(\vect{\ell}\cdot\vect{\ell}',\epsilonNu) \fNe(\omegaNu',\epsilonNu)\,d\omegaNu'
- \frac{\epsilonNu^2}{\SOL(\Planck\SOL)^3} \fNe(\omegaNu,\epsilonNu) \int_{\bbS^2} R^{\IS}(\vect{\ell}\cdot\vect{\ell}',\epsilonNu) \,d\omegaNu'
\nonumber \\
&+\frac{1}{\SOL(\Planck\SOL)^3}\big(1- \fNe(\omegaNu,\epsilonNu)\big)
\int_{\bbR^{+}}\int_{\bbS^{2}}R^{\IN}(\vect{\ell}\cdot\vect{\ell}', \epsilonNu, \epsilonNu')\, 
\fNe (\omegaNu',\epsilonNu')\,d\omegaNu'\,dV_{\epsilonNu'}\nonumber\\
&-\frac{1}{\SOL(\Planck\SOL)^3}\fNe (\omegaNu,\epsilonNu)
\int_{\bbR^{+}}\int_{\bbS^{2}}R^{\OUT}(\vect{\ell}\cdot\vect{\ell}', \epsilonNu, \epsilonNu')\,
\big(1-\fNe (\omegaNu',\epsilonNu')\big)\,d\omegaNu'\,dV_{\epsilonNu'}\\
&+\frac{1}{\SOL(\Planck\SOL)^3}\big(1 - \fNe(\omegaNu,\epsilonNu)\big)
\int_{\bbR^{+}}\int_{\bbS^{2}}R^{\PROD}(\vect{\ell}\cdot\vect{\ell}', \epsilonNu,\epsilonNu')\,
\big(1-\fNeb(\omegaNu',\epsilonNu')\big)\,d\omegaNu'\,dV_{\epsilonNu'} \nonumber\\
&- \frac{1}{\SOL(\Planck\SOL)^3} \fNe(\omegaNu,\epsilonNu)
\int_{\bbR^{+}}\int_{\bbS^{2}}R^{\ANN}(\vect{\ell}\cdot\vect{\ell}', \epsilonNu,\epsilonNu')\,
\fNeb(\omegaNu',\epsilonNu')\,d\omegaNu'\,dV_{\epsilonNu'},\nonumber
\end{align}
with $\Planck$ the Planck constant, $\hat{\eta}$ the emissivity, $\hat{\chi}$ the absorption opacity, $R^{\IS}$ the elastic (isoenergetic) scattering kernel (due to scattering with nucleons and nuclei), $R^{\IN}$ and $R^{\OUT}$ the neutrino-electron scattering kernels, and $R^{\PROD}$ and $R^{\ANN}$ the thermal production and annihilation kernels due to pair processes.  
The antineutrino collision operator $\bar{\cC}$ is defined analogously with $\hat{\eta}$, $\hat{\chi}$, and the opacity kernels $R^{\op}$, $\op=\IS,\IN,\OUT,\PROD,\ANN$, replaced by their antineutrino counterparts.

For thermal emission and absorption, we follow the approach in, e.g., \cite{burrows_etal_2006a}, and rewrite
\begin{equation}
\big(1-\fNe\big)\,\hat{\eta} - \hat{\chi}\,\fNe
={\chi}\,\big(\fNe_0 - \fNe\big) \:,
\end{equation}
where the effective opacity and the equilibrium distribution are defined as 
\begin{equation}
{\chi} := \hat{\eta} + \hat{\chi} \:,\quand
\fNe_0 := \f{\hat{\eta}}{\hat{\eta} + \hat{\chi}}\:,
\end{equation}
respectively.
Similarly, the antineutrino emission and absorption term is written as $\bar{\chi}\,\big(\fNeb_0 - \fNeb\big)$, with $\bar{\chi}$ and $\fNeb_0$ defined analogously.
The equilibrium distributions $\fNe_0$ and $\fNeb_0$ are given by the Fermi-Dirac distribution, which is isotropic in angle $\omegaNu$.
Specifically,
\begin{equation}\label{eq:FermiDirac}
  \fNe_0(\epsilonNu)=\f{1}{e^{(\epsilonNu-\mu)/\Boltzmann T}+1}\:,\quand
  \fNeb_0(\epsilonNu)=\f{1}{e^{(\epsilonNu-\bar{\mu})/\Boltzmann T}+1}\:,
\end{equation}
where $\Boltzmann$ is the Boltzmann constant, $T$ is the matter temperature, $\mu := \mu_{e^{-}}+(\mu_{p}-\mu_{n})$ the neutrino chemical potential, and $\bar{\mu} := \mu_{e^{+}}-(\mu_{p}-\mu_{n})$ the antineutrino chemical potential.  
Here, $\mu_{e^{-(+)}}$ is the electron (positron) chemical potential, $\mu_{n}$ the neutron chemical potential, and $\mu_{p}$ the proton chemical potential, which are evaluated from an appropriate EoS.  
Since $\mu_{e^{+}} = - \mu_{e^{-}}$, we have $\bar{\mu} = - \mu$.  

Following \cite{bruenn_1985}, the neutrino scattering and pair process kernels are approximated with $L$-term Legendre expansions in the cosine of the scattering angle $\alpha:=\vect{\ell}\cdot\vect{\ell}'$, with expansion coefficients $\Phi$ depending on the neutrino energy.
Specifically,
\begin{equation}\label{eq:Op_approx}
R^{\IS}(\alpha,\epsilonNu)
\approx\sum_{l=0}^{L}\Phi^{\IS}_{l}(\epsilonNu)\,P_{l}(\alpha)\:,\quand
R^{\op}(\alpha,\epsilonNu,\epsilonNu')
\approx\sum_{l=0}^{L}\Phi^{\op}_{l}(\epsilonNu,\epsilonNu')\,P_{l}(\alpha)\:,
\end{equation}
where $\op=\IN,\OUT,\PROD,\ANN$, $P_{l}$ denotes the Legendre polynomial of degree $l$, and the expansion coefficients of degree $l$ are given by
\begin{equation}
\Phi^{\IS}_{l}(\epsilonNu)
=\f{1}{C_{l}}\int_{-1}^{1}R^{\IS}(\alpha,\epsilonNu)\,P_{l}(\alpha)\,d \alpha\:,\quand
\Phi^{\op}_{l}(\epsilonNu,\epsilonNu')
=\f{1}{C_{l}}\int_{-1}^{1}R^{\op}(\alpha,\epsilonNu,\epsilonNu')\,P_{l}(\alpha)\,d \alpha\:.
\end{equation}
Here $C_{l}:=\int_{-1}^{1} P_{l}^2(\alpha)\,d \alpha$, $l=0,\ldots,L$, are normalization constants.  
In this work, we use $P_{0}(\alpha) = 1$ and $P_{1}(\alpha) = \alpha$, thus the normalization constants are $C_0 = 2$ and $C_1 = \frac{2}{3}$.
The antineutrino opacity kernels are approximated analogously.
Due to particle conservation and detailed balance (e.g., \cite{cernohorsky_1994}), the neutrino-electron scattering and pair process kernels satisfy 
\begin{equation}\label{eq:kernelSymmetries}
\begin{alignedat}{2}
R^{\IN}(\alpha,\epsilonNu,\epsilonNu') &= R^{\OUT}(\alpha,\epsilonNu',\epsilonNu)\:,\quad
R^{\OUT}(\alpha,\epsilonNu,\epsilonNu') &&= R^{\OUT}(\alpha,\epsilonNu',\epsilonNu) e^{(\epsilonNu - \epsilonNu')/\Boltzmann T}\:,\\
R^{\ANN}(\alpha,\epsilonNu,\epsilonNu') &= \bar{R}^{\ANN}(\alpha,\epsilonNu',\epsilonNu)\:,\qquad
R^{\PROD}(\alpha,\epsilonNu,\epsilonNu') &&= R^{\ANN}(\alpha,\epsilonNu,\epsilonNu') e^{-(\epsilonNu + \epsilonNu')/\Boltzmann T}\:.
\end{alignedat}
\end{equation}
Note that these symmetry properties are enforced in the energy space, thus they are preserved in the angular approximation for the kernels in Eq.~\eqref{eq:Op_approx}. 

In the following, as a first approximation, we will only include the isotropic part of the kernels (i.e., $L=0$).  
More realistic treatments would include linear corrections ($L=1$), while further corrections ($L>1$) have been shown to result in only minor differences for neutrino-electron scattering (e.g, \cite{smitCernohorsky_1996}).  

\subsection{Angular moment equations}
\label{subsec:moments}

The need for high spatial resolution and unconstrained spatial dimensionality, e.g., to capture fluid dynamics in our target applications, renders direct solutions of the Boltzmann equation too expensive.  
However, neutrino heating rates are sensitive to the neutrino energy distribution, which demands retention of the energy dimension of momentum space.  
Therefore, to balance computational cost with physical fidelity, we settle for solving for a finite number of angular moments of the distribution function by adopting a two-moment model.  
Two-moment models are widely used to model neutrino transport in core-collapse supernovae (e.g., \cite{oConnor_2015,just_etal_2015,kuroda_etal_2016,roberts_etal_2016,skinner_etal_2019}).
In the spectral two-moment model, we solve for the zeroth and first moments of the neutrino distribution function, while the second moments are obtained from a closure procedure.  
These moments are defined respectively as
\begin{equation}
  \big\{\,\cJ,\cH^{i},\cK^{ij}\,\big\}(\vect{x},\epsilonNu,t)=\f{1}{4\pi}\int_{\bbS^{2}}f(\vect{x},\omegaNu,\epsilonNu,t)\,\big\{\,1,\ell^{i}(\omegaNu), \ell^{i}(\omegaNu)\ell^{j}(\omegaNu)\,\big\}\,d\omegaNu,
\end{equation}
with moments for antineutrinos ($\bar{\cJ}$, $\bar{\cH}^{i}$, and $\bar{\cK}^{ij}$) defined analogously.  
Then, the zeroth moment $\cJ$ ($\bar{\cJ}$) is the spectral number density of neutrinos (antineutrinos), the first moment $\cH^{i}$ ($\bar{\cH}^{i}$) is the spectral number flux density of neutrinos (antineutrinos), while the second moment $\cK^{ij}$ ($\bar{\cK}^{ij}$) is proportional to the spectral pressure tensor of neutrinos (antineutrinos).  
Since the neutrino distribution function is bounded between zero and one, it can be shown that the moments must satisfy the bounds \citep{lareckiBanach_2011}
\begin{equation}
  0\le\JNe\le 1
  \quand
  |\bcH|\le\big(1-\JNe\big)\,\JNe\:,
  \label{eq:momentBounds}
\end{equation}
where $\bcH:=(\cH_{1},\cH_{2},\cH_{3})$ and $|\bcH|=\sqrt{\cH_{i}\cH^{i}}$, with $\cH_{i}$ associated to $\cH^{k}$ via the spatial metric $\gamma_{ik}$, i.e., $\cH_{i}=\gamma_{ik}\cH^{k}$.  
(For notational convenience in this section we sometimes use Einstein's summation convention where repeated Latin indices imply summation from $1$ to $3$.)  
Bounds equivalent to those in Eq.~\eqref{eq:momentBounds} also hold for the antineutrino moments.

Taking the zeroth and first moments of Eq.~\eqref{eq:neutrino_transport} leads to 
\begin{subequations}\label{eq:momentmodel_abstract}
\begin{align}
  \frac{1}{\SOL}\p_{t}{\JNe} &+ \f{1}{4\pi}\int_{\bbS^{2}} \cT(\fNe) \,d\omegaNu
  = \f{1}{4\pi}\int_{\bbS^{2}} \cC(\fNe,\fNeb) \,d\omegaNu\:,\\
  \frac{1}{\SOL}\p_{t}{\cH_{j}} &+ \f{1}{4\pi}\int_{\bbS^{2}} \cT(\fNe) \ell_{j}\,d\omegaNu
  = \f{1}{4\pi}\int_{\bbS^{2}} \cC(\fNe,\fNeb) \ell_{j}\,d\omegaNu\:,
\end{align}
\end{subequations}
which, after plugging in the definitions of $\cT$ and $\cC$ in Eqs.~\eqref{eq:transportOperator} and \eqref{eq:collisionTerms}, results in the moment equations for the number density and number flux
\begin{subequations}	\label{eq:momentmodel}
\begin{align}\label{eq:momentNumber}
\frac{1}{\SOL}\p_{t}{\JNe} + \frac{1}{\sqrt{\gamma}}\sum_{i=1}^3 \p_{x^i} \big(\sqrt{\gamma}\, {{\cH}^{i}}\big)
&= {\eta_{\TOTAL}} - {\chi_{\TOTAL}} \,\JNe\:,\\
\label{eq:momentFlux}
\frac{1}{\SOL}\p_{t}{\cH_{j}} + \frac{1}{\sqrt{\gamma}}\sum_{i=1}^3 \p_{x^i}  \big(\sqrt{\gamma}\, {\cK^{i}_{~j}}\big)
&= \f{1}{2}\,\cK^{ik}\,\p_{x^{j}}\gamma_{ik}-({\chi}_{\TOTAL} + \sigma_{\IS})\,\cH_{j}\:,
\end{align}
\end{subequations}
where we have used the facts that the position space flux $\sfrac{\vect{p}}{|\vect{p}|} \fNe=\vect{\ell}(\omegaNu)\fNe$ and that contributions from momentum space fluxes vanish due to boundary conditions.  
(An analogous set of moment equations is derived for antineutrinos.)  
Eqs.~\eqref{eq:momentmodel} hold for general, time-independent curvilinear spatial coordinates encoded in the spatial metric $\gamma_{ij}$, including the commonly used Cartesian, spherical polar, and cylindrical coordinates.  
The metric tensor is used to raise and lower indices on vectors and tensors; e.g., $\cH_{j}=\gamma_{jk}\cH^{k}$, $\cK^{i}_{~j}=\gamma_{jk}\cK^{ik}$.  

\begin{remark}
We note that Eqs.~\eqref{eq:momentmodel} can also be obtained from the non-relativistic (i.e., zero fluid velocity and no gravitational fields) limit of the moment equations in \cite{shibata_etal_2011} and \cite{cardall_etal_2013a} (see also the number conservative two-moment model discussed by \citet{mezzacappa_etal_2020}; their Eqs.~(123) and (125)).  
Moreover, when including relativistic effects, one is, among other things, confronted with choosing appropriate momentum space coordinates.  
The most common (and perhaps the most natural) choice is to use momentum space coordinates in the frame of reference of the inertial observer instantaneously comoving with the fluid (i.e., the comoving frame), as opposed to the so-called laboratory frame \citep[see, e.g.,][]{mihalasMihalas_1999}.  
(In the absence of fluid motion and gravitational fields, which is assumed here, there is no distinction between the comoving and laboratory frames.)  
Specifically, the choice of comoving frame momentum coordinates provides the most straightforward framework for describing neutrino-matter interactions, and this is the choice we intend to make when including relativistic effects in the future.  
However, this choice complicates the advection operator associated with the moment equations, which then includes Doppler and/or gravitational frequency shift terms.  
The inclusion of these terms is beyond the scope of the present paper, but will be considered in a future study.  
\end{remark}

In the right-hand side of Eqs.~\eqref{eq:momentmodel}, the elastic scattering opacity is given by $\sigma_{\IS}=\frac{4\pi\epsilonNu^2}{\SOL(\Planck\SOL)^3}\Phi^{\IS}_0(\epsilonNu)$, and we define the total emissivity and total opacity as
\begin{equation}\label{eq:totalEmissivityandOpacity}
{\eta}_{\TOTAL}(\JNe,\JNeb)
={\chi}\,\JNe_{0} + \eta_{\SC}(\JNe) + \eta_{\TP}(\JNeb)\:,\quand
{\chi}_{\TOTAL}(\JNe,\JNeb)
={\chi} + \chi_{\SC}(\JNe) + \chi_{\TP}(\JNeb)\:,  
\end{equation}
where $\JNe_{0}=\frac{1}{4\pi}\int_{\bbS^2}f_0\,d\omegaNu=f_0$. 
Let $d\tilde{V}_{\epsilonNu}:=4\pi\epsilonNu^{2}d\epsilonNu$ denote the scaled spherical shell energy volume element, then the opacity terms in Eq.~\eqref{eq:totalEmissivityandOpacity} are defined respectively as the scattering emissivity
\begin{equation}\label{eq:NESemissitivity}
\eta_{\SC}(\JNe)
=\frac{1}{\SOL(\Planck\SOL)^3} \int_{\bbR^{+}}\Phi_{0}^{\IN}(\epsilonNu,\epsilonNu')\,\JNe(\epsilonNu')\,d\tilde{V}_{\epsilonNu'},
\end{equation}
the scattering opacity
\begin{equation}\label{eq:NESOpacity}
\chi_{\SC}(\JNe)
=\frac{1}{\SOL(\Planck\SOL)^3} \int_{\bbR^{+}}\big[\,\Phi_{0}^{\IN}(\epsilonNu,\epsilonNu')\,\JNe(\epsilonNu')+\Phi_{0}^{\Out}(\epsilonNu,\epsilonNu')\,\big(1-\JNe(\epsilonNu')\big)\,\big]\,d\tilde{V}_{\epsilonNu'},
\end{equation}
the emissivity due to thermal pair processes
\begin{equation}\label{eq:pairEmissivity}
\eta_{\TP}(\JNeb)
=\frac{1}{\SOL(\Planck\SOL)^3} \int_{\bbR^{+}}\Phi_{0}^{\PROD}(\epsilonNu,\epsilonNu')\,\big(1-\JNeb(\epsilonNu')\big)\,d\tilde{V}_{\epsilonNu'},
\end{equation}
and the opacity due to thermal pair processes
\begin{equation}\label{eq:pairOpacity}
\chi_{\TP}(\JNeb)
=\frac{1}{\SOL(\Planck\SOL)^3} \int_{\bbR^{+}}\big[\,\Phi_{0}^{\PROD}(\epsilonNu,\epsilonNu')\,\big(1-\JNeb(\epsilonNu')\big)+\Phi_{0}^{\ANN}(\epsilonNu,\epsilonNu')\,\JNeb(\epsilonNu')\,\big]\,d\tilde{V}_{\epsilonNu'}.  
\end{equation}
We make the following remarks on Eqs.~\eqref{eq:momentmodel}: (i) the scattering and pair processes opacities depend on $\JNe$ and $\JNeb$, respectively, due to the Fermi blocking factors; (ii) since $\Phi_0^{\op}\geq0$ and $\JNe$, $\JNeb$ are between zero and one, we have $\sigma_{\IS},\eta_{\SC},\chi_{\SC},\eta_{\TP},\chi_{\TP}\ge0$; and (iii) the emissivities and opacities depend on the neutrino energy $\varepsilon$ and local matter states (e.g., density $\rho$, temperature $T$, and electron fraction $Y_{e}$).

To close the two-moment model in Eqs.~\eqref{eq:momentmodel}, we adopt an algebraic closure of the form \citep{levermore_1984}
\begin{equation}\label{eq:moment_closure}
  \cK^{i}_{~j} = \f{1}{2}\,\big[\,\big(1-\psi\big)\,\delta^{i}_{~j}+\big(3\,\psi-1\big)\,\widehat{h}^{i}\,\widehat{h}_{j}\,\big]\,\cJ,
\end{equation}
where $\widehat{h}^{i}=\sfrac{\cH^{i}}{|\bcH|}$ are components of a unit vector parallel to $\bcH$, and the Eddington factor $\psi=\psi(\cJ,h)$ with $h=\sfrac{|\bcH|}{\cJ}$.  
We use the maximum entropy Eddington factor of \cite{cernohorskyBludman_1994}
\begin{equation}
  \psi(\cJ,h) = \f{1}{3} + \f{2\,(1-\cJ)\,(1-2\,\cJ)}{3}\,\Theta\Big(\f{h}{1-\cJ}\Big),
\end{equation}
where the closure polynomial is given by
\begin{equation}\label{eq:closure_polynomial}
  \Theta(x) = \f{1}{5}\,(3-x+3\,x^{2})\,x^{2}.  
\end{equation}
We point out that this closure is based on Fermi-Dirac statistics, and is suitable for designing numerical methods for the two-moment model satisfying the bounds in Eq.~\eqref{eq:momentBounds} (e.g., \cite{CHU201962}).  

For the antineutrino transport equation \eqref{eq:antineutrino_transport}, a two-moment model on the antineutrino number density $\JNeb$ and number flux $\HNeb$ can be derived following similar procedure as in Eqs.~\eqref{eq:momentmodel_abstract}--\eqref{eq:pairOpacity}.
We then close the resulting two-moment model using a closure analogous to Eq.~\eqref{eq:moment_closure}.
To simplify the notation, we denote the neutrino and antineutrino moments as $\bcM:=(\JNe,\HNe, \JNeb, \HNeb)$ and write the coupled two-moment models in operator form as
\begin{equation}\label{eq:MomentModelClosed}
\frac{1}{\SOL}\pd{\vect{\cM}}{t}+\frac{1}{\sqrt{\gamma}}\sum_{i=1}^3\p_{x^i} \,\big(\sqrt{\gamma}\,\vect{\cF}^i(\vect{\cM})\big)\,
=  \vect{\cG}(\vect{\cM}) + \vect{\cC}(\vect{\cM})\:,
\end{equation}
where $\bcF$, $\bcG$, and $\bcC$ are the position flux operator, geometry source operator, and collision operator, respectively. 
In particular, 
\begin{equation}\label{eq:collision_operator}
\bcC(\bcM) := 
\big( {\eta_{\TOTAL}} - {\chi_{\TOTAL}} \,\JNe,\,
({\chi}_{\TOTAL} + \sigma_{\IS})\,\HNe,\,
 {\etab_{\TOTAL}} - {\chib_{\TOTAL}} \,\JNeb,\,
 ({\chib}_{\TOTAL} + \sigmab_{\IS})\,\HNeb \big)\:.
\end{equation}

\subsection{Coupling to the matter equations}
\label{subsec:conservations}

Neutrino-matter interactions mediate exchange of lepton number, momentum, and energy between matter and neutrinos.  
As a first approximation, we assume that the fluid remains static and momentum exchange is ignored.  
Under these assumptions, the matter is described by the mass density $\rho(\vect{x})$ (fixed in time), temperature $T(\vect{x},t)$, and electron fraction $Y_{e}(\vect{x},t)$, and neutrino-matter interactions result in changes to the electron fraction 
\begin{equation}\label{eq:ElectronFractionEvolution} 
\f{1}{\SOL}\, \deriv{Y_{e}}{t} = - \bigg(\f{\baryonmass}{\rho}\bigg) \bigg(\f{1}{(\Planck\SOL)^3}\bigg) \int_{\bbR^{+}} \big( ({\eta_{\TOTAL}} - {\chi_{\TOTAL}} \,\JNe) - ({\etab_{\TOTAL}} - {\chib_{\TOTAL}} \,\JNeb ) \big) \,d\tilde{V}_{\epsilonNu}\:,
\end{equation}
and specific internal energy
\begin{equation}\label{eq:EnergyEvolution}
\f{1}{\SOL} \,\deriv{\epsilon}{t}  = - \bigg(\f{1}{\rho}\bigg) \bigg(\f{1}{(\Planck\SOL)^3}\bigg) \int_{\bbR^{+}}\big( ({\eta_{\TOTAL}} - {\chi_{\TOTAL}} \,\JNe) + ({\etab_{\TOTAL}} - {\chib_{\TOTAL}} \,\JNeb ) \big)\varepsilon\,d\tilde{V}_{\epsilonNu}\:,
\end{equation}
where $\baryonmass$ is the average baryon mass.
The specific internal energy $\epsilon(\vect{x},t)$ is defined such that $\rho\epsilon$ is the internal energy density.
We note that, given any $\rho$ and $Y_e$, the specific internal energy $\epsilon$ is an one-to-one (injective) function of the temperature $T$, i.e., one can map a given $T$ to a unique $\epsilon$, and vice versa; specifically, $(\p\epsilon/\p T)_{\rho,Y_{e}}>0$.

Together with the number density evolution equations (cf. Eq.~\eqref{eq:momentNumber}) for neutrinos and antineutrinos, Eqs.~\eqref{eq:ElectronFractionEvolution} and \eqref{eq:EnergyEvolution} lead to conservation of lepton number 
\begin{equation}\label{eq:LeptonConservation}
\int_D \bigg[ \f{\rho}{\baryonmass}  \deriv{Y_{e}}{t} +  \f{1}{(\Planck\SOL)^3} \deriv{}{t} \int_{\bbR^{+}} \left(\JNe (\epsilonNu) - \JNeb(\epsilonNu)\right) \,d\tilde{V}_{\epsilonNu} \bigg]d\vect{x}
= - \oint_{\partial D} \bigg[\f{\SOL}{(\Planck\SOL)^3} \int_{\bbR^{+}} (\HNe - \HNeb)  \, d\tilde{V}_{\epsilonNu}\bigg] d\vect{x}\:,
\end{equation}
and energy
\begin{equation}\label{eq:EnergyConservation}
 \int_D \bigg[ \rho \, \deriv{\epsilon}{t} +  \f{1}{(\Planck\SOL)^3} \deriv{}{t}  \int_{\bbR^{+}}\left( \JNe (\epsilonNu) + \JNeb (\epsilonNu) \right)\varepsilon\,d\tilde{V}_{\epsilonNu} \bigg] d\vect{x} 
 = - \oint_{\partial D} \bigg[\f{\SOL}{(\Planck\SOL)^3} \int_{\bbR^{+}} (\HNe + \HNeb)\, \epsilonNu\, d\tilde{V}_{\epsilonNu}\bigg] d\vect{x}\:,
\end{equation}
respectively.  
Note that $(\rho/\baryonmass)\,Y_{e}=n_{e}$, is the electron density (technically the electron minus positron density), and that neutrinos and electrons have lepton number $1$, while antineutrinos and positrons have lepton number $-1$.  
Eq.~\eqref{eq:LeptonConservation} implies that the total lepton number in the domain $D$ (left-hand side of Eq.~\eqref{eq:LeptonConservation}) only changes due to fluxes through the domain boundary $\partial D$ (right-hand side of Eq.~\eqref{eq:LeptonConservation}).
A similar conservation statement holds for the total energy in $D$ as given in Eq.~\eqref{eq:EnergyConservation}.

\section{DG-IMEX scheme for solving the moment equations}
\label{sec:DG_IMEX}

\subsection{Nodal Discontinuous Galerkin (DG) space and energy discretization}
\label{subsec:space_energy_DG}

We apply the nodal DG discretization (see, e.g., \cite{hesthavenWarburton_2008} for an overview) to Eq.~\eqref{eq:MomentModelClosed} in space $\vect{x}\in\bbR^3$ and energy $\epsilonNu\in\bbR^+$,
in which a logically Cartesian mesh with coordinate aligned elements is considered.
To derive the discretized equation from the nodal DG method, we first divide the computational domain $D=D_{\vect{x}}\times D_{\epsilonNu}\subset\bbR^3\times\bbR^+$ into a disjoint union $\cK$ of open elements $\vect{K}$, where each element takes the form
\begin{equation}
\vect{K}=\{\,(\vect{x},\epsilonNu) \colon x^{i} \in K^{i} := (\xLo^{i},\xHi^{i}),\,i=1,2,3,\,\epsilonNu \in K^\epsilonNu:=(\epsilonNuL,\epsilonNuH)\,\}
\end{equation}
with $\dx^{i}=\xHi^{i}-\xLo^{i}$ and $\deps = \epsilonNuH-\epsilonNuL$ denoting the side-lengths of $\vect{K}$.
For $i=1,2,3$, the spatial surface elements in direction $x^i$ are denoted as $\tilde{\vect{K}}^{i}=\times_{j\neq i}K^{j}$, with space variables $\tilde{\vect{x}}^{i}:=\{x^{j}\in\vect{x}\colon j\neq i\}$ on $\tilde{\vect{K}}^{i}$, where $\times$ is the Cartesian product operator. 
We use $V_{\vect{K}}$ to denote the proper volume of the element
\begin{equation}
V_{\vect{K}} = \int_{\vect{K}}dV, \quad\text{where}\quad 
dV =  \sqrt{\gamma}\, 
d\vect{x} \,d V_{\epsilonNu}
= \sqrt{\gamma}\, d\vect{x}\,\epsilonNu^2 d\epsilonNu\:.  
\end{equation}

We let the approximation space for the DG method, $\bbV^{N_{\vect{x}},N_{\epsilonNu}}$, be constructed from the tensor product of one-dimensional polynomials of maximal degrees $N_{\vect{x}}$ and $N_{\epsilonNu}$ in space and energy, respectively.
Note that functions in $\bbV^{N_{\vect{x}},N_{\epsilonNu}}$ can be discontinuous across element interfaces.  
The semi-discrete DG problem is to find $\bcM_{\DG}\in\bbV^{N_{\vect{x}},N_{\epsilonNu}}$ (which approximates $\bcM$ in Eq.~\eqref{eq:MomentModelClosed}) such that (cf. \cite{cockburnShu_2001})
\begin{equation}
\begin{alignedat}{2}
&\frac{1}{\SOL}\partial_{t}\int_{\vect{K}}\bcM_{\DG}\,v\,dV
+\sum_{i=1}^3\int_{K^\epsilonNu} \int_{\tilde{\vect{K}}^{i}}
\big(\,
\sqrt{\gamma}\,\widehat{\bcF}^{i}(\bcM_{\DG})\,v\big|_{\xHi^{i}}
-\sqrt{\gamma}\,\widehat{\bcF}^{i}(\bcM_{\DG})\,v\big|_{\xLo^{i}}\,\big)\,d\tilde{\vect{x}}^{i}\,d V_{\epsilonNu}  \\
&\hspace{24pt}
-\sum_{i=1}^3\int_{\vect{K}}\big(\bcF^{i}(\bcM_{\DG})\,\pderiv{v}{x^{i}}\big)\,d V
=\int_{\vect{K}}\bcG(\bcM_{\DG})\,v\,dV+\int_{\vect{K}}\bcC(\bcM_{\DG})\,v\,dV,
\label{eq:semidiscreteDG}
\end{alignedat}
\end{equation}
for all $v\in\bbV^{N_{\vect{x}},N_{\epsilonNu}}$ and all $\vect{K}\in\cK$.
Here the numerical flux approximating the flux on the spatial surface element $\tilde{\vect{K}}^{i}$ is denoted as $\widehat{\bcF}^{i}(\bcM_{\DG})$.
In this work, we consider the Lax-Friedrichs (LF) flux
\begin{equation}
\begin{alignedat}{2}
\widehat{\bcF}^{i}(\bcM_{\DG})|_{x^i} &= \vect{f}^{\mbox{\tiny LF},i}(\bcM_{\DG}|_{x^{i,-}},\bcM_{\DG}|_{x^{i,+}})\\
&:=\f{1}{2}\,\big(\,\bcF^{i}(\bcM_{\DG}|_{x^{i,-}})+\bcF^{i}(\bcM_{\DG}|_{x^{i,+}})-\alpha^{i}\,(\,\bcM_{\DG}|_{x^{i,+}}-\bcM_{\DG}|_{x^{i,-}}\,)\,\big)\:,
\end{alignedat}
\end{equation}
where $\bcM_{\DG}|_{x^{i,\pm}}=\lim_{\delta\to0}\bcM_{\DG}|_{x^i\pm\delta}$ are the evaluations of $\bcM_{\DG}$ at the immediate right/left of $x^{i}$, which thus are functions of $(\tilde{\vect{x}}^i, \epsilonNu, t)$.
The parameter $\alpha^{i}=||\mbox{eig}\big(\partial\bcF^{i}/\partial\bcM\big)||_{\infty}$ is the largest eigenvalue of the flux Jacobian.  
For massless neutrinos, which propagate at the speed of light, we can take $\alpha^{i}=\SOL$ (i.e., the global LF flux).

\begin{figure}[h]
	\captionsetup[subfigure]{justification=centering}
	\subfloat[Local coordinate]
	{\includegraphics[width=0.4\linewidth]{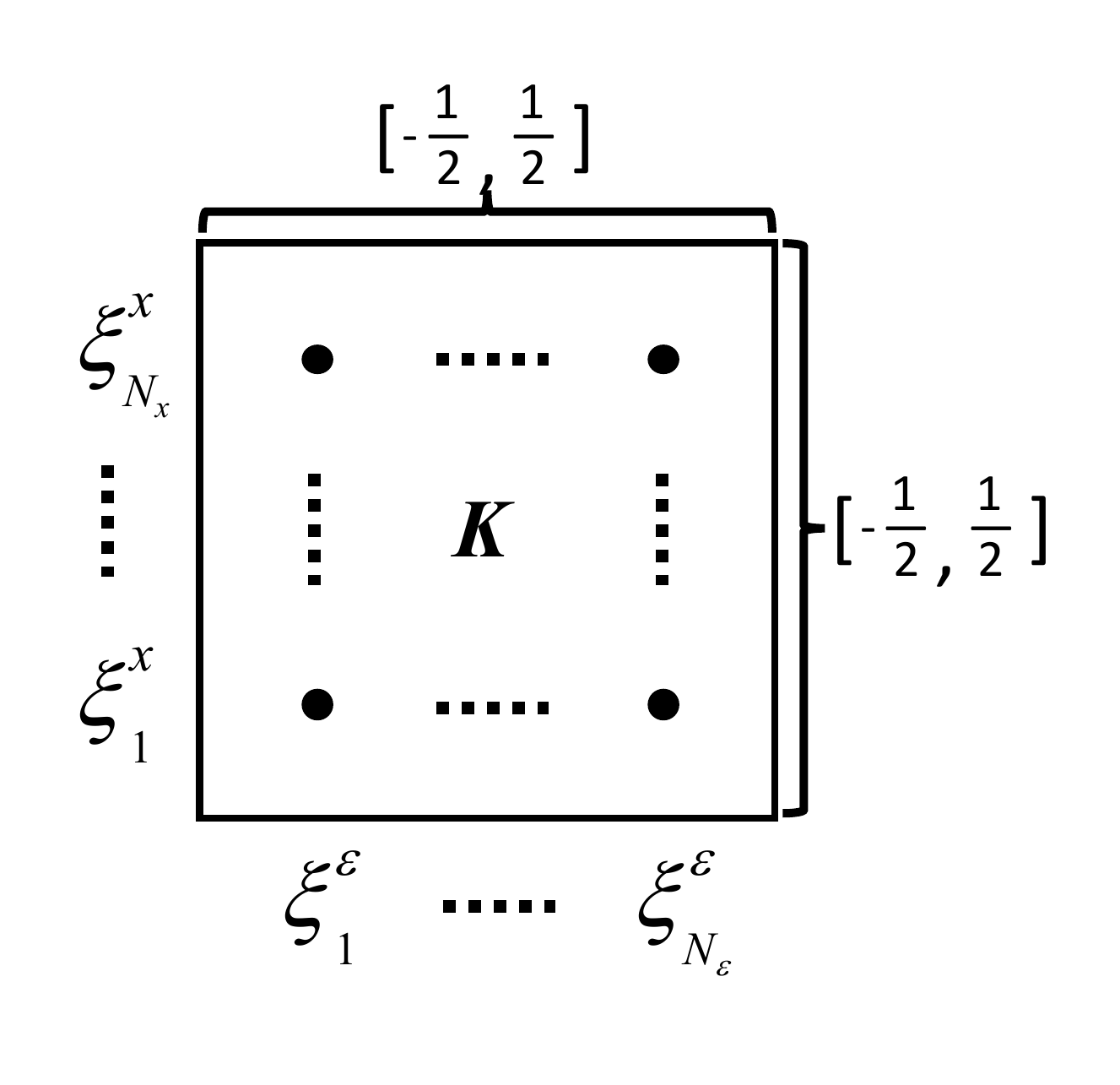}
		\label{fig:LocalElement}}~
	\subfloat[Global coordinate]	
	{\includegraphics[width=0.6\linewidth]{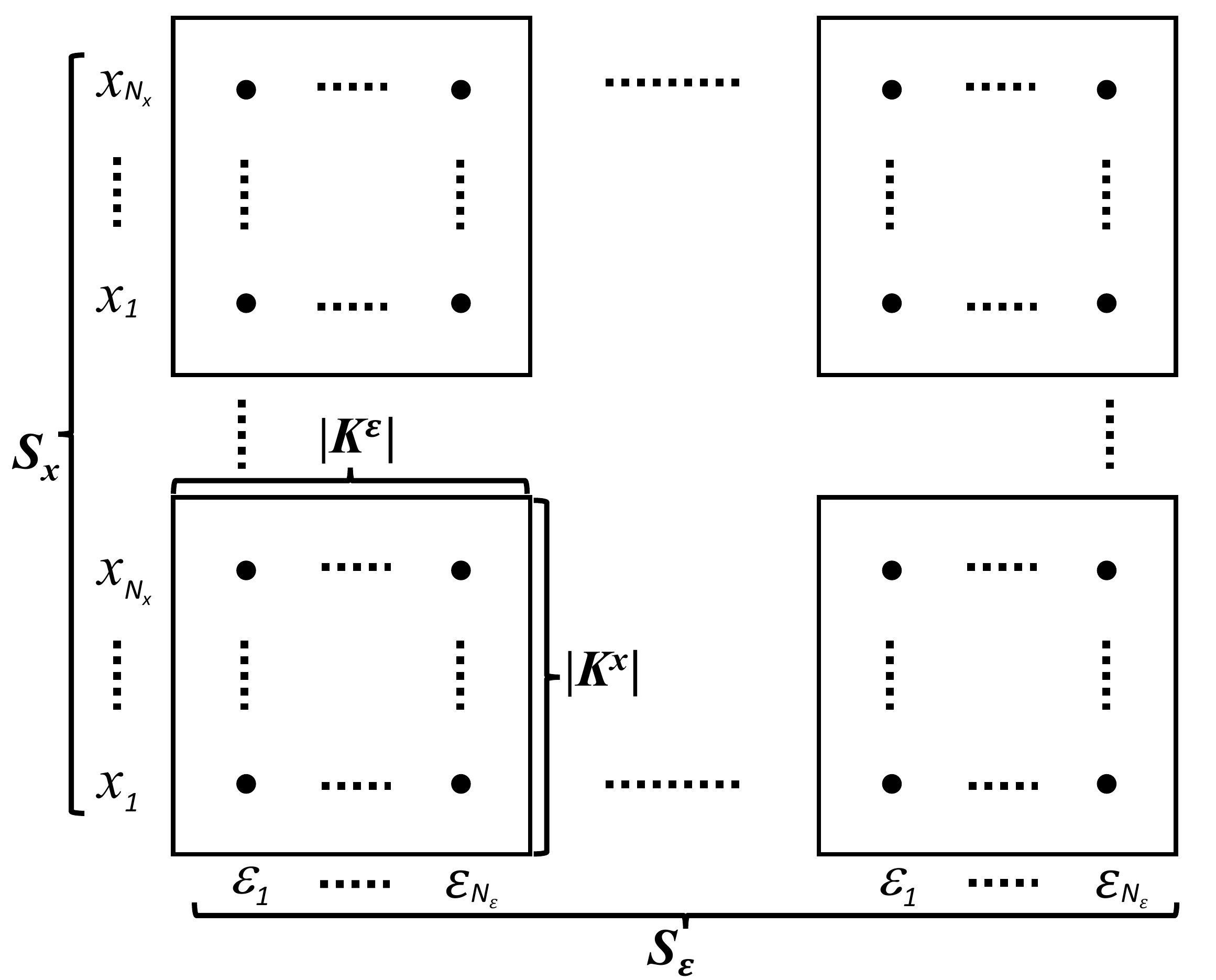}
		\label{fig:GlobalElement}}		
	\caption{Example of nodal DG elements in a computational domain in $\bbR\times\bbR^+$ -- Figure~\ref{fig:LocalElement} shows the interpolation points $\{\xi_j^{i}\}_{j=1}^{N_{\vect{x}}}$ and $\{\xi_j^{\epsilonNu}\}_{j=1}^{N_{\epsilonNu}}$ in a nodal DG element $\vect{K}$ in the local coordinate. Figure~\ref{fig:GlobalElement} shows the union $\cK$ of all elements $\vect{K}$ and the space and energy nodes -- the values of interpolation points in the global coordinate. Figure~\ref{fig:GlobalElement} also illustrates the collection of space nodes $\Sx$ and the collection energy nodes $\Se$, which are defined in Eqs.~\eqref{eq:S_def}--\eqref{eq:SxSe_def}.}	
	\label{fig:DGElementDiagram}	
\end{figure}

In each element $\vect{K}$, we approximate the conserved variables $\bcM$ by 
\begin{equation}
\bcM(\vect{x},\epsilonNu,t)\approx
\bcM_{\DG}(\vect{x},\epsilonNu,t)=
\sum_{\ne=1}^{N_{\epsilonNu}} \sum_{\nx=\vect{1}}^{\bN_{\vect{x}}} \bcM_{\nx,\ne}(t)\,\ell_{\nx}(\vect{x}) \ell_{\ne}(\epsilonNu)\:,
\label{eq:conservedNodalExpansion}
\end{equation}
where $\{\ell_{\ne}\}_{\ne=1}^{N_{\epsilonNu}}$ is chosen to be a collection of Lagrange polynomials in energy of degree $N_{\epsilonNu}$, $\nx:=\{n^1,n^2,n^3\}$ is a multi-index that goes from $\vect{1}=\{1,1,1\}$ to $\bN_{\vect{x}}=\{N_{\vect{x}},N_{\vect{x}},N_{\vect{x}}\}$, and $\{\ell_{\nx}\}_{\nx=\vect{1}}^{\bN_{\vect{x}}}$ is the collection of multidimensional polynomials defined as $\ell_{\nx}(\vect{x}):=\prod_{i=1}^{3}\ell_{n^{i}}(x^{i})$, with $\{\ell_{n^{i}}\}_{n^{i}=1}^{N_{\vect{x}}}$ one-dimensional Lagrange polynomials in $x^i$ of degree $N_{\vect{x}}$.
It then follows from these definitions that $\{\ell_{\nx}(\vect{x})\,\ell_{\ne}(\epsilonNu)\}_{\nx=\vect{1},\ne=1}^{\bN_{\vect{x}},N_{\epsilonNu}}$ forms a basis of $\bbV^{N_{\vect{x}},N_{\epsilonNu}}$ on $\vect{K}$. 
Motivated by the numerical experiments reported in \cite{bassi_etal_2013}, we consider here the Lagrange polynomials with Gauss-Legendre interpolation points (instead of Gauss-Legendre-Lobatto points), e.g., $\ell_\ne$ on interval $K^{\epsilonNu}$ is defined as
\begin{equation}
\ell_{\ne}(\epsilonNu) = \ell^{\textup{loc}}_{N_{\epsilonNu},\ne}(\xi) :=
\prod_{\substack{j=1\\j\neq \ne}}^{N_{\epsilonNu}}\f{\xi-\xi_{j}^{\epsilonNu}}{\xi_{\ne}^{\epsilonNu}-\xi_{j}^{\epsilonNu}}\:, \quad\forall \xi\in I:=[-1/2,1/2]\:,\quad \ne=1,\dots,N_{\epsilonNu}\:.
\end{equation}
The local variable $\xi(\epsilonNu):=\f{\epsilonNu-\epsilonNuC}{\deps}$ with the center $\epsilonNuC:=(\epsilonNuL+\epsilonNuH)/2$, and interpolation points $\{\xi_j^{\epsilonNu}\}_{j=1}^{N_{\epsilonNu}}$ are given by the $N_{\epsilonNu}$-point Gauss-Legendre quadrature abscissas on $I$. 
On interval $K^{i}$, $\ell_{n^{i}}$ is defined analogously as $\ell_{n^{i}}(x^i) = \ell^{\textup{loc}}_{N_{\vect{x}},n^{i}}(\xi)$, with Gauss-Legendre interpolation points $\{\xi_j^{i}\}_{j=1}^{N_{\vect{x}}}$ on $I$. 
With this choice of basis, it follows that the expansion in Eq.~\eqref{eq:conservedNodalExpansion} becomes a nodal representation of $\bcM_{\DG}$, i.e., $\bcM_{\DG}(\vect{x}_{\nx},\epsilonNu_\ne,t)=\bcM_{\nx,\ne}(t)$, where $\vect{x}_\nx=\{x^{1}_{n^1},x^{2}_{n^2},x^{3}_{n^3}\}$ and $\epsilonNu_\ne$ are the global space and energy variables corresponding to the local interpolation points $\{\xi_{j}^{i}\}_{j=1}^{N_{\vect{x}}}$  and $\{\xi_{j}^{\epsilonNu}\}_{j=1}^{N_{\epsilonNu}}$ on element $\vect{K}$, respectively.
Figure~\ref{fig:DGElementDiagram} shows an example of nodal DG elements in a reduced space $\vect{x}\in\bbR$ and energy $\epsilonNu\in\bbR^+$ with the interpolations points in both local and global coordinates.

We then follow the standard practice and approximate Eq.~\eqref{eq:semidiscreteDG} by a semi-discrete system consisting of $(N_{\vect{x}})^3 N_{\epsilonNu}$ equations, each on a nodal value $\bcM_{\nx,\ne}$, $\nx=\vect{1},\dots,\bN_{\vect{x}}$, $\ne=1,\dots,N_{\epsilonNu}$.
To derive these equations, we approximate the integrals in Eq.~\eqref{eq:semidiscreteDG} using the $N_{\vect{x}}$-points and $N_{\epsilonNu}$-points Gauss-Legendre quadrature rules in the space and energy, respectively, with the associated weights $\{w_{n^i}\}_{n^i=1}^{N_{\vect{x}}}$ and $\{w_\ne\}_{\ne=1}^{N_{\epsilonNu}}$ normalized such that $\sum_{n^i=1}^{N_{\vect{x}}} w_{n^i} = 1$ and $\sum_{\ne=1}^{N_{\epsilonNu}} w_\ne = 1$.
Specifically, for $v(\vect{x},\epsilonNu) = \ell_\nx(\vect{x})\ell_\ne(\epsilonNu)$, we have
\begin{equation}\label{eq:DGterm1}
\pd{}{t}\int_{\vect{K}}\bcM_{\DG}\,v\,dV
\approx w_{\nx,\ne}\,\epsilonNu_{\ne}^2\,\sqrt{\gamma}_{\nx}\,|\vect{K}|\,\pd{}{t}\bcM_{\nx,\ne}\:,
\end{equation}
and
\begin{equation}\label{eq:DGterm2}
\int_{\vect{K}}\big(\bcG(\bcM_{\DG})+\bcC(\bcM_{\DG})\big)\,v\,dV\approx
w_{\nx,\ne}\,\epsilonNu_{\ne}^2\,\sqrt{\gamma}_{\nx}\,|\vect{K}|\,\big(\bcG(\bcM_{\DG})_{\nx,\ne}+\bcC(\bcM_{\DG})_{\nx,\ne}\big)\:,
\end{equation}
where $|\vect{K}|=(\prod_{i=1}^{3}\dx^{i})\deps$, $w_{\nx,\ne} = (\prod_{i=1}^{3}w_{n^i})w_\ne$, and $\gamma_{\nx}$ denotes the value of $\gamma$ at $\vect{x}_{\nx}$.
For the surface integrals and the `volume terms', we denote the index $\tnxi:=\{n^j\in \nx\colon j\neq i\}$ and obtain, e.g., 
\begin{equation}\label{eq:DGterm3}
\int_{K^\epsilonNu}\int_{\tilde{\vect{K}}^{i}}\sqrt{\gamma}\,\widehat{\bcF}^{i}(\bcM_{\DG})\,v|_{\xHi^{i}}\,d\tilde{\vect{x}}^{i}\, d V_{\epsilonNu}
\approx w_{\tnxi,\ne}\,\epsilonNu_{\ne}^2\,\big(\sqrt{\gamma}_{\tnxi}\,
\widehat{\bcF}^{i}_{\tnxi,\ne}\,\ell_{n^{i}}\big)\bigr\rvert_{\xHi^{i}}\,|\tilde{\vect{K}}^{i}|\,|K^{\epsilonNu}|\:,
\end{equation}
and 
\begin{equation}\label{eq:DGterm4}
\int_{\vect{K}}\bcF^{i}(\bcM_{\DG})\,\pderiv{v}{x^{i}}\,dV
\approx w_{\tnxi,\ne}\,\epsilonNu_{\ne}^2\,\sum_{k^{i}=1}^{N_{\vect{x}}}w_{k^{i}}\,\big(\sqrt{\gamma}_{\tnxi}\,
\bcF^{i}_{\tnxi,\ne}\,\pderiv{\ell_{k^{i}}}{x^{i}}\big)\bigr\rvert_{x_{k^{i}}^{i}}\,|\tilde{\vect{K}}^{i}|\,|K^{\epsilonNu}|\:,
\end{equation}
where $|K^{\epsilonNu}|=\deps$, $|\tilde{\vect{K}}^{i}|=\prod_{j\neq i}\dx^{j}$, $w_{\tnxi,\ne} = (\prod_{j\neq i}w_{n^j})w_\ne$, ${\gamma}_{\tnxi}$ is the evaluations of ${\gamma}$ at $\tilde{\vect{x}}^{i}_{\tnxi}$, and $\widehat{\bcF}^{i}_{\tnxi,\ne}$ and ${\bcF}^{i}_{\tnxi,\ne}$ are the evaluations of $\widehat{\bcF}^{i}(\bcM_{\DG})$ and ${\bcF}^{i}(\bcM_{\DG})$ at $(\tilde{\vect{x}}^{i}_{\tnxi},\epsilonNu_\ne)$, respectively.
Here ${\gamma}_{\tnxi}$, $\widehat{\bcF}^{i}_{\tnxi,\ne}$, and ${\bcF}^{i}_{\tnxi,\ne}$ are functions of $x^{i}$.

Plugging the terms in Eqs.~\eqref{eq:DGterm1}--\eqref{eq:DGterm4} into Eq.~\eqref{eq:semidiscreteDG} and dividing through by $w_{\nx,\ne}\,\sqrt{\gamma}_{\nx}\,\epsilonNu_{\ne}^2\,|\vect{K}|$ leads to the semi-discrete form of the moment equations when the test function is $v = \ell_\nx \,\ell_\ne$:
\begin{align}\label{eq:semidiscreteDiscretized}
\f{1}{\SOL}\pd{\bcM_{\nx,\ne}}{t}
&=
-\f{1}{\sqrt{\gamma}_{\nx}}\sum_{i=1}^{3} 
\f{1}{w_{n^{i}}\Delta x^{i}}\,
\big(\sqrt{\gamma}_{\tnxi}\,\widehat{\bcF}^{i}_{\tnxi,\ne}\,\ell_{n^{i}}\big)\bigr\rvert^{\xHi^{i}}_{\xLo^{i}}\,
 \\
&\hspace{12pt}
+\f{1}{\sqrt{\gamma}_{\nx}}\sum_{i=1}^{3}\f{1}{w_{n^{i}}\Delta x^{i}}
\sum_{k^{i}=1}^{N_{\vect{x}}}w_{k^{i}}\,\big(\sqrt{\gamma}_{\tnxi}\,
\bcF^{i}_{\tnxi,\ne}\,\pderiv{\ell_{k^{i}}}{x^{i}}\big)\bigr\rvert_{x_{k^{i}}^{i}}\,
+\bcG(\bcM_{\DG})_{\nx,\ne}+\bcC(\bcM_{\DG})_{\nx,\ne}\:,
\nonumber
\end{align}
for $\nx=\vect{1},\dots,\bN_{\vect{x}}$ and $\ne = 1,\dots, N_{\epsilonNu}$ in all $\vect{K}\in\cK$.
Eq.~\eqref{eq:semidiscreteDiscretized} defines the spatial and energy discretization of the moment equations and provides the basis for implementation in {\thornado}.
Before discussing the time discretization, we further simplify Eq.~\eqref{eq:semidiscreteDiscretized} utilizing the structure of the collision operator $\bcC$.
Here we introduce the collection of space and energy nodes from all elements, denoted as
\begin{equation}\label{eq:S_def}
S
:=\cup_{\vect{K}\in\cK} \{(\vect{x}_{\nx},\epsilonNu_{\ne})\in \vect{K}\colon \nx = \vect{1},\dots, \vect{N}_{\vect{x}},\,\ne = 1,\dots, N_{\epsilonNu}\}\:,
\end{equation}
and we define the spatial component of $S$ as $\Sx$ and the energy component of $S$ as $\Se$, i.e.,
\begin{equation}\label{eq:SxSe_def}
\vect{x}\in\Sx\:,\,\, \epsilonNu\in\Se \quad \text{ if and only if } \quad (\vect{x},\epsilonNu) \in S\:.
\end{equation}
A simplified illustration of $\Sx$ and $\Se$ is given in Figure~\ref{fig:GlobalElement}, in which $\vect{x}\in\bbR$.
At each $\vect{x}_{\mx}\in\Sx$, the nodal value of $\bcM_{\DG}$ on all $\epsilonNu\in\Se$ is then denoted as 
$
\bcM_{\mx}(t):=\{\bcM_{\DG}(\vect{x}_{\mx},\epsilonNu_{\me},t)\colon \epsilonNu_{\me} \in \Se\}\:.
$
With these notations, we write Eq.~\eqref{eq:semidiscreteDiscretized} in the operator form in the remainder of this paper as
\begin{equation}\label{eq:DiscretizedOperatorForm}
\f{1}{\SOL}\pd{\bcM_{\mx}}{t}
=\FMh(\bcM_{\DG})_{\mx} + \GMh(\bcM_{\DG})_{\mx} + \CMh(\bcM_{\mx})\:,\quad \forall \mx\text{ such that }\vect{x}_{\mx}\in\Sx\:,
\end{equation}
where $\FMh$, $\GMh$, and $\CMh$ denote respectively the discrete position space flux operator, the discrete geometry source, and the discrete collision operator.
Here the collision term $\CMh(\bcM_{\mx})$ depends only on $\bcM_{\mx}$ instead of the full discretized solution $\bcM_{\DG}$, since the physical interactions modeled in the collision operator (see Eq.~\eqref{eq:collisionTerms}) are independent of the position $\vect{x}$, while coupled in the energy domain.
Specifically, let $(\bJNe_{\mx},\bHNe_{\mx},\bJNeb_{\mx},\bHNeb_{\mx})$ denote the discretized moments $\bcM_{\mx}$, then it follows from Eq.~\eqref{eq:collision_operator} that
\begin{equation}\label{eq:discrete_collision}
\CMh(\bcM_{\mx}) := 
( {\etah_{\TOTAL}} - {\chih_{\TOTAL}} \,\bJNe_{\mx},\,
({\chih}_{\TOTAL} + \sigmah_{\IS})\,\bHNe_{\mx},\,
{\etabh_{\TOTAL}} - {\chibh_{\TOTAL}} \,\bJNeb_{\mx},\,
({\chibh}_{\TOTAL} + \sigmabh_{\IS})\,\bHNeb_{\mx} )\:,
\end{equation}
where $\sigmah_{\IS}$ and $\sigmabh_{\IS}$ are the values of $\sigma_{\IS}$, $\sigmab_{\IS}$ on the energy nodes $\Se$, respectively, and $\etah_{\TOTAL}$, $\chih_{\TOTAL}$, $\etabh_{\TOTAL}$, and $\chibh_{\TOTAL}$ are the discrete counterparts of $\eta_{\TOTAL}$, $\chi_{\TOTAL}$, $\etab_{\TOTAL}$, and $\chib_{\TOTAL}$.
These discrete opacities are given by replacing the energy integrals in Eqs.~\eqref{eq:totalEmissivityandOpacity}--\eqref{eq:pairOpacity} with the numerical integrals using $\Se$ as quadrature points, for example,
\begin{equation}\label{eq:discreteOpacity}
\eta_{\SC}(\JNe)
=\frac{1}{\SOL(\Planck\SOL)^3}\int_{\bbR^{+}}\Phi_{0}^{\IN}(\epsilonNu,\epsilonNu')\,\JNe(\epsilonNu')\,d\tilde{V}_{\epsilonNu'}\approx
\frac{1}{\SOL(\Planck\SOL)^3}\sum_{\me=1}^{|\Se|} w_{\me}^{\epsilonNu} \Phi_{0}^{\IN}(\epsilonNu,\epsilonNu_{\me})\,\bJNe_{\me}
=:\etah_{\SC}(\bJNe)\:,
\end{equation}
where $w_{\me}^{\epsilonNu}$ denotes the weight associated to energy node $\epsilonNu_{\me}$, and $|\Se|$ denotes the total number of energy nodes in $\Se$.
Specifically, for $\epsilonNu_{\me}\in\Se$ in some energy element $K_\epsilonNu$, 
\begin{equation}\label{eq:weights}
w_{\me}^{\epsilonNu}:=  4\pi\epsilonNu_{\me}^2 w_{\ne} |K_\epsilonNu|\:,
\end{equation}
where $w_{\ne}$ is the local Gauss-Legendre weight at $\epsilonNu_{\me}$ in $K_\epsilonNu$ defined earlier in this section.

Finally, we apply this same nodal DG discretization on the electron fraction and specific internal energy evolution equations \eqref{eq:ElectronFractionEvolution}--\eqref{eq:EnergyEvolution}.
Augmenting Eq.~\eqref{eq:DiscretizedOperatorForm} to the resulting semidiscrete equations then gives 
\begin{equation}\label{eq:DiscretizedOperatorFormFull}
\f{1}{\SOL}\pd{\bcV_{\mx}}{t}
=\Fh(\bcV_{\DG})_{\mx} + \Gh(\bcV_{\DG})_{\mx} + \Ch(\bcV_{\mx})\:,\quad \forall \mx\text{ such that }\vect{x}_{\mx}\in\Sx\:,
\end{equation}
where ${\bcV}:=(Y_e,\epsilon,\bcM)$, ${\bcV}_{\DG}:=(Y_{e,\DG},\epsilon_{\DG},\bcM_{\DG})$ is the fully discretized version of $\bcV$, and ${\bcV}_{\mx}$ denotes the nodal value of ${\bcV}_{\DG}$ at point $\vect{x}_{\mx}\in\Sx$.
Here the operators $\Fh:=(\mathsf{0},\mathsf{0},\FMh)$, $\Gh:=(\mathsf{0},\mathsf{0},\GMh)$, and $\Ch$ is given by the discrete version of the right-hand sides in Eqs.~\eqref{eq:ElectronFractionEvolution}--\eqref{eq:EnergyEvolution} and $\CMh$.
Specifically, in $\Ch$, the energy integrals in Eqs.~\eqref{eq:ElectronFractionEvolution}--\eqref{eq:EnergyEvolution} are evaluated using a quadrature with the energy nodes $\epsilonNu_{\me}\in\Se$ as abscissas and weights defined in Eq.~\eqref{eq:weights}.

\begin{remark}
In multidimensional CCSN simulations, the use of curvilinear coordinates, e.g., spherical-polar coordinates, suffers from an excessively stringent Courant-Friedrichs-Lewy (CFL) time step restriction due to singularities at the origin and the poles (see, e.g., \citet[Section~2.1.1]{muller_2020} and references therein). 
Although these singularities are not an issue in the spherically symmetric CCSN models considered here, they will become a problem when extending the nodal DG scheme to multidimensional CCSN models.
Several techniques have been developed to address this issue, such as mesh coarsening \citep{skinner_etal_2019}, element averaging/merging \citep{Asaithambi_2017,muller_2019}, and spectral filtering \citep{muller_2019}. 
For future multidimensional simulations, we will consider (i) adopting one of the aforementioned techniques or (ii) using Cartesian coordinates in combination with adaptive mesh refinement. 
\end{remark}

\subsection{Implicit-explicit time integration scheme}
\label{subsec:IMEX}

An implicit-explicit (IMEX) time integration scheme \citep{ascher_etal_97,pareschiRusso_2005} is considered here for solving the two-moment model in Eq.~\eqref{eq:MomentModelClosed}. 
When applied to transport equations with collision terms, IMEX schemes usually handle the collision term with an implicit method, while applying an explicit method on the advection term \citep{Hu-Shu-Zhang-2018} (see also \cite{oConnor_2015,just_etal_2015,kuroda_etal_2016,skinner_etal_2019} for applications of IMEX-type schemes to neutrino transport).
This approach relaxes the excessive time-step restriction from an explicit and stiff collision term, and avoids the spatially coupled nonlinear solves from an implicit advection term.
While the class of IMEX schemes considered in this paper is detailed in \cite{CHU201962}, we include it in the following paragraph for completeness. 
We also stress that even though the nonlinear solution strategies given in Section~\ref{sec:techniques} are motivated from the implicit part of this class of IMEX schemes, they are general enough to be used with any IMEX scheme that treats the collision term implicitly.

To perform time integration, we discretize the time interval $[t_0,\,t_\textup{f}]$ into $N$ time steps $t_0 = t^0 < t^1 < \dots < t^N = t_\text{f}$ and denote $\bcV_{\mx}(t^n)$ as $\bcV_{\mx}^n$, $n=1,\dots,N$.
At each spatial node $\vect{x}_{\mx}$ and time $t^n$, the IMEX scheme integrates the semi-discrete equation \eqref{eq:DiscretizedOperatorFormFull} in time via
\begin{subequations}\label{eq:imexXNodes}
\begin{align}
&\bcV_{\mx}^{(0)} = \bcV_{\mx}^{n},  \\
\label{eq:imexStages}
&\bcV_{\mx}^{(i)} = \sum_{j=0}^{i-1}c_{ij}\,\bcV_{\mx}^{(ij)}
+a_{ii}\,{\SOL}\,\dt\,\Ch(\,\bcV_{\mx}^{(i)}\,), \quad i=1,\ldots,s,\\
&\bcV_{\mx}^{n+1} = \bcV_{\mx}^{(s)},  
\end{align}
\end{subequations}
where $s$ is the number of stages, the parameters $c_{ij}\ge0$, $\sum_{j=0}^{i-1}c_{ij}=1$, $a_{ii}>0$, and $\bcV_{\mx}^{(ij)}$ is given by an explicit update
\begin{equation}\label{eq:imexExplicit}
\bcV_{\mx}^{(ij)}
=\bcV_{\mx}^{(j)}+\hat{c}_{ij}\,{\SOL}\,\dt\,\big(\,\Fh(\bcV^{(j)}_{\DG})_{\mx} + \Gh(\bcV^{(j)}_{\DG})_{\mx}\big)
\end{equation}
with parameters $\hat{c}_{ij}\ge0$. 
Thus, in each time step, the $s$-stage IMEX scheme requires $s$ evaluations of the discrete flux and geometry operators $\Fh$ and $\Gh$, and $s$ inversions of the discrete collision operator $\Ch$. 
Here the number of evaluations of $\Fh$ and $\Gh$ is identical to the number of stages due to the fact that, by reusing values of $\Fh$ and $\Gh$ from earlier stages, each stage only requires one additional evaluation of $\Fh$ and $\Gh$.
In general, the inversion of $\Ch$ is the dominant cost in the IMEX scheme.
In the remainder of this section, we present the details of the nonlinear system arising from inverting $\Ch$.

At each stage $i$ of the IMEX scheme, Eq.~\eqref{eq:imexStages} can be considered as the nonlinear system
\begin{equation}\label{eq:IMStep}
\bcV_{\mx}^{\new}
=\bcV_{\mx}^{\old}
+\tau\,\Ch(\,\bcV_{\mx}^{\new}\,)\:,
\end{equation}
where $\tau\,:=a_{ii}\,{\SOL}\,\dt>0$ denotes the effective time step, $\bcV_{\mx}^{\old}$ denotes the weighted sum of explicit updates $\bcV_{\mx}^{(ij)}$, and $\bcV_{\mx}^{\new}$ denotes the unknown nodal values to be solved.

Let $(\mathsf{Y}_{e,\mx}, \upepsilon_{\mx})$ denote the nodal value of $(Y_e, \epsilon)$ at $\vect{x}_{\mx}$, and let $(\bJNe_{\mx},\bHNe_{\mx},\bJNeb_{\mx},\bHNeb_{\mx})$ denote the discrete moment $\bcM_{\mx}$, which collects values of moments $(\JNe,\HNe,\JNeb,\HNeb)$ at $\vect{x}_{\mx}$ on all energy nodes in $\Se$.
Eq.~\eqref{eq:IMStep} can then be considered as a nonlinear system on $\bcV_{\mx}:=(\mathsf{Y}_{e,\mx}, \upepsilon_{\mx}, \bJNe_{\mx},\bHNe_{\mx},\bJNeb_{\mx},\bHNeb_{\mx})$.
For notational simplicity, we suppress all $\mx$ subscripts when denoting the nodal values of electron fraction, specific internal energy, and moments in the remainder of this paper.
It follows from the definition of $\Ch$ that, at each $\vect{x}_\mx$, the resulting nonlinear system is given by
\begin{subequations}\label{eq:IMAll}
\begin{align}
\label{eq:IMYe}
\mathsf{Y}_e^{\new}
&= 	\mathsf{Y}_e^{\old} - \tau \, {\textstyle \f{\baryonmass}{\rho}  \,\sum_{\me=1}^{|\Se|} } w_{\me}^{(2)} \big( ({\etah_{\TOTAL}} - {\chih_{\TOTAL}} \,\bJNe^{\new}_{\me}) - ({\etabh_{\TOTAL}} - {\chibh_{\TOTAL}} \,\bJNeb^{\new}_{\me} ) \big) \:,\\
\label{eq:IMeps}
{\upepsilon}^{\new}
&= 	{\upepsilon}^{\old} - \tau \, {\textstyle \f{1}{\rho} \, \sum_{\me=1}^{|\Se|} } w_{\me}^{(3)} \big( ({\etah_{\TOTAL}} - {\chih_{\TOTAL}} \,\bJNe^{\new}_{\me}) + ({\etabh_{\TOTAL}} - {\chibh_{\TOTAL}} \,\bJNeb^{\new}_{\me} ) \big)\:,\\
\label{eq:IMJ}
{\bJNe}^{\new}
&= 	{\bJNe}^{\old} + \tau \big( {\etah_{\TOTAL}} - {\chih_{\TOTAL}} \,\bJNe^{\new} \big)\:,\\
\label{eq:IMJbar}
{\bJNeb}^{\new} 
&= 	{\bJNeb}^{\old} + \tau \big( {\etabh_{\TOTAL}} - {\chibh_{\TOTAL}} \,\bJNeb^{\new} \big)\:,\\
\label{eq:IMH}
{\bHNe}^{\new}
&={\bHNe}^{\old} - \tau ({\chih}_{\TOTAL} + {\sigmah}_{\IS})\,\bHNe^{\new}\:,\\
\label{eq:IMHbar}
{\bHNeb}^{\new}
&={\bHNeb}^{\old} - \tau ({\chibh}_{\TOTAL} + {\sigmabh}_{\IS})\,\bHNeb^{\new}\:.
\end{align}
\end{subequations}
In Eqs.~\eqref{eq:IMYe} and \eqref{eq:IMeps}, a quadrature is used to evaluate the energy integrals in Eqs.~\eqref{eq:ElectronFractionEvolution} and \eqref{eq:EnergyEvolution}, as described when defining $\Ch$ in Eq.~\eqref{eq:DiscretizedOperatorFormFull}.
Here $(\bJNe_{\me},\bJNeb_{\me})$ denotes the value of $(\bJNe,\bJNeb)$ at energy node $\epsilonNu_{\me}\in\Se$. 
The physical constants are absorbed into the weights, i.e., 
\begin{equation}
w_{\me}^{(2)}:= \f{1}{(\Planck\SOL)^3}  w_{\me}^{\epsilonNu} \:,\quand
w_{\me}^{(3)}:= \f{1}{(\Planck\SOL)^3}  \epsilonNu_{\me} w_{\me}^{\epsilonNu}\:,
\end{equation}
with $w_{\me}^{\epsilonNu}$ defined in Eq.~\eqref{eq:weights}.

We take a two-step approach to solve Eq.~\eqref{eq:IMAll}, which first solves the fully coupled nonlinear system in Eqs.~\eqref{eq:IMYe}--\eqref{eq:IMJbar}, plug the solution $(\mathsf{Y}_e^{\new},\upepsilon^{\new},\bJNe^{\new}, \bJNeb^{\new})$ into Eqs.~\eqref{eq:IMH}--\eqref{eq:IMHbar} to compute $({\chih}_{\TOTAL},{\sigmah}_{\IS},{\chibh}_{\TOTAL}, {\sigmabh}_{\IS})$, and then update $(\bHNe^{\new},\bHNeb^{\new})$.  
We note that, once $({\chih}_{\TOTAL},{\sigmah}_{\IS},{\chibh}_{\TOTAL}, {\sigmabh}_{\IS})$ are known, solving Eqs.~\eqref{eq:IMH}--\eqref{eq:IMHbar} is straightforward.
Thus, we focus on the solution procedure of the coupled system in Eqs.~\eqref{eq:IMYe}--\eqref{eq:IMJbar}, where the opacities $(\etah_{\TOTAL},\chih_{\TOTAL},\etabh_{\TOTAL},\chibh_{\TOTAL})$ are functions of $(\mathsf{Y}_e,\upepsilon,\bJNe, \bJNeb)$.
Specifically, while the opacities are written explicitly as functions of $(\bJNe, \bJNeb)$ in Eqs.~\eqref{eq:totalEmissivityandOpacity}--\eqref{eq:pairOpacity}, they also depend on the matter state $(\mathsf{Y}_e,\upepsilon)$ through the opacity kernels $\Phi_{0}^{\IN}$, $\Phi_{0}^{\OUT}$, $\Phi_{0}^{\PROD}$, and $\Phi_{0}^{\ANN}$ in Eqs.~\eqref{eq:NESemissitivity}--\eqref{eq:pairOpacity}.
In a fully implicit approach, these opacities need to be updated in the solution procedure of Eqs.~\eqref{eq:IMYe}--\eqref{eq:IMJbar} in order to remain consistent with $(\mathsf{Y}_e^{\new},\upepsilon^{\new},\bJNe^{\new}, \bJNeb^{\new})$.  

We proceed to investigate various approaches to solve the coupled nonlinear system in Eqs.~\eqref{eq:IMYe}--\eqref{eq:IMJbar}.

\section{Nonlinear solution strategies}
\label{sec:techniques}

In this section, we discuss two approaches for solving the system in Eqs.~\eqref{eq:IMYe}--\eqref{eq:IMJbar}, which couples the evolution of the matter states $(\mathsf{Y}_e,\upepsilon)$ to the neutrino and antineutrino spectral distributions $(\bJNe,\bJNeb)$.
To start, we first rewrite Eqs.~\eqref{eq:IMYe}--\eqref{eq:IMJbar} as
\begin{subequations}\label{eq:CoupledSystem}
	\begin{align}
	\label{eq:CoupledSystem_Ye}	
	\mathsf{Y}_{e}^{\new} 
	&= \mathsf{Y}_{e}^{\old} - {\textstyle \f{\baryonmass}{\rho}\sum_{\me=1}^{|\Se|} } w_{\me}^{(2)} \big(\bJNe^{\new}_{\me} - \bJNeb^{\new}_{\me}\big) +
	 {\textstyle \f{\baryonmass}{\rho}\sum_{\me=1}^{|\Se|} } w_{\me}^{(2)} \big(\bJNe^{\old}_{\me} - \bJNeb^{\old}_{\me} \big)\:,\\
	\label{eq:CoupledSystem_eps}	 
	{\upepsilon}^{\new} 
	&= {\upepsilon}^{\old} - {\textstyle \f{1}{\rho}\sum_{\me=1}^{|\Se|} } w_{\me}^{(3)} \big( \bJNe^{\new}_{\me} + \bJNeb^{\new}_{\me} \big) 
	+  {\textstyle \f{1}{\rho}\sum_{\me=1}^{|\Se|} } w_{\me}^{(3)} \big( \bJNe^{\old}_{\me} + \bJNeb^{\old}_{\me} \big)\:,\\
	\label{eq:CoupledSystem_JNe}
	{\bJNe}^{\new} 
	&= 	{\bJNe}^{\old} + \tau  \,{\etah_{\TOTAL}}\big(\mathsf{Y}_{e}^{\new}, \upepsilon^{\new},{\bJNe}^{\new},{\bJNeb}^{\new}\big) - \tau\, {\chih_{\TOTAL}}\big(\mathsf{Y}_{e}^{\new}, \upepsilon^{\new},{\bJNe}^{\new},{\bJNeb}^{\new}\big) \,\,\bJNe^{\new} \:,\\
	\label{eq:CoupledSystem_JNeb}
	{\bJNeb}^{\new} 
	&= 	{\bJNeb}^{\old} + \tau \,{\etabh_{\TOTAL}}\big(\mathsf{Y}_{e}^{\new}, \upepsilon^{\new},{\bJNe}^{\new},{\bJNeb}^{\new}\big) - \tau\, {\chibh_{\TOTAL}}\big(\mathsf{Y}_{e}^{\new}, \upepsilon^{\new},{\bJNe}^{\new},{\bJNeb}^{\new}\big) \,\,\bJNeb^{\new} \:,
	\end{align}
\end{subequations}
with unknowns $\mathsf{Y}_{e}^{\new}\in\bbR$, $\upepsilon^{\new}\in\bbR$, ${\bJNe}^{\new}\in\bbR^{|\Se|}$, and ${\bJNeb}^{\new}\in\bbR^{|\Se|}$.
Here Eqs.~\eqref{eq:CoupledSystem_Ye}--\eqref{eq:CoupledSystem_eps} are derived by substituting Eqs.~\eqref{eq:IMJ}--\eqref{eq:IMJbar} into the right-hand sides of Eqs.~\eqref{eq:IMYe}--\eqref{eq:IMeps} to remove the explicit dependency on opacities, and Eqs.~\eqref{eq:CoupledSystem_JNe}--\eqref{eq:CoupledSystem_JNeb} are identical to Eqs.~\eqref{eq:IMJ}--\eqref{eq:IMJbar}, with explicit expression of the dependency of opacities $(\etah_{\TOTAL},\chih_{\TOTAL},\etabh_{\TOTAL},\chibh_{\TOTAL})$ on matter states (through opacity kernels) and neutrino (antineutrino)  distributions.
Specifically,
\begin{equation}\label{eq:discreteTotalOpacity}
{\etah}_{\TOTAL}(\mathsf{Y}_{e}, \upepsilon, {\bJNe},{\bJNeb})
={\chih}\,\bJNe_{0} + \etah_{\SC}(\bJNe) + \etah_{\TP}(\bJNeb)\:,\quand
{\chih}_{\TOTAL}(\mathsf{Y}_{e}, \upepsilon, {\bJNe},{\bJNeb})
={\chih} + \chih_{\SC}(\bJNe) + \chih_{\TP}(\bJNeb)\:,  
\end{equation}
where the discrete opacities $\chih$, $\chih_{\op}$, and $\etah_{\op}$ are computed as in Eq.~\eqref{eq:discreteOpacity}, using opacity kernels $\Phi^{\op}$ evaluated at $(\mathsf{Y}_{e}, \upepsilon)$, and the Fermi-Dirac distribution $\bJNe_0$ evaluated at chemical potential $\mu(\mathsf{Y}_{e}, \upepsilon)$ and matter temperature $T(\mathsf{Y}_{e}, \upepsilon)$.

We propose two approaches for solving the nonlinear system in Eq.~\eqref{eq:CoupledSystem} --  a coupled approach and a nested approach.
The former directly considers Eq.~\eqref{eq:CoupledSystem} as a fully coupled system, while the latter formulates Eq.~\eqref{eq:CoupledSystem} as a nested system with Eqs.~\eqref{eq:CoupledSystem_Ye}--\eqref{eq:CoupledSystem_eps} in the outer layer and Eqs.~\eqref{eq:CoupledSystem_JNe}--\eqref{eq:CoupledSystem_JNeb} in the inner. 
Opacity kernel evaluations, i.e., evaluating $\Phi^{\op}$ at given $(\mathsf{Y}_{e}, \upepsilon)$, are needed when solving Eqs.~\eqref{eq:CoupledSystem_Ye}--\eqref{eq:CoupledSystem_eps}.
Since the tabulated opacity kernels are used, evaluating $\Phi^{\op}$ requires opacity table interpolations, which are the dominant cost in solving Eq.~\eqref{eq:CoupledSystem}.
The nested approach aims to reduce the number of opacity kernel evaluations by giving a better prediction on $(\bJNe, \bJNeb)$ through the inner solver on Eqs.~\eqref{eq:CoupledSystem_JNe}--\eqref{eq:CoupledSystem_JNeb}.

We consider a fixed-point iteration method with Anderson acceleration and Newton's method as the nonlinear system solvers in both the coupled and nested approach.
When solving the systems considered here, fixed-point methods are often more attractive than Newton's method because they (1) do not require the Jacobian matrix, which can be difficult to compute accurately with tabulated opacities; and (2) avoid inversion of dense linear systems.  
However, the rate of convergence can be slower for fixed-point methods than that of Newton-based methods.  
The performance of these two types of solvers on systems arising from each approach is compared in the numerical results reported in Section~\ref{sec:num_results}.
In the following subsections, we state the coupled fixed-point algorithm (section~\ref{subsec:coupled_FP}), the coupled Newton's method (section~\ref{subsec:coupled_Newton}), the nested fixed-point algorithm (section~\ref{subsec:nested_FP}), and the nested Newton's method (section~\ref{subsec:nested_Newton}).

\subsection{Coupled fixed-point algorithm}
\label{subsec:coupled_FP}

To simplify the notation, we denote the matter states as $\bu:=(\mathsf{Y}_{e}^{\new}, \upepsilon^{\new})$, the discretized neutrino and antineutrino distributions as $\bcU:=({\bJNe}^{\new}, {\bJNeb}^{\new})$, and the collection of all the unknowns as $\bU:=(\bu,\bcU)$ in the remainder of the paper.
To formulate the system in Eq.~\eqref{eq:CoupledSystem} as a fixed-point problem, we write it as
\begin{equation}\label{eq:FP}
\bU = \bG(\bU) := \left(\begin{array}{c}
\bg(\bcU)\\\bcG(\bu,\bcU)
\end{array}\right)\:,
\end{equation}
where 
\begin{equation}\label{eq:OuterFP_operator}
\bg(\bcU):=
\left(\begin{array}{c}
- \f{\baryonmass}{\rho}\sum_{\me=1}^{|\Se|} w_{\me}^{(2)} \big(\bJNe^{\new}_{\me} - \bJNeb^{\new}_{\me}\big) + C_{\mathsf{Y}_e} \\
- \f{1}{\rho}\sum_{\me=1}^{|\Se|} w_{\me}^{(3)} \big( \bJNe^{\new}_{\me} + \bJNeb^{\new}_{\me} \big) + C_{\upepsilon}
\end{array}\right)
\end{equation}
with 
$C_{\mathsf{Y}_e} = \mathsf{Y}_{e}^{\old} +
\f{\baryonmass}{\rho}\sum_{\me=1}^{|\Se|} w_{\me}^{(2)} \big(\bJNe^{\old}_{\me} - \bJNeb^{\old}_{\me} \big)$,
$C_{\upepsilon} = {\upepsilon}^{\old} + \f{1}{\rho}\sum_{\me=1}^{|\Se|} w_{\me}^{(3)} ( \bJNe^{\old}_{\me} + \bJNeb^{\old}_{\me} \big)$, and 
\begin{equation}\label{eq:InnerFP_operator}
\bcG(\bu,\bcU):=
\left(\begin{array}{c}
\big( {\bJNe}^{\old} + \tau  \,{\etah_{\TOTAL}}(\bu,\bcU) \big)
\,/\, \big(  1 + \tau\, {\chih_{\TOTAL}}(\bu,\bcU) \big)\\
\big( {\bJNeb}^{\old} + \tau \,{\etabh_{\TOTAL}}(\bu,\bcU) \big)
\,/\, \big( 1 + \tau\, {\chibh_{\TOTAL}}(\bu,\bcU) \big) 
\end{array}\right)\:.
\end{equation}
Here $\bcG$ is an equivalent form of Eqs.~\eqref{eq:CoupledSystem_JNe}--\eqref{eq:CoupledSystem_JNeb}, which ensures that $\bG$ is a contraction map, i.e., the Lipschitz constant of $\bG$ is strictly less than one.

The coupled fixed-point algorithm considers Eq.~\eqref{eq:FP} as a fixed-point problem with unknowns $\bU$, e.g., applying Picard iteration on Eq.~\eqref{eq:FP} leads to
\begin{equation}
\bU^{[k+1]} = \bG(\bU^{[k]})\:,
\end{equation}
where $\bU^{[k]}$ denotes the $k$-th iterate of unknowns $\bU$, starting from an initial guess $\bU^{[0]}$. When $\bG$ is a contraction mapping, Picard iteration guarantees that, as $k\to \infty$, the iterate $\bU^{[k]}$ converges to $\bU = (\mathsf{Y}_{e}^{\new}, \upepsilon^{\new},{\bJNe}^{\new}, {\bJNeb}^{\new})$, the solution to Eq.~\eqref{eq:CoupledSystem}.
Here the opacities $(\etah,\chih,\etabh,\chibh)$ in $\bG$ are updated at each iteration $k$ using $\bU^{[k]}$ and thus are consistent with the solution.
While the Picard iteration guarantees convergence when $\bG$ is a contraction, the convergence could be slow.  
To achieve faster convergence, we implement Anderson acceleration \citep{Anderson-1965,Walker-Ni-2011} to solve Eq.~\eqref{eq:FP}.
Anderson acceleration utilizes information from previous iterations to update the unknowns, which is expected to give faster convergence than Picard iteration, but at a cost of additional memory usage.
Specifically, in iteration $k$, Anderson acceleration on the coupled problem first computes the residual
\begin{equation}
{\br}^{[k]} :=\bG(\bU^{[k]}) - \bU^{[k]}\:,
\end{equation}
then solves a least-squares problem $ \alpha^*:={\displaystyle\argmin_{\alpha\in\bbR^{m_k+1}}} \big\{ \big\|\sum_{i=0}^{m_k} \alpha_i\,\br^{[k-i]}  \big\|_2^2\,\,
\colon\, \sum_{i=0}^{m_k} \alpha_i = 1 \big\} $
with $m_k:=\min\{m,k\}$, and finally updates
\begin{equation}\label{eq:AA_step}
\bU^{[k+1]} = \sum\nolimits_{i=0}^{m_k} \alpha^{*}_{i}\,\bG(\bU^{[k-i]})\:.
\end{equation}
Here the truncation parameter $m\geq0$ is an integer that indicates the ``memory" of Anderson acceleration, i.e., the maximum number of residuals kept in memory.
When $m=0$, the solver reduces to Picard iteration.
For $m>0$, Anderson acceleration updates $\bU$ using a linear combination of the last $m_k$ iterates that leads to the minimum residual.
In the numerical tests in Section~\ref{sec:num_results}, we use $m=2$, which we have found to significantly reduce the number of iterations when compared to Picard.  
For $m=2$, the additional memory required for Anderson acceleration is small since each implicit solve is local in space.  
In addition, the least-squares problem for $\alpha_{i}^{*}$ is small, and can be written out explicitly or solved using LAPACK's DGELS.

\subsection{Coupled Newton's method}
\label{subsec:coupled_Newton}

The other solver we considered for the nonlinear coupled system in Eq.~\eqref{eq:CoupledSystem} is Newton's method, which formulates Eq.~\eqref{eq:CoupledSystem} as a root-finding problem 
\begin{equation}\label{eq:Newton}
\left(\begin{array}{c}
\bff(\bu,\bcU)\\\bcF(\bu,\bcU)
\end{array}\right)=:\bF(\bU) = \mathbf{0}\:,
\end{equation}
where 
\begin{equation}\label{eq:OuterNewtonOperator}
\bff(\bu,\bcU):=
\left(\begin{array}{c}
\bff_{\mathsf{Y}_{e}}(\bu,\bcU) \\ \bff_{\upepsilon}(\bu,\bcU)
\end{array}\right):=
\left(\begin{array}{c}
\mathsf{Y}_{e}^{\new} + \f{\baryonmass}{\rho}\sum_{\me=1}^{|\Se|} w_{\me}^{(2)} \big(\bJNe^{\new}_{\me} - \bJNeb^{\new}_{\me}\big) - C_{\mathsf{Y}_{e}} \\
\upepsilon^{\new} + \f{1}{\rho}\sum_{\me=1}^{|\Se|} w_{\me}^{(3)} \big( \bJNe^{\new}_{\me} + \bJNeb^{\new}_{\me} \big) - C_{\upepsilon}
\end{array}\right)
\end{equation}
with constants $C_{\mathsf{Y}_{e}}$ and $C_{\upepsilon}$ defined as in Eq.~\eqref{eq:OuterFP_operator}, and 
\begin{equation}\label{eq:InnerNewtonOperator}
\bcF(\bu,\bcU):=
\left(\begin{array}{c}
\bcF_{\bJNe}(\bu,\bcU) \\ \bcF_{\bJNeb}(\bu,\bcU)
\end{array}\right):=
\left(\begin{array}{c}
\big(  1 + \tau\, {\chih_{\TOTAL}}(\bu,\bcU) \big) {\bJNe}^{\new} - \big( {\bJNe}^{\old} + \tau  \,{\etah_{\TOTAL}}(\bu,\bcU) \big)
\\
\big( 1 + \tau\, {\chibh_{\TOTAL}}(\bu,\bcU) \big) {\bJNeb}^{\new} - \big( {\bJNeb}^{\old} + \tau \,{\etabh_{\TOTAL}}(\bu,\bcU) \big)
\end{array}\right)\:.
\end{equation}
Applying Newton's method to solve Eq.~\eqref{eq:Newton} leads to the following update of $\bU$ in iteration $k$, \begin{equation}
\bU^{[k+1]}=\bU^{[k]} + \delta\bU^{[k]}\:,
\end{equation}
where the Newton step is given by
\begin{equation}\label{eq:Newton_step}
\delta\bU^{[k]} = - \bJ(\bU^{[k]})^{-1}\,\bF(\bU^{[k]}), \quad\text{with}\quad \bJ(\bU^{[k]}):=\bF'(\bU^{[k]}).
\end{equation}
Eq.~\eqref{eq:Newton_step} shows that two key components are needed in Newton's method -- evaluating the Jacobian $\bJ$ and solving the linear system for the Newton step.
For the coupled problem in Eq.~\eqref{eq:Newton}, Jacobian evaluation at a given $\hat{\bU}$ requires computing $\pderiv{\bff}{\bu}$, $\pderiv{\bff}{\bcU}$, $\pderiv{\bcF}{\bu}$, and $\pderiv{\bcF}{\bcU}$ at $(\hat\bu,\hat\bcU)=:\hat{\bU}$.
From Eq.~\eqref{eq:OuterNewtonOperator}, it is clear that evaluating $\pderiv{\bff}{\bu}$ and $\pderiv{\bff}{\bcU}$ at $(\hat\bu,\hat\bcU)$ is straightforward with minimal cost.
However, it follows from Eq.~\eqref{eq:InnerNewtonOperator} that evaluating $\pderiv{\bcF}{\bu}$ and $\pderiv{\bcF}{\bcU}$ at $(\hat\bu,\hat\bcU)$ requires gradients of the opacities $(\etah_{\TOTAL},\chih_{\TOTAL},\etabh_{\TOTAL},\chibh_{\TOTAL})$ with respect to the matter state $\bu$ and the neutrino and antineutrino number densities $\bcU$, respectively.
In particular, Eqs.~\eqref{eq:discreteOpacity} and \eqref{eq:totalEmissivityandOpacity}--\eqref{eq:pairOpacity}
imply that, $\pderiv{\Phi^{\op}}{\bu}$, which is the gradient of opacity kernels with respect to the matter state, is involved in the computation of $\pderiv{\bcF}{\bu}(\hat\bu,\hat\bcU)$. 
As discussed earlier, tabulated opacity kernels are considered in this paper. 
Thus, we can only obtain approximate $\pderiv{\Phi^{\op}}{\bu}$ from the tabulated quantities.
Further, when the opacity kernels are not tabulated in terms of $\bu$, the gradient $\pderiv{\Phi^{\op}}{\bu}$  has to be approximated using the chain rule.
The detailed calculations of these kernel derivatives are given in Appendix~\ref{appendix:kernelDerivatives}.
Once the approximate Jacobian is obtained, the linear solve in Eq.~\eqref{eq:Newton_step} is performed via LAPACK's DGESV.

\subsection{Nested fixed-point algorithm}
\label{subsec:nested_FP}

The next approach we consider is a nested algorithm, which formulates Eq.~\eqref{eq:FP} as a nested fixed-point problem with two layers
\begin{subequations}\label{eq:Nested}
	\begin{align}
	\label{eq:NestedOuter}
	\bu &= \bg(\hat{\bcU}(\bu))\:,\\
	\label{eq:NestedInner}
	\hat{\bcU}(\bu) &= \bcG(\bu, \hat{\bcU}(\bu))\:,
	\end{align}
\end{subequations}
where the outer layer, Eq.~\eqref{eq:NestedOuter}, is a fixed-point problem on the matter states $\bu$, and the inner, Eq.~\eqref{eq:NestedInner}, is on the distributions $\hat{\bcU}$ for fixed matter states $\bu$. 
These two problems are nested in the sense that evaluating the right-hand side of Eq.~\eqref{eq:NestedOuter} at a given $\bu$ requires solving Eq.~\eqref{eq:NestedInner}.
For example, applying Picard iteration on both Eqs.~\eqref{eq:NestedOuter} and \eqref{eq:NestedInner} gives the following iterative scheme
\begin{subequations}
	\begin{equation}
	\bu^{[k+1]} = \bg(\hat{\bcU}(\bu^{[k]}))\:,
	\end{equation}
	where $\hat{\bcU}(\bu^{[k]}) = \bcU^{[k,*]}$, the limit point of the inner Picard iteration
	\begin{equation}
	\bcU^{[k,\ell+1]} = \bcG(\bu^{[k]}, \bcU^{[k,\ell]})\:.
	\end{equation}
\end{subequations}
In practice, we use Anderson acceleration with $m=2$, as described in Section~\ref{subsec:coupled_FP}, to accelerate both the outer and inner solves separately.  

The nested approach was considered in \cite{Laiu-2020-solver} for relaxing the nonlinear coupling between the electric field and electron concentration when solving implicit systems for semiconductor models.
Here the nested approach is motivated by the fact that in solving Eq.~\eqref{eq:FP}, the most costly part is evaluating the opacity kernels $\Phi$ at a given matter state $\bu$, which is performed in $\bcG$ whenever $\bu$ is updated.
Therefore, while the coupled approach seems simple and straightforward, the nested structure in Eq.~\eqref{eq:Nested} justifies the additional complexity by reducing the number of updates (on $\bu$) in Eq.~\eqref{eq:NestedOuter} via a more accurate distribution update given by solving Eq.~\eqref{eq:NestedInner} at the current matter state.
Note that the matter state $\bu$ is fixed in the solution procedure of the inner problem in Eq.~\eqref{eq:NestedInner}, which does not require opacity kernel evaluations and results in much cheaper inner iterations.

\subsection{Nested Newton's method}
\label{subsec:nested_Newton}

The nested Newton's method formulates the inner layer of the nested system in Eq.~\eqref{eq:Nested} as a root-finding problem, resulting in 
\begin{subequations}\label{eq:NestedNewton}
	\begin{align}
	\label{eq:NestedNewtonOuter}
	\bu &= \bg(\hat{\bcU}(\bu))\:,\\
	\label{eq:NestedNewtonInner}
	\bcF&(\bu, \hat{\bcU}(\bu)) = \mathbf{0}\:,
	\end{align}
\end{subequations}
where the outer layer in Eq.~\eqref{eq:NestedNewtonOuter} is still a fixed-point problem on $\bu$, and the inner layer in Eq.~\eqref{eq:NestedNewtonInner} is a root-finding problem on $\hat{\bcU}$ for fixed $\bu$, with $\bcF$ defined in Eq.~\eqref{eq:InnerNewtonOperator}.
Here Eq.~\eqref{eq:NestedNewtonOuter} is solved using Anderson acceleration, and, whenever the right-hand side of Eq.~\eqref{eq:NestedNewtonOuter} is evaluated at some given $\bu$, the inner problem in Eq.~\eqref{eq:NestedNewtonInner} is solved via Newton's method to obtain $\hat{\bcU}(\bu)$ that is used to evaluate $\bg(\hat{\bcU}(\bu))$.
This nested solver is identical to the nested fixed-point algorithm in Section~\ref{subsec:nested_FP}, except that the inner problem is solved via Newton's method instead of Anderson acceleration.
Specifically, let $\bu^{[k]}$ be the $k$th iterate in the outer layer, then $\hat{\bcU}(\bu^{[k]}) = \bcU^{[k,*]}$, which is the limit point of the Newton iterate
\begin{equation}
\bcU^{[k,\ell+1]}=\bcU^{[k,\ell]} + \delta\bcU^{[k,\ell]}\:,
\quad\text{with}\quad
\delta\bcU^{[k,\ell]} = - \big[\pderiv{\bcF}{\bcU}(\bu^{[k]},\bcU^{[k,\ell]})\big]^{-1}\,\bcF(\bu^{[k]},\bcU^{[k,\ell]})\:.
\end{equation}
Since the matter state $\bu$ is fixed in the inner iterations, Eqs.~\eqref{eq:discreteOpacity} and \eqref{eq:totalEmissivityandOpacity}--\eqref{eq:pairOpacity} imply that the Jacobian $\pderiv{\bcF}{\bcU}$ can be calculated with no additional opacity kernel evaluation, which justifies this nested approach.
The reason we choose not to formulate the outer layer as a root-finding problem and solve it with Newton's method is to avoid the costly opacity kernel gradient approximation discussed in Section~\ref{subsec:coupled_Newton}.

\section{Numerical experiments}
\label{sec:num_results}

The four iterative solvers introduced in Section~\ref{sec:techniques} are compared in this section.  
First, to investigate the iterative solvers in isolation, we report on results obtained on relaxation problems under conditions expected in CCSNe.  
Then we compare the iterative solvers in the context of the IMEX scheme in Eqs.~\eqref{eq:imexXNodes}-\eqref{eq:imexExplicit} on proto-neutron star deleptonization problems using matter conditions from spherically symmetric CCSN simulations at various times after core-bounce.

\subsection{Implementation details}
\label{subsec:implementation}

In the numerical tests discussed in this section, the DG-IMEX scheme and the nonlinear solvers are implemented 
following the specifics below, unless otherwise noted.

\begin{itemize}
\item{DG scheme --}
We consider problems with one spatial dimension (imposing spherical symmetry). 
For the relaxation problems in Section~\ref{subsec:relaxation}, we solve the space-homogeneous problem in Eq.~\eqref{eq:IMStep} for a single spatial element.  
For the proto-neutron star deleptonization problems in Section~\ref{subsec:deleptonization}, the spatial domain $r\in[0,300]$~{km} is divided into $128$ geometrically progressing elements with the first element of size $\dx=1$~{km} and the last element of size $\dx\approx4.54$~km.
In both tests, the energy domain covering $\varepsilon\in[0,300]$~{MeV} is divided into $16$ geometrically progressing elements, where the first element has $\Delta\varepsilon=4$~{MeV} and the last element has $\Delta\varepsilon\approx50$~{MeV}.
The spatial and energy DG elements considered here are linear ($k=1$). 

\item{IMEX scheme --}
In the proto-neutron star deleptonization problem in Section~\ref{subsec:deleptonization} we use the IMEX scheme in Eqs.~\eqref{eq:imexXNodes}--\eqref{eq:imexExplicit} with two stages ($s=2$).  
When written in the so-called Shu-Osher form (as in Eqs.~\eqref{eq:imexXNodes}--\eqref{eq:imexExplicit}), the coefficients are given by \cite{CHU201962}
\begin{equation}\label{eq:imexCoefficients}
  \left[
  \begin{array}{cc}
    c_{10} & \\
    c_{20} & c_{21}
  \end{array}
  \right]
  =
  \left[
  \begin{array}{cc}
    1 & \\
    1/2 & 1/2
  \end{array}
  \right],
  \quad
  \left[
  \begin{array}{cc}
    \hat{c}_{10} & \\
    \hat{c}_{20} & \hat{c}_{21}
  \end{array}
  \right]
  =
  \left[
  \begin{array}{cc}
    1 & \\
    0 & 1
  \end{array}
  \right],
  \quad\mbox{and}\quad
  \left[
  \begin{array}{c}
    a_{11} \\
    a_{22}
  \end{array}
  \right]
  =
  \left[
  \begin{array}{c}
    1 \\
    1/2
  \end{array}
  \right].
\end{equation}
This scheme consists of two evaluations of the explicit part and two implicit solves.  
We use the realizability-enforcing limiter in \cite{CHU201962} to enforce realizable moments (cf. Eq.~\eqref{eq:momentBounds}) after each stage.  

\item{Equation of state and neutrino opacity tables --}
In all the tests we use a tabulated version of the SFHo EoS \citep{steiner_etal_2013}.  
Thermodynamic (dependent) variables are tabulated as a function of mass density, temperature, and electron fraction ($\rho$, $T$, and $Y_{e}$).  
The EoS table covers the ranges $\rho\in[1.66\times10^{3},3.16\times10^{15}]$~g~cm$^{-3}$, using $N_{\rho}=185$ points (logarithmically spaced to achieve about $15$ points per decade), $T\in[1.16\times10^{9},1.84\times10^{12}]$~K, using $N_{T}=81$ points (logarithmically spaced to achieve about $25$ points per decade), and $Y_{e}\in[0.01,0.6]$, using $N_{Y_{e}}=30$ points (linearly spaced).  
The neutrino opacities are taken from \cite{bruenn_1985}, with all the input thermodynamic quantities computed with the SFHo EoS.  
The absorption and scattering opacities ($\chi$ and $\sigma_{\IS}$) are tabulated in terms of the neutrino energy $\varepsilon$ in addition to $\rho$, $T$, and $Y_{e}$ (using the same resolution as the EoS table).  
The neutrino energy range covers $\varepsilon\in[0.1,300]$~MeV, using $N_{\varepsilon}=40$ logarithmically spaced points.  
The neutrino-electron scattering and pair creation and annihilation kernels are tabulated in terms of neutrino energy pairs $\varepsilon$ and $\varepsilon'$ (using the same points as is used for $\chi$ and $\sigma_{\IS}$), $T$ (using the same points as in the EoS table), and the degeneracy parameter $\eta=\mu_{e}/(\Boltzmann T)\in[1\times10^{-3},2.5\times10^{3}]$, using $N_{\eta}=60$ logarithmically spaced points.  
To evaluate dependent variables from the table, following, e.g., \cite{mezzacappaMesser_1999}, we use bilinear interpolation (or the higher-dimensional equivalent), while derivatives with respect to any of the independent variables are computed by taking the derivative of the interpolation formula.  
When interpolating the opacity kernels, we enforce the symmetries in Eq.~\eqref{eq:kernelSymmetries}.

\item{Nonlinear solvers --}
The four nonlinear solvers are implemented following the description in Section~\ref{sec:techniques}, with one exception that, in the implementation, the effective emission and absorption opacity $\chi$ in Eq.~\eqref{eq:totalEmissivityandOpacity} is lagged. In other words, when solving Eq.~\eqref{eq:CoupledSystem}, $\chi$ is evaluated at the starting matter state $(\mathsf{Y}_e^{\old},\upepsilon^{\old})$ and is not being updated in the solution procedure. We choose to lag $\chi$ in the nonlinear solvers to simplify the Jacobian calculation in Newton's method. Since $\chi$ is usually varying slowly in time, lagging $\chi$ has minimal impact on the solution accuracy. For the fixed-point solvers, $\chi$ can be updated at each iteration at a minor additional cost.
As mentioned in Section~\ref{sec:techniques}, we choose the truncation parameter in Anderson acceleration to be $m=2$, unless otherwise specified. In this case, the least-squares problem for determining $\alpha^*$ in Eq.~\eqref{eq:AA_step} becomes an inversion of a $3\times3$ matrix and is thus solved analytically. A numerical justification of this choice of $m$ is given in Section~\ref{subsec:relaxation}.
In the Newton's method, the Jacobian matrix is constructed using the derivatives given in Appendix~\ref{appendix:kernelDerivatives}.

We also note that Jacobian-free Newton-Krylov (JFNK) methods, where the Newton step is computed by solving an approximate Newton system with a Krylov solver, are not well suited for this problem (see, e.g., \cite{knoll2004jacobian} for a comprehensive survey on these methods).
The reason is that, in JFNK methods, one evaluation of the opacity kernels is needed in every Krylov iteration to approximate the Jacobian matrix. Thus, JFNK methods require several opacity evaluations per Newton iteration, while the standard Newton's method only requires one, which makes JFNK methods more expensive for solving these problem, where the opacity evaluation is a dominant computational cost.

\item{Nonlinear solver initial guess --}
When solving the coupled nonlinear system inn Eq.~\eqref{eq:CoupledSystem}, a natural choice of initial guess for the unknowns $(\mathsf{Y}_e^{\new},\upepsilon^{\new},\bJNe^{\new}, \bJNeb^{\new})$ is $(\mathsf{Y}_e^{\old},\upepsilon^{\old},\bJNe^{\old}, \bJNeb^{\old})$, the weighted sum from explicit steps defined in Eq.~\eqref{eq:IMStep}. 
In this work, we add a ``presolve" step that aims to provide a better starting point for the iterative solvers and speedup the computation.
Specifically, for given $(\mathsf{Y}_e^{\old},\upepsilon^{\old},\bJNe^{\old}, \bJNeb^{\old})$, the presolve step solves a subsystem of Eq.~\eqref{eq:CoupledSystem}, which is obtained by setting the opacities $\etah_{\SC}$, $\etah_{\TP}$, $\chih_{\SC}$, and $\chih_{\TP}$ in Eq.~\eqref{eq:discreteTotalOpacity} to be zero; i.e., with only emission, absorption, and isoenergetic scattering.
The solution of this simplified system then serves as the initial guess for $(\mathsf{Y}_e^{\new},\upepsilon^{\new},\bJNe^{\new}, \bJNeb^{\new})$ in the nonlinear solvers for system in Eq.~\eqref{eq:CoupledSystem}.
This presolve step is computationally inexpensive (no opacity table interpolations are needed, since the emission and absorption opacity $\chi$ is lagged as discussed in the previous paragraph), while giving a reasonable initial guess for the full system. 
We observe that, without the presolve step, the Coupled Newton's method and the Nested Newton's method in Sections~\ref{subsec:coupled_Newton} and \ref{subsec:nested_Newton} could diverge if the initial guess $(\mathsf{Y}_e^{\old},\upepsilon^{\old},\bJNe^{\old}, \bJNeb^{\old})$ is too far away from the solution.

\item{Nonlinear solver convergence criteria --}
We set the convergence criteria for the iterative solvers based on the relative residual of the system in Eq.~\eqref{eq:CoupledSystem} at the current iterate.
Specifically, for the coupled solvers in Sections~\ref{subsec:coupled_FP} and \ref{subsec:coupled_Newton}, the convergence criteria are
\begin{subequations}
	\label{eq:ConvergenceCriteria}
	\begin{align}
	\label{eq:C1}
	\big|\bff_{\mathsf{Y}_{e}}(\bu^{[k]},\bcU^{[k]})\big|\leq \texttt{tol}\,\big|\mathsf{Y}_{e}^{[0]}\big|\:,& \quad \big|\bff_{\upepsilon}(\bu^{[k]},\bcU^{[k]})\big|\leq \texttt{tol}\, \big|\upepsilon^{[0]}\big|\:, \\
	\label{eq:C2}
	\big\|\bcF_{\bJNe}(\bu^{[k]},\bcU^{[k]})\big\|\leq \texttt{tol}\,\big\|\bJNe^{[0]}\big\|\:,& \quad \big\|\bcF_{\bJNeb}(\bu^{[k]},\bcU^{[k]})\big\|\leq \texttt{tol}\, \big\|\bJNeb^{[0]}\big\|\:,
	\end{align}
\end{subequations}
where $\texttt{tol}>0$ is a constant relative tolerance, and $(\bff_{\mathsf{Y}_{e}}, \bff_{\upepsilon}, \bcF_{\bJNe}, \bcF_{\bJNeb})$ are the residuals as defined in Eqs.~\eqref{eq:OuterNewtonOperator} and \eqref{eq:InnerNewtonOperator}.
As for the nested solvers in Sections~\ref{subsec:nested_FP} and \ref{subsec:nested_Newton}, the outer layer (Eqs.~\eqref{eq:NestedOuter} and \eqref{eq:NestedNewtonOuter}) uses the convergence criteria in Eq.~\eqref{eq:C1}, while the convergence criteria in the inner layer (Eqs.~\eqref{eq:NestedInner} and \eqref{eq:NestedNewtonInner}) are given by 
\begin{equation}
\big\|\bcF_{\bJNe}(\bu^{[k]},\bcU^{[k,\ell]})\big\|\leq \texttt{tol}\,\big\|\bJNe^{[k,0]}\big\|\:, \quad \big\|\bcF_{\bJNeb}(\bu^{[k]},\bcU^{[k,\ell]})\big\|\leq \texttt{tol}\, \big\|\bJNeb^{[k,0]}\big\|\:.
\end{equation}
When all convergence criteria are satisfied, the solvers return the current iterate as the solution.
In all numerical experiments, we choose the norm to be the discrete $L^2$ norm on the energy domain, i.e., $\|\bJNe\|:=\big(\sum_{q=1}^{|\Se|} w_q^{(2)} \bJNe_q^2\big)^{\sfrac{1}{2}}$, and we set the relative tolerance to be $\texttt{tol}=10^{-8}$.

\end{itemize}

\begin{remark}
In the coupled fixed-point and the two nested solvers, the basic version of Anderson acceleration outlined in Section~\ref{subsec:coupled_FP} is implemented. 
It is known (see, e.g., \cite{Walker-Ni-2011} and references therein) that the iterations in Anderson acceleration may suffer from ``stagnation" when the least-squares problem is ill-conditioned. Due to the choice of small truncation parameter (e.g., $m=2$) and the contractive property of the collision operator, we do not encounter the stagnation issue in any of the numerical tests presented in this section.
Stagnation may potentially become a practical concern when applying Anderson acceleration to solve more complicated systems, e.g., fully-coupled neutrino radiation hydrodynamics. 
A common approach to mitigate the stagnation issue is to control the condition number of the least-squares problems. This can be achieved by (i) 
solving a proper reformulation of the least-squares problems using QR factorization \citep{Ni-Walker-2010}, 
(ii) modifying the truncation parameter $m$ adaptively \citep{Yang-etal-2009}, and/or (iii) regularizing the least-squares problems \citep{Scieur-d'Aspremont-Bach-2016}.
Alternatively, one may consider the recently proposed globally convergent variant of Anderson acceleration \citep{Zhang-etal-2020}, in which ill-conditioning is handled via regularization and global convergence is guaranteed using safeguarding steps.
\end{remark}

\subsection{Relaxation problem}
\label{subsec:relaxation}

The first class of test problems we consider here is the \textit{relaxation} problem, where the neutrino and antineutrino transport equations \eqref{eq:neutrino_transport} and \eqref{eq:antineutrino_transport} are solved with only the collision terms considered, i.e., the space-homogeneous case where the advection terms $\cT(\fNe)$ and $\cT(\fNeb)$ are zero.
In these relaxation problems, the collision operator \textit{relaxes} the distributions $\fNe$ and $\fNeb$ to the equilibrium Fermi-Dirac distributions $\fNe_0$ and $\fNeb_0$ in Eq.~\eqref{eq:FermiDirac}, respectively, as time evolves.  
Due to the lack of advection terms, there is no spatial coupling. 
Thus, these problems can be solved independently in space, which makes them ideal test cases for the nonlinear collision system solvers considered in this paper.
In this setup, the semidiscrete moment and matter equations reduce from \eqref{eq:DiscretizedOperatorFormFull} to
\begin{equation}\label{eq:relaxation}
\f{1}{\SOL}\pd{\bcV_{\mx}}{t}
= \Ch(\bcV_{\mx})\:,\quad \forall \mx\text{ such that }\vect{x}_{\mx}\in\Sx\:.
\end{equation}
For the relaxation problems, we discretize Eq.~\eqref{eq:relaxation} in time with the backward Euler method.
At each time step, this time discretization results in a coupled system that takes the form of Eq.~\eqref{eq:IMAll} with the effective time step $\tau = \dt$.
To solve this system, we apply the nonlinear solvers in Section~\ref{sec:techniques} on the subsystem in Eq.~\eqref{eq:CoupledSystem} (with $\tau = \dt$) to obtain $(\mathsf{Y}_e^{\new},\upepsilon^{\new},\bJNe^{\new}, \bJNeb^{\new})$, which are then used to compute $(\bHNe^{\new},\bHNeb^{\new})$.  
We start with trivial initial states for $\bHNe$ and $\bHNeb$, which remain unchanged with time.

\begin{figure}[h]
	\captionsetup[subfigure]{justification=centering}
	\subfloat[$\rho = 1.084\times10^{14}$~g~cm$^{-3}$, $T = 1.845\times10^{11}$~K, $Y_e = 0.2728$]
	{\begin{minipage}{0.45\textwidth}
			\includegraphics[width=\linewidth]{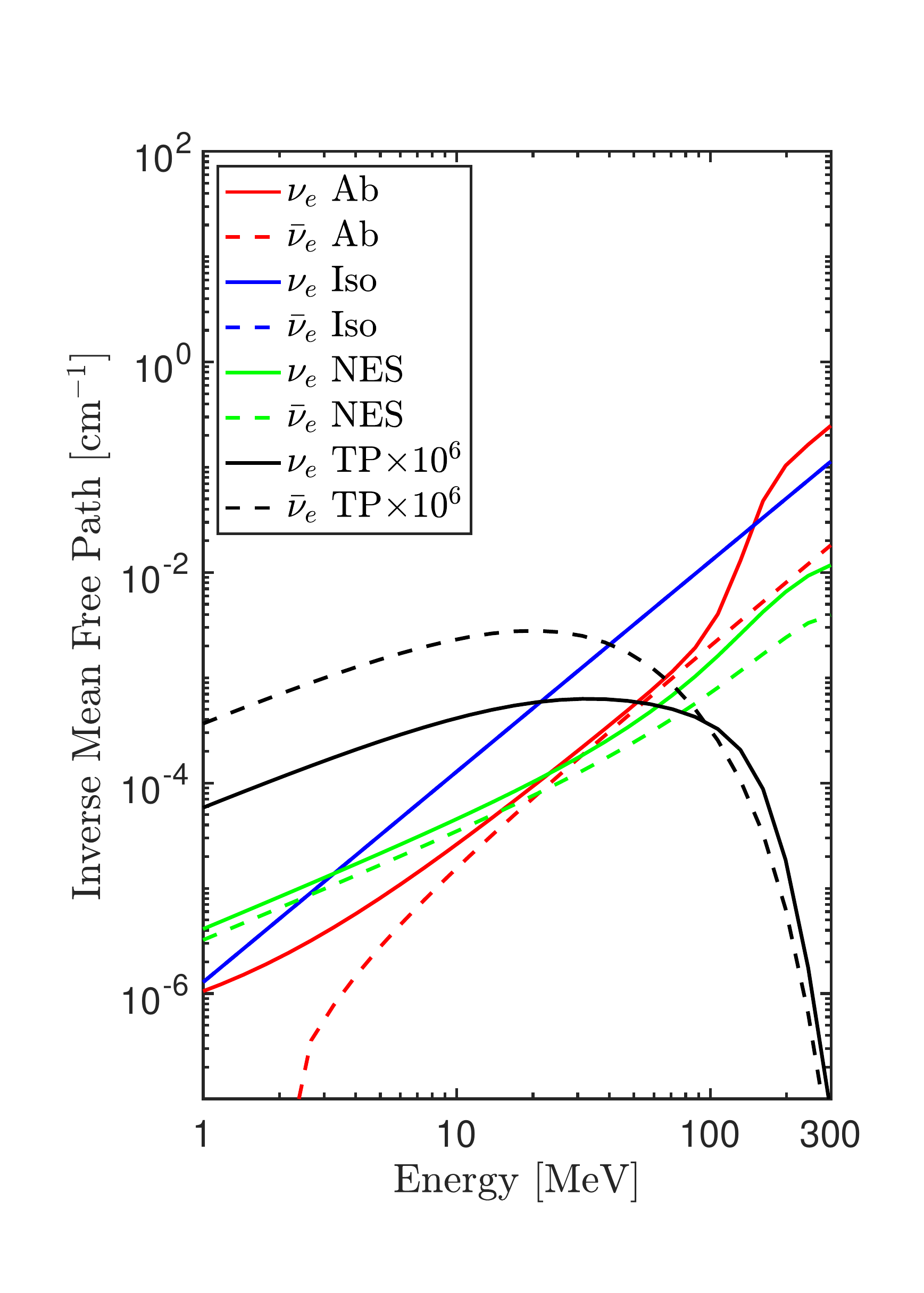}
			\label{fig:OpacitiesHigh}
		\end{minipage}
	}~~~~~~~
	\subfloat[$\rho = 1.032\times10^{12}$~g~cm$^{-3}$, $T = 8.806\times10^{10}$~K, $Y_e = 0.1347$]	
	{\begin{minipage}{0.45\textwidth}
			\includegraphics[width=\linewidth]{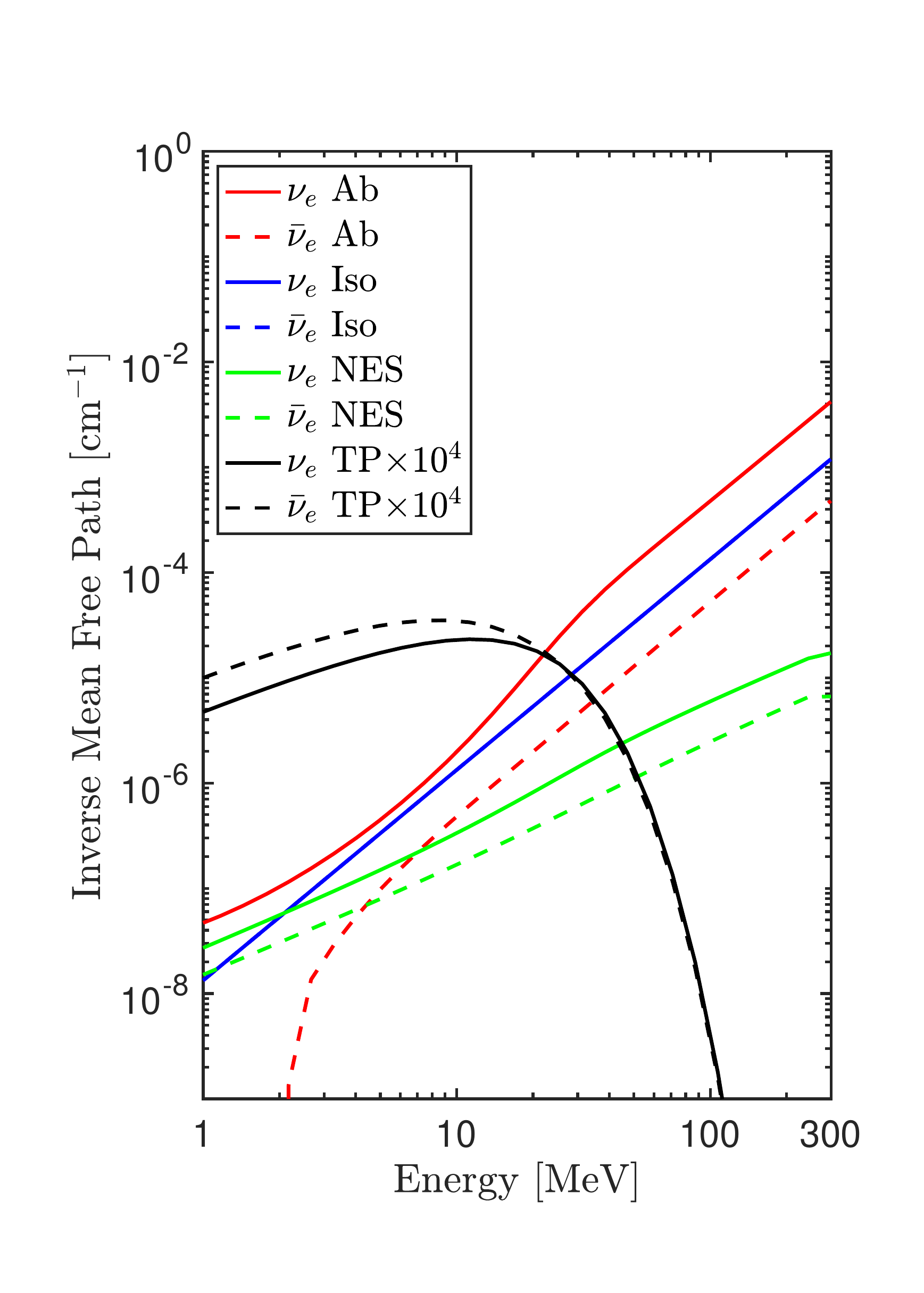}
			\label{fig:OpacitiesLow}
		\end{minipage}
	}	
	
	\caption{Neutrino opacities for high (left panel) and moderate (right panel) mass density.  In each panel, for neutrinos (solid lines) and antineutrinos (dashed lines), we plot, versus neutrino energy $\varepsilon$, the absorption opacity ($\chi$ and $\bar{\chi}$; red lines), the elastic (isoenergetic) scattering opacity ($\sigma_{\IS}$ and $\bar{\sigma}_{\IS}$; blue lines), the neutrino-electron scattering opacity ($\chi_{\SC}$ and $\bar{\chi}_{\SC}$; green lines), and the opacity due to pair creation and annihilation ($\chi_{\TP}$ and $\bar{\chi}_{\TP}$; black lines).  The neutrino-electron scattering and pair creation and annihilation opacities where computed with the neutrino and antineutrino number densities set to zero; cf. Eqs.~\eqref{eq:NESOpacity} and \eqref{eq:pairOpacity}.}
	\label{fig:Opacities}
\end{figure}

We test the nonlinear solvers on the relaxation problem in Eq.~\eqref{eq:relaxation} with two initial matter states that present problems with different degrees of collisionality.
The first state, which represents the high density, strongly collisional region inside a proto-neutron star, is sampled at radius $r^{(1)}=9.756$~km from the center of a collapsed stellar core, and the second state, which represents the lower density regions around the surface of a proto-neutron star with relatively weaker collisionality, is sampled at radius $r^{(2)}=39.52$~km.  
We obtained the matter states at these two locations from a spherically symmetric core-collapse supernova simulation $50$~ms after core bounce \citep{liebendorfer_etal_2005} (results from the \textsc{vertex} code using a 15~$M_{\odot}$ progenitor).  
Specifically, the matter state at $r^{(1)}$ is given by
$\rho^{(1)} = 1.084\times10^{14}$~g~cm$^{-3}$, 
$T^{(1)} = 1.845\times10^{11}$~K, 
and $Y_e^{(1)} = 0.2728$; 
and the matter state at $r^{(2)}$ is 
$\rho^{(2)} = 1.032\times10^{12}$~g~cm$^{-3}$, 
$T^{(2)} = 8.806\times10^{10}$~K, 
and $Y_e^{(2)} = 0.1347$. 
The inverse mean free path associated with the neutrino opacities for these matter states are plotted versus neutrino energy in Figure~\ref{fig:Opacities}.  
Considering the isoenergetic scattering opacity (which is equal for neutrinos and antineutrinos, and increases with neutrino energy as $\varepsilon^{2}$), the mean free path $\lambda_{\IS}=\sigma_{\IS}^{-1}$ varies from about $10$~km to about $10^{-4}$~km in the high collisional state (left panel) in the energy range $\varepsilon\in[1,300]$~MeV.  
In the low collisional state, the scattering mean free path varies from $\lambda_{\IS}\approx10^{3}$~km to $\lambda_{\IS}\approx10^{-2}$~km.  
Correspondingly, the collision time $\tau_{\IS}=\lambda_{\IS}/\SOL$ varies from about $3\times10^{-2}$~ms to $3\times10^{-7}$~ms in the high collisional state case, and from about $3$~ms to $3\times10^{-5}$~ms in the low collisional state case.  

We run the simulations from initial time $t_0=0$ to final time $t_{f}=0.5$~ms for the high collisional case, and from $t_0=0$ to final time $t_{f}= 30$~ms for the low collisional case. 
This is to guarantee that the distributions have relaxed to the equilibrium distributions by the end of simulations.
Each simulation is solved on a single spatial element (the relaxation problem is space-homogeneous) with $16$ geometrically progressing energy elements that divide the energy domain $[0,300]$~{MeV}, where the first element has $\Delta\varepsilon=4$~{MeV} and the last element has $\Delta\varepsilon\approx50$~{MeV}.
In the simulations, the initial matter states take electron fraction $Y_{e,0} = Y_e^{(i)}$ and specific internal energy $\epsilon_0$ calculated at $(\rho^{(i)}, T^{(i)}, Y_e^{(i)})$ from the EoS, with $i=1,2$.  
The initial neutrino and antineutrino moments are given by
\begin{equation}\label{eq:Relaxation_init}
\JNe_0(r^{(i)},\epsilonNu) = \JNeb_0(r^{(i)},\epsilonNu) =  0.99\times\exp \big( - \frac{ (\epsilonNu - 2\, \Boltzmann T^{(i)} )^2 }{ 2\times10^2 } \big)\:,\,\text{and}\,
\HNe_0(r^{(i)},\epsilonNu) = \HNeb_0(r^{(i)},\epsilonNu) = 0\:, 
\end{equation}
for $i = 1,2$, which were chosen such that the initial moments are away from the expected final equilibrium distributions, while making sure the initial moments are realizable, i.e., satisfy Eq.~\eqref{eq:momentBounds}.

\begin{figure}[h]
	\subfloat[Number density $\JNe(r^{(1)},\epsilonNu,t)$ at $t=0$ and $t=t_f=0.5$~ms. At $r^{(1)}$, with initial matter state $\rho = 1.084\times10^{14}$~g~cm$^{-3}$, $T = 1.845\times10^{11}$~K, $Y_e = 0.2728$.]
	{\includegraphics[width=0.45\linewidth]{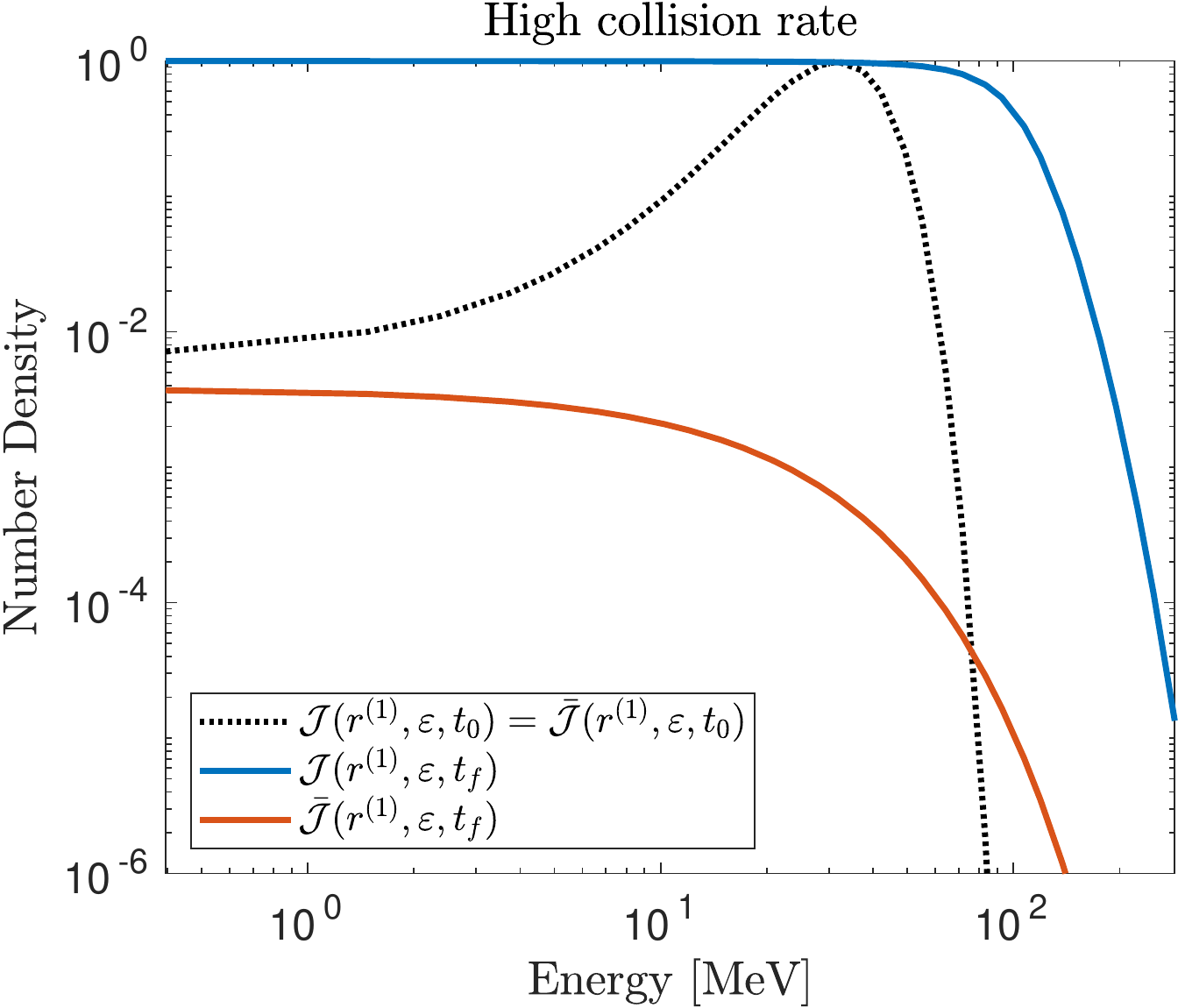}
		\label{fig:High_J_Jbar}
	}~~~~~~~
	\subfloat[Number density $\JNe(r^{(2)},\epsilonNu,t)$ at $t=0$ and $t=t_f=30$~ms. At $r^{(2)}$, with initial matter state $\rho = 1.032\times10^{12}$~g~cm$^{-3}$, $T = 8.806\times10^{10}$~K, $Y_e = 0.1347$.]	
	{\includegraphics[width=0.45\linewidth]{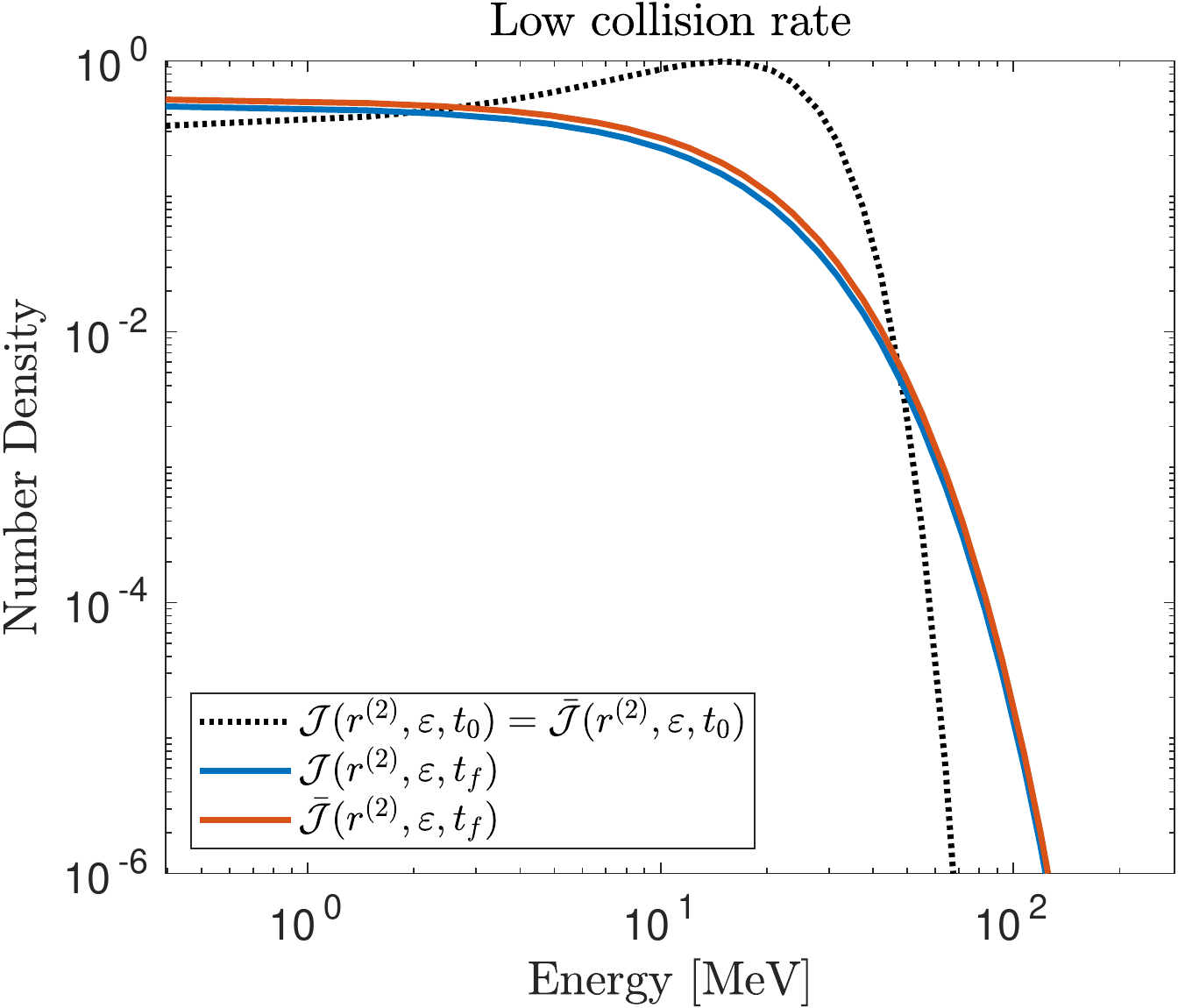}
		\label{fig:Low_J_Jbar}
	}	
	\caption{Initial and final number densities versus neutrino energy for relaxation problems with high (left column) and low (right column) collision rates. Each plot shows the initial number density given in Eq.~\eqref{eq:Relaxation_init} for both neutrinos and antineutrinos (black dashed line), the final neutrino number density (blue solid line), and the final antineutrino number density (red solid line).}
	\label{fig:Relaxation_init_final}	
\end{figure}

\begin{figure}[h]
	\subfloat[Changes in lepton number and energy relative to their initial values in the relaxation simulation at $r^{(1)}$ from $t=0$ to $t=t_f=0.5$~ms. The initial matter state is $\rho = 1.084\times10^{14}$~g~cm$^{-3}$, $T = 1.845\times10^{11}$~K, $Y_e = 0.2728$.]
	{\includegraphics[width=0.45\linewidth]{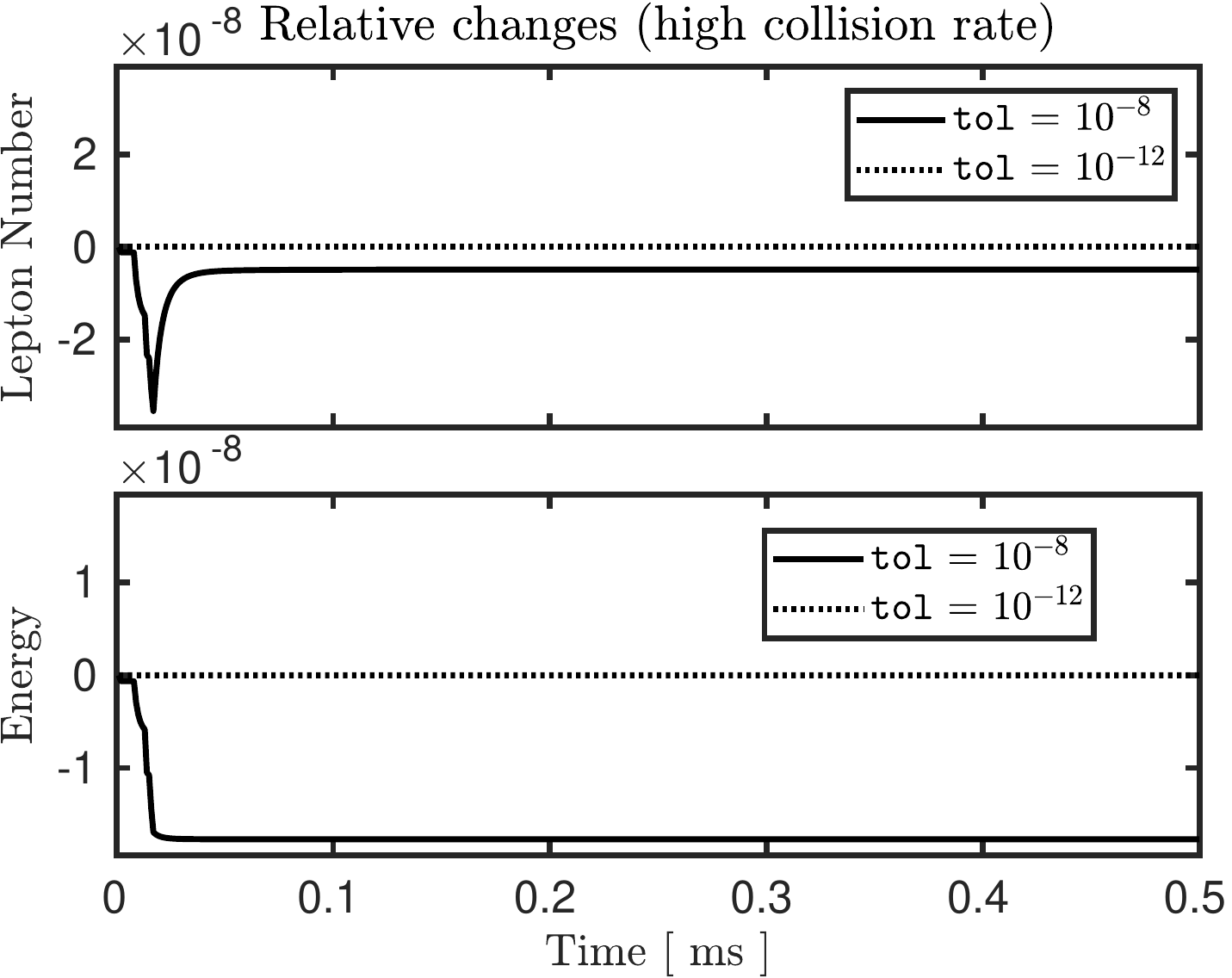}
		\label{fig:Relaxation_High_Con}
	}~~~~~~~
	\subfloat[Changes in lepton number and energy relative to their initial values in the relaxation simulation at $r^{(2)}$ from $t=0$ to $t=t_f=30$~ms. The initial matter state is $\rho = 1.032\times10^{12}$~g~cm$^{-3}$, $T = 8.806\times10^{10}$~K, $Y_e = 0.1347$.]	
	{\includegraphics[width=0.45\linewidth]{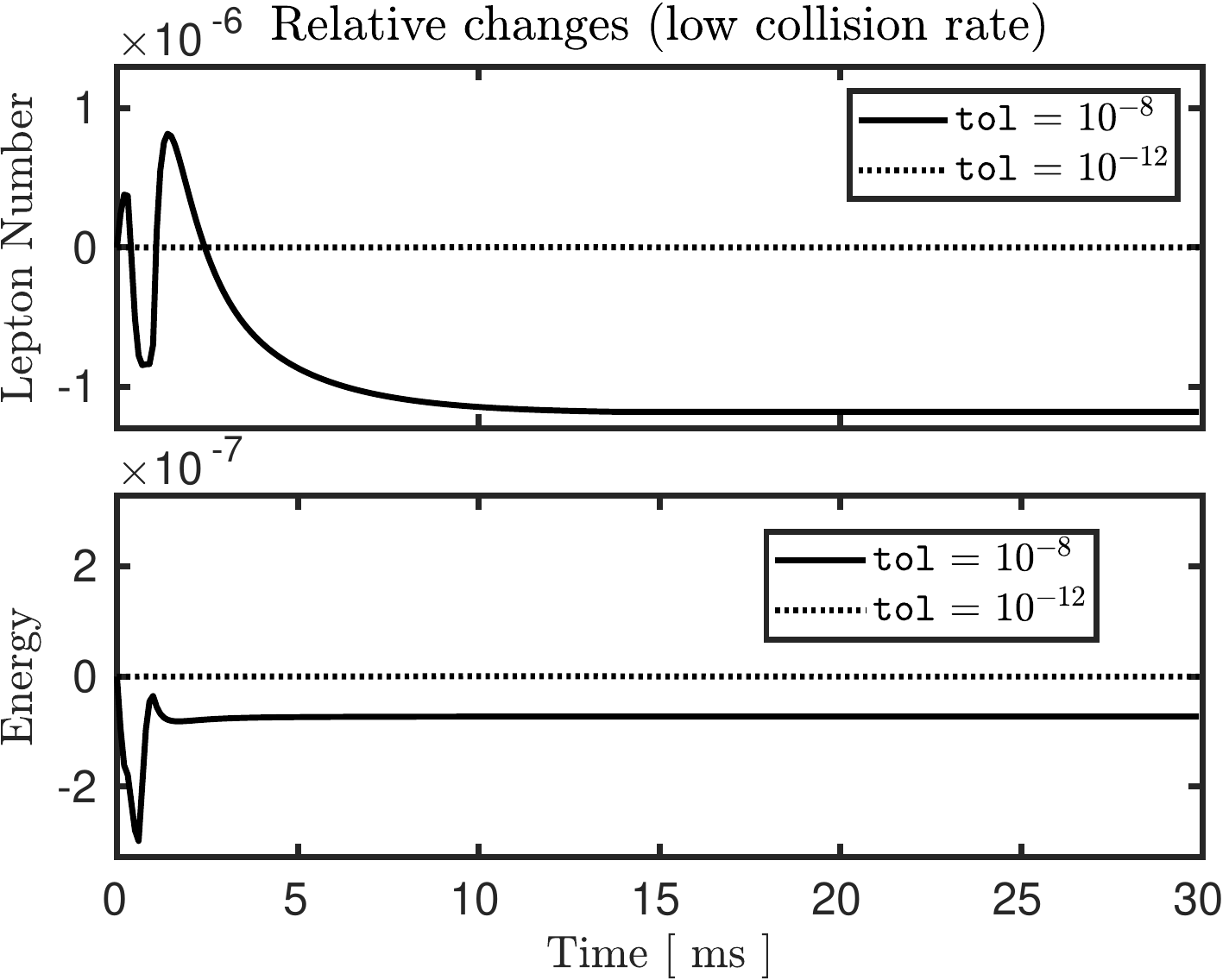}
		\label{fig:Relaxation_Low_Con}
	}

	\caption{Lepton number and energy conservation (see Eqs.~\eqref{eq:LeptonConservation} and \eqref{eq:EnergyConservation}, respectively) in relaxation problems with high (left column) and low (right column) collision rate. Each plot shows the relative changes in lepton number and energy in the relaxation simulations with the nonlinear solver tolerance \texttt{tol} set to $10^{-8}$ (black solid line) and $10^{-12}$ (black dashed line). Tightening the tolerance from  $10^{-8}$ to $10^{-12}$ leads to approximately four orders of magnitude smaller relative changes in both lepton number and energy.}
	\label{fig:Relaxation_Conservation}	
\end{figure}

Figure~\ref{fig:Relaxation_init_final} illustrates the initial and final equilibrium electron neutrino and electron antineutrino number densities for the two relaxation problems.  
Here the initial number densities are given in Eq.~\eqref{eq:Relaxation_init}, and the final number densities were generated by solving Eq.~\eqref{eq:relaxation} using the nodal DG and backward Euler discretization, together with the proposed nonlinear solvers. 
We validated these results by comparing them to number densities from simulations on finer energy-temporal meshes, in which no noticeable differences were observed.
The number densities for the high collision case are shown in Figure~\ref{fig:High_J_Jbar}, and the ones for the low collision case are shown in Figure~\ref{fig:Low_J_Jbar}.  
For the high collision case, the temperature is $1.845\times10^{11}$~K at $t_0=0$, and increases by $7.3\%$ to $1.979\times10^{11}$~K at $t_f = 0.5$~ms, while the electron fraction drops by $14.0\%$ from 0.2728 to 0.2347.  
For the low collision case, the temperature raises by $18.8\%$ from $8.806\times10^{10}$~K at $t_0=0$ to $1.046\times10^{11}$~K at $t_f = 30$~ms, and meanwhile the electron fraction increases by  $2.2\%$ from 0.1347 to 0.1376. 
From these plots, we observe that, in the high collision case, the neutrino number density $\JNe$ is at least two orders of magnitude higher than the antineutrino number density $\JNeb$ at the final equilibrium, thus they are further away from the initial densities than the ones in the low collision case.  
This is one of the reasons that the high collision rate problems are more challenging than the low collision rate problems in the earlier stage, as discussed in the following paragraphs.  

The conservation of lepton number and energy (see Eqs.~\eqref{eq:LeptonConservation} and \eqref{eq:EnergyConservation}, respectively) in the relaxation tests is shown in Figure~\ref{fig:Relaxation_Conservation}. Since the relaxation problem is space-homogeneous, the right-hand sides of Eqs.~\eqref{eq:LeptonConservation} and \eqref{eq:EnergyConservation} are both zero. It then follows from the nonlinear system formulation in Eq.~\eqref{eq:CoupledSystem} that the lepton number and energy are conserved if and only if Eqs.~\eqref{eq:CoupledSystem_Ye} and \eqref{eq:CoupledSystem_eps} are satisfied. Thus, the lepton number and energy are expected to be conserved up to the nonlinear solver tolerance at each point on the space-time grid, which is confirmed in the results reported in Figure~\ref{fig:Relaxation_Conservation}. In addition, we also observe from Figure~\ref{fig:Relaxation_Conservation} that, in both the high and low collision rate cases, tightening the nonlinear solver tolerance from $10^{-8}$ to $10^{-12}$ indeed improves the conservation results. 

\begin{figure}[h]
	\captionsetup[subfigure]{justification=centering}
	\subfloat[High collision rate, $\rho = 1.084\times10^{14}$~g~cm$^{-3}$, $T = 1.845\times10^{11}$~K, $Y_e = 0.2728$]	
	{\begin{minipage}{0.45\textwidth}
		\includegraphics[width=\linewidth]{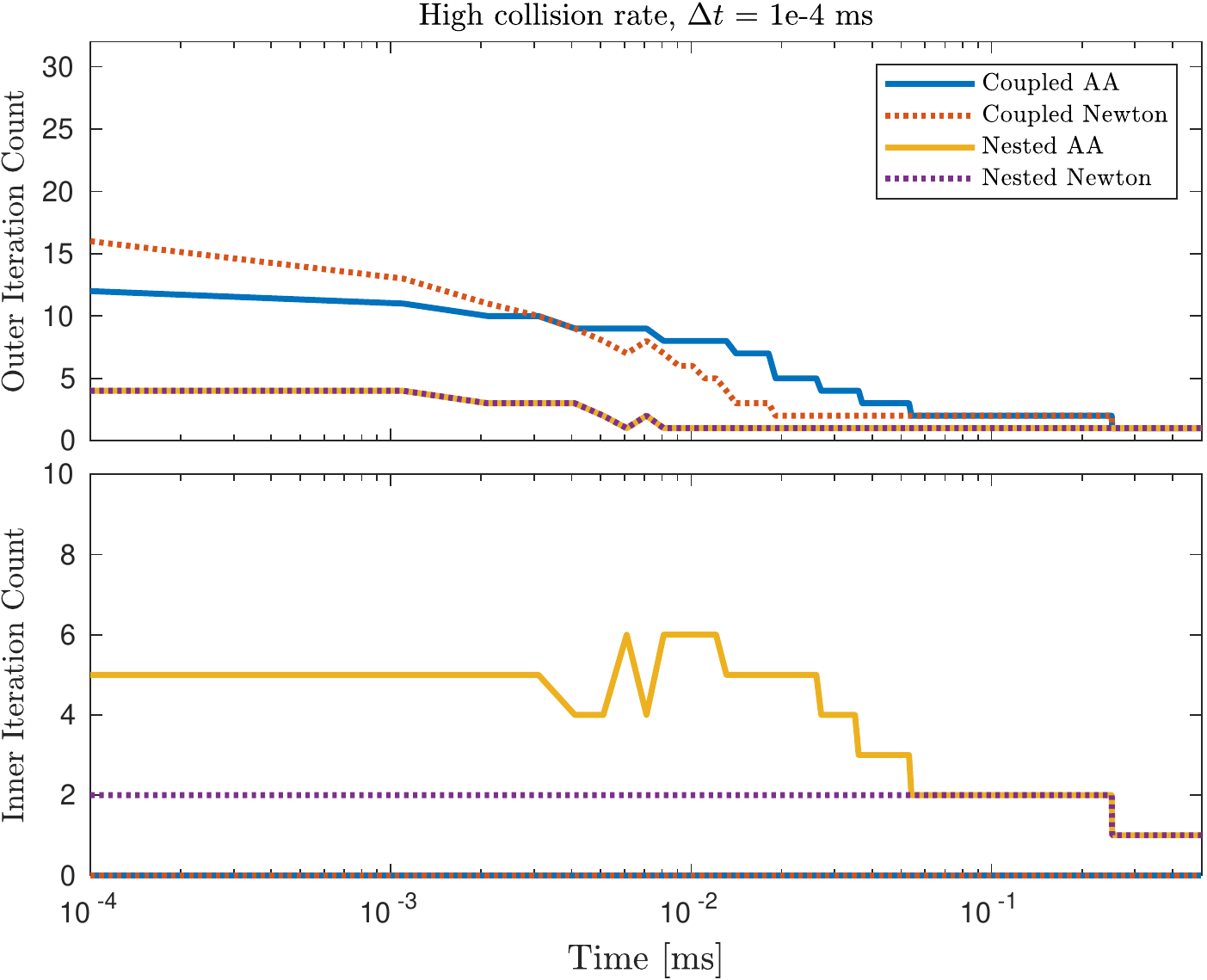}\\
		
		\includegraphics[width=\linewidth]{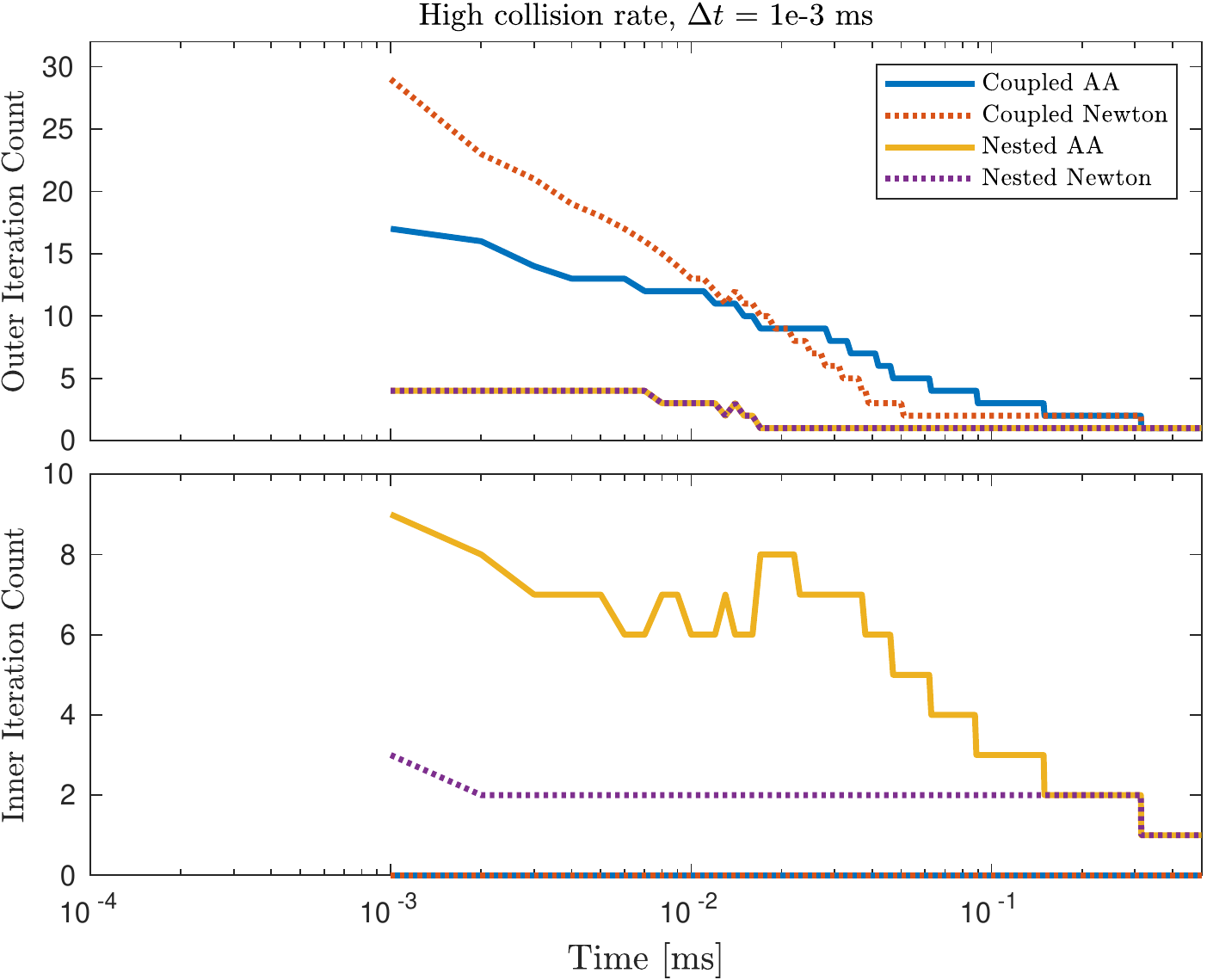}\\

		\includegraphics[width=\linewidth]{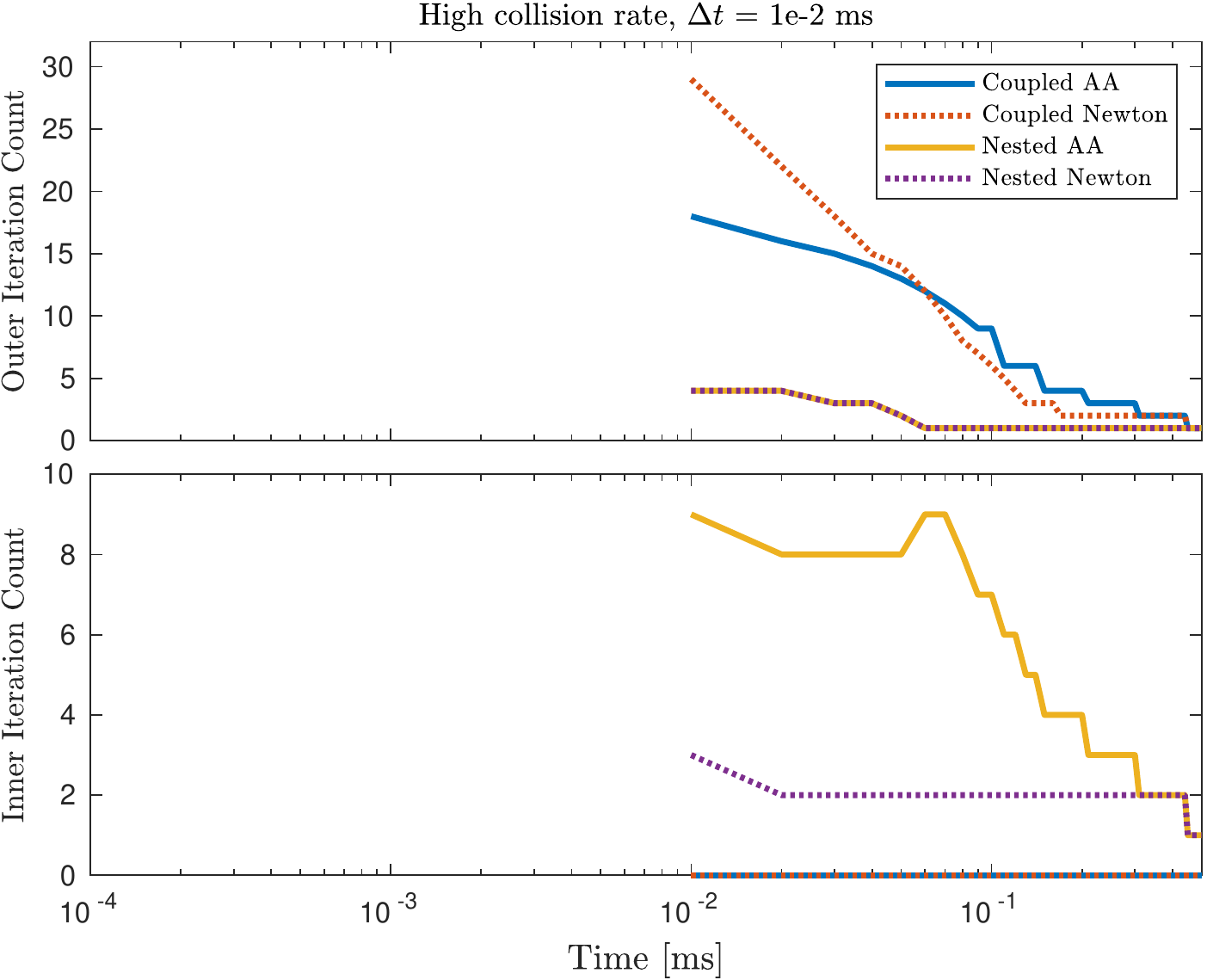}
		\label{fig:Relaxation_1}
	\end{minipage}
	}~~~~~~~
	\subfloat[Low collision rate, $\rho = 1.032\times10^{12}$~g~cm$^{-3}$, $T = 8.806\times10^{10}$~K, $Y_e = 0.1347$]	
	{\begin{minipage}{0.45\textwidth}
			\includegraphics[width=\linewidth]{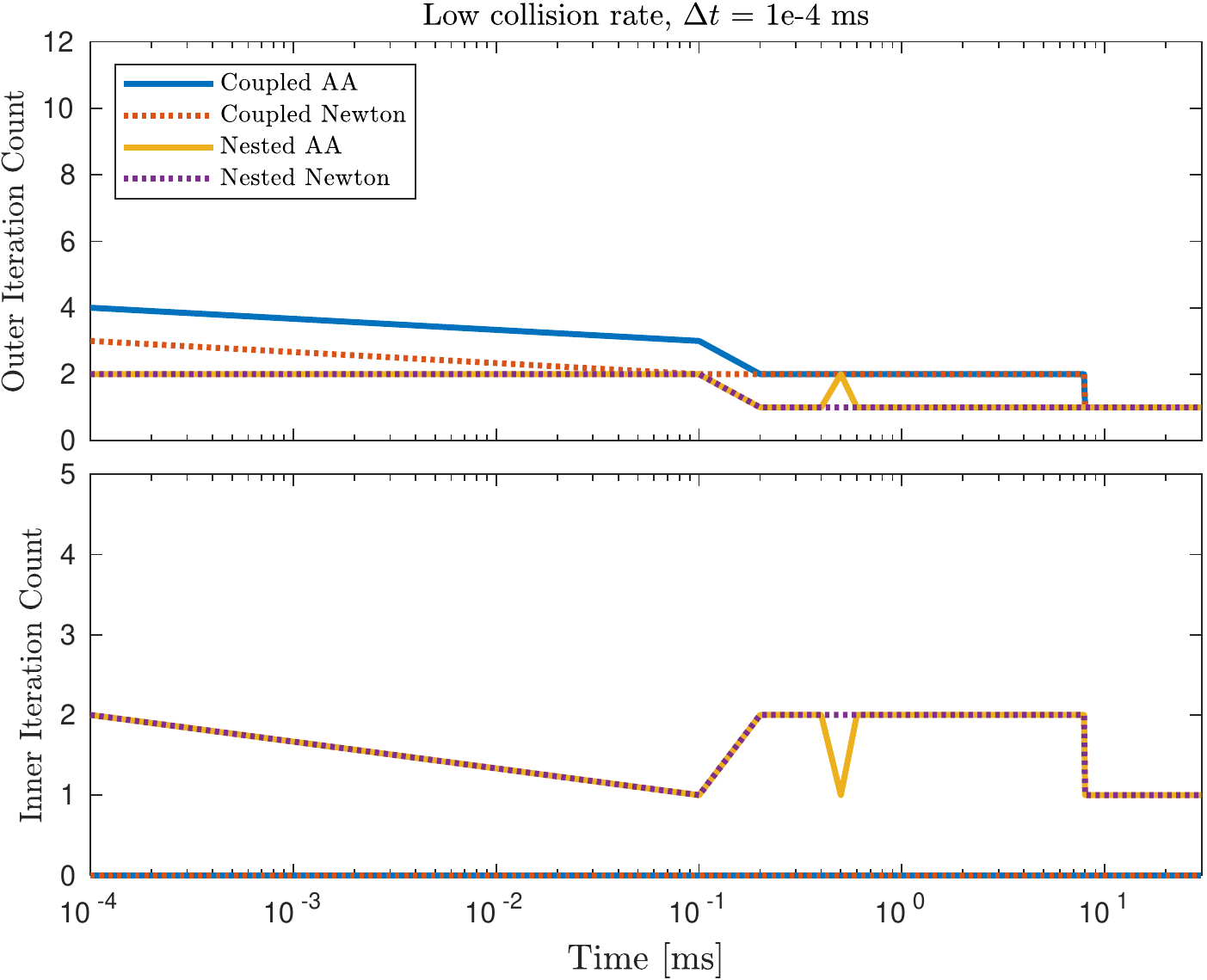}\\
			
			\includegraphics[width=\linewidth]{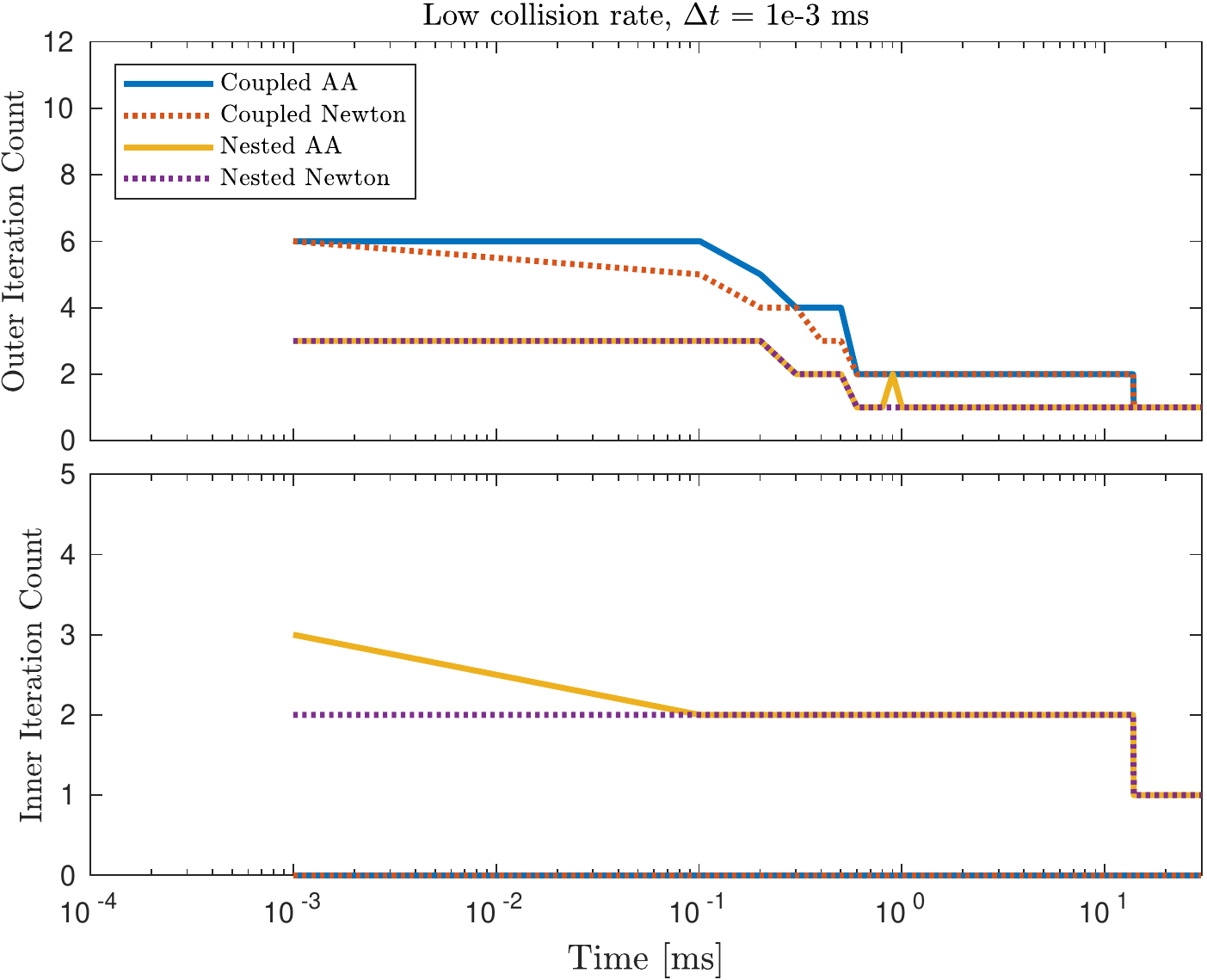}\\
			
			\includegraphics[width=\linewidth]{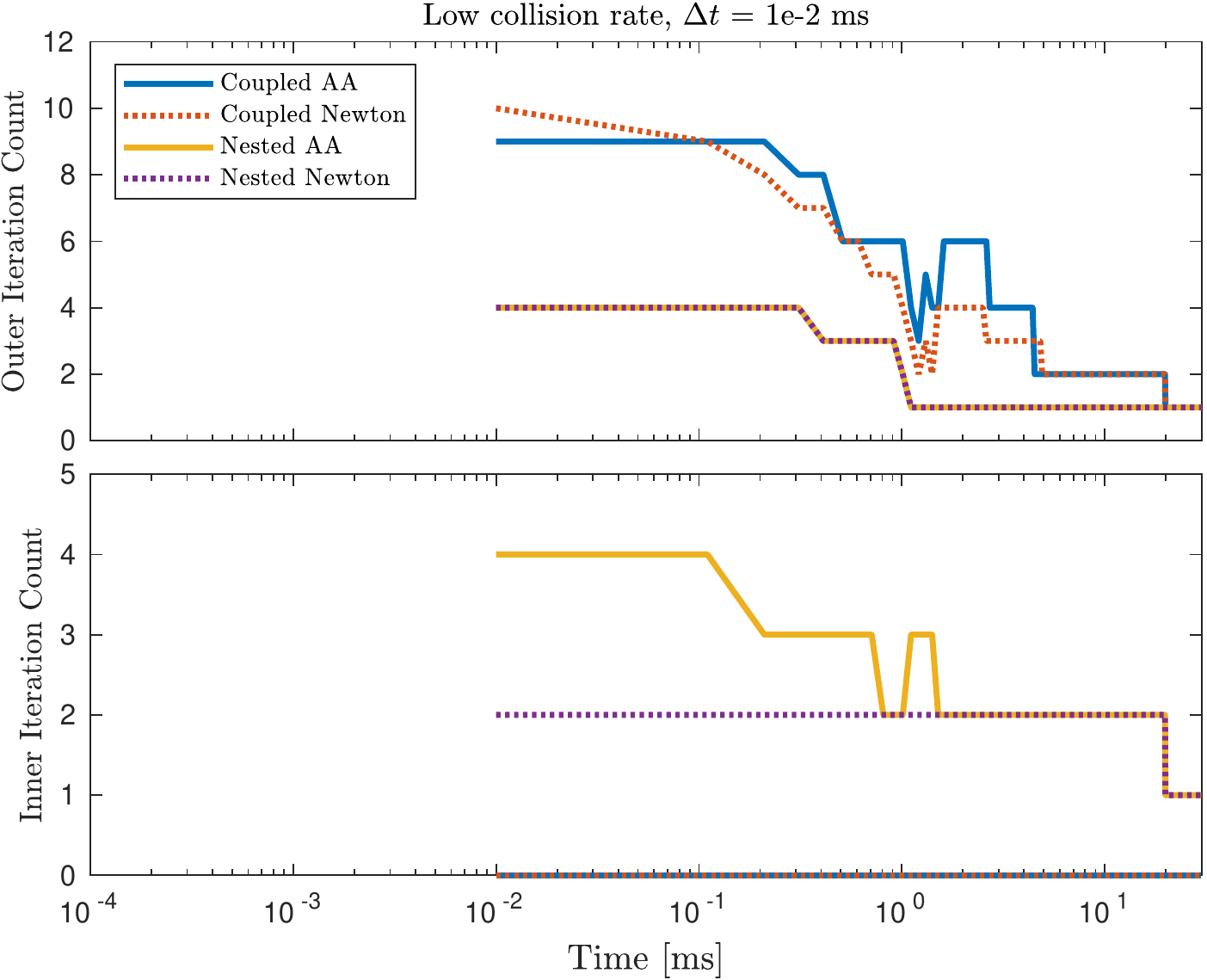}
			\label{fig:Relaxation_2}
		\end{minipage}
	}~~~	

	\caption{Iteration counts of the nonlinear solvers on relaxation problems. The iteration counts of the Coupled AA, Coupled Newton, Nested AA, and Nested Newton solvers on relaxation problems in the high (left column) and low (right column) collision rate regimes with time step sizes varying from $10^{-4}$ to $10^{-2}$~ms.}
	\label{fig:Relaxation}
\end{figure}

Figure~\ref{fig:Relaxation} shows iteration counts versus time for each nonlinear solver on the two relaxation problems, using various time step sizes: $\dt = 10^{-4}$, $10^{-3}$, and $10^{-2}$~ms.  
These time step sizes are motivated by the fact that in the context of the IMEX scheme in Eqs.~\eqref{eq:imexXNodes}-\eqref{eq:imexExplicit}, the maximum stable time step is $\dt=\CFL\times3\times10^{-3}\,\big(\dx/\mbox{1~km}\big)$~ms, where $\CFL\lesssim1$ is the CFL number.  
For a spatial resolution $\dx=\cO(\mbox{1~km})$, our chosen time steps bracket what is typically used in core-collapse supernova simulations.  
The results for the high collision problem are shown in Figure~\ref{fig:Relaxation_1}, while the ones for the low collision problem are shown in Figure~\ref{fig:Relaxation_2}.  
In these figures, the top plot shows the ``outer" iteration counts for each solver, while the bottom plot shows the averaged ``inner" iteration counts for the two nested solvers. 
Here the outer iteration counts represent the number of iterations needed for solving Eqs.~\eqref{eq:FP}, \eqref{eq:Newton}, \eqref{eq:NestedOuter}, and \eqref{eq:NestedNewtonOuter} in the Coupled AA (Anderson acceleration), Coupled Newton, Nested AA, and Nested Newton solvers, respectively; the inner iteration counts are the number of iterations needed for solving Eqs.~\eqref{eq:NestedInner} and \eqref{eq:NestedNewtonInner} in the Nested AA and Nested Newton solvers, respectively.
Since the inner equations \eqref{eq:NestedInner} and \eqref{eq:NestedNewtonInner} are solved in every outer iteration, the reported inner iteration counts in Figures~\ref{fig:Relaxation_1} and \ref{fig:Relaxation_2} are averaged over the number of times that Eqs.~\eqref{eq:NestedInner} and \eqref{eq:NestedNewtonInner} were solved, i.e., the total number of inner iterations taken in the solver is the product of the outer iteration count and the averaged inner iteration count.

Before comparing the solvers, we first observe from the results that the implicit system in Eq.~\eqref{eq:CoupledSystem} in the high collision rate case indeed requires more iterations to reach convergence than the same system does in the low collision rate case.  
Also, the overall iteration counts grows as $\dt$ increases.  
These observations agree with our expectation, since increasing $\dt$ effectively increases the collision rates, and higher collision rates result in stronger coupling of the number densities, which makes the system in Eq.~\eqref{eq:CoupledSystem} harder to solve.

For the nonlinear solver performance, we observe that the Coupled Newton solver requires more iterations than the Coupled AA solver on harder problems, e.g., problems with higher collision rates, larger time step, or at an earlier stage (further away from equilibrium); on easier problems, the Coupled Newton solver converges in fewer iterations than the Coupled AA solver.  
As expected, the nested solvers indeed reduce the number of outer iterations when compared to the coupled solvers, presumably by providing a more accurate update of the neutrino and antineutrino number densities from the inner iteration.  
The Nested AA and Nested Newton solvers share nearly identical outer iteration counts.  
This is due to the fact that both nested solvers use Anderson acceleration in the outer layer (Eqs.~\eqref{eq:NestedOuter} and \eqref{eq:NestedNewtonOuter}), which takes number densities $\hat{\bcU}(\bu)$ from the inner layers (Eqs.~\eqref{eq:NestedInner} and \eqref{eq:NestedNewtonInner}) of the two solvers.
Since the problems in the inner layers are equivalent, the solutions $\hat{\bcU}(\bu)$ are identical up to the residual tolerance. 
The Nested AA solver generally requires more (inner) iterations to converge than the Nested Newton solver does, especially on harder problems. 
We also note that the comparison of iteration counts here does not fully reflect the performance of the solvers in terms of computational time.
In particular, as discussed in Section~\ref{subsec:nested_FP}, opacity kernel evaluations make outer iterations much more computationally expensive than the inner iterations.  
In addition, the iterations in Anderson acceleration are computationally cheaper than the ones in Newton's method, since they do not require constructing and inverting the Jacobian matrix.  
We defer the comparison of computational times to Section~\ref{subsec:deleptonization}, where a more realistic test problem is considered.

\begin{figure}[h]
	\captionsetup[subfigure]{justification=centering}
	\subfloat[High collision rate, $\rho = 1.084\times10^{14}$~g~cm$^{-3}$, $T = 1.845\times10^{11}$~K, $Y_e = 0.2728$]
	{\includegraphics[width=0.49\linewidth]{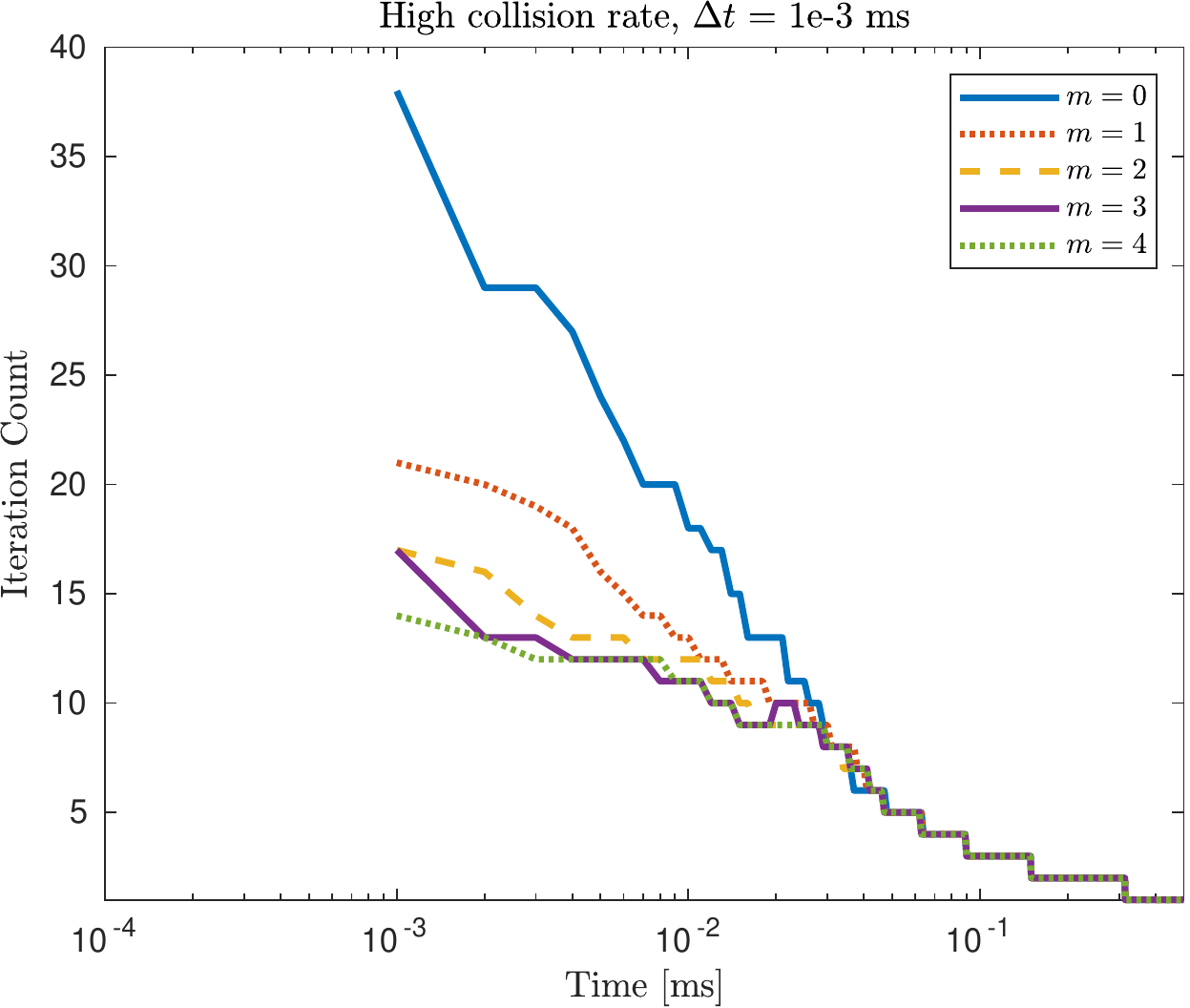}
	\label{fig:AA_1}}~~~~~~
	\subfloat[Low collision rate, $\rho = 1.032\times10^{12}$~g~cm$^{-3}$, $T = 8.806\times10^{10}$~K, $Y_e = 0.1347$]	
	{\includegraphics[width=0.47\linewidth]{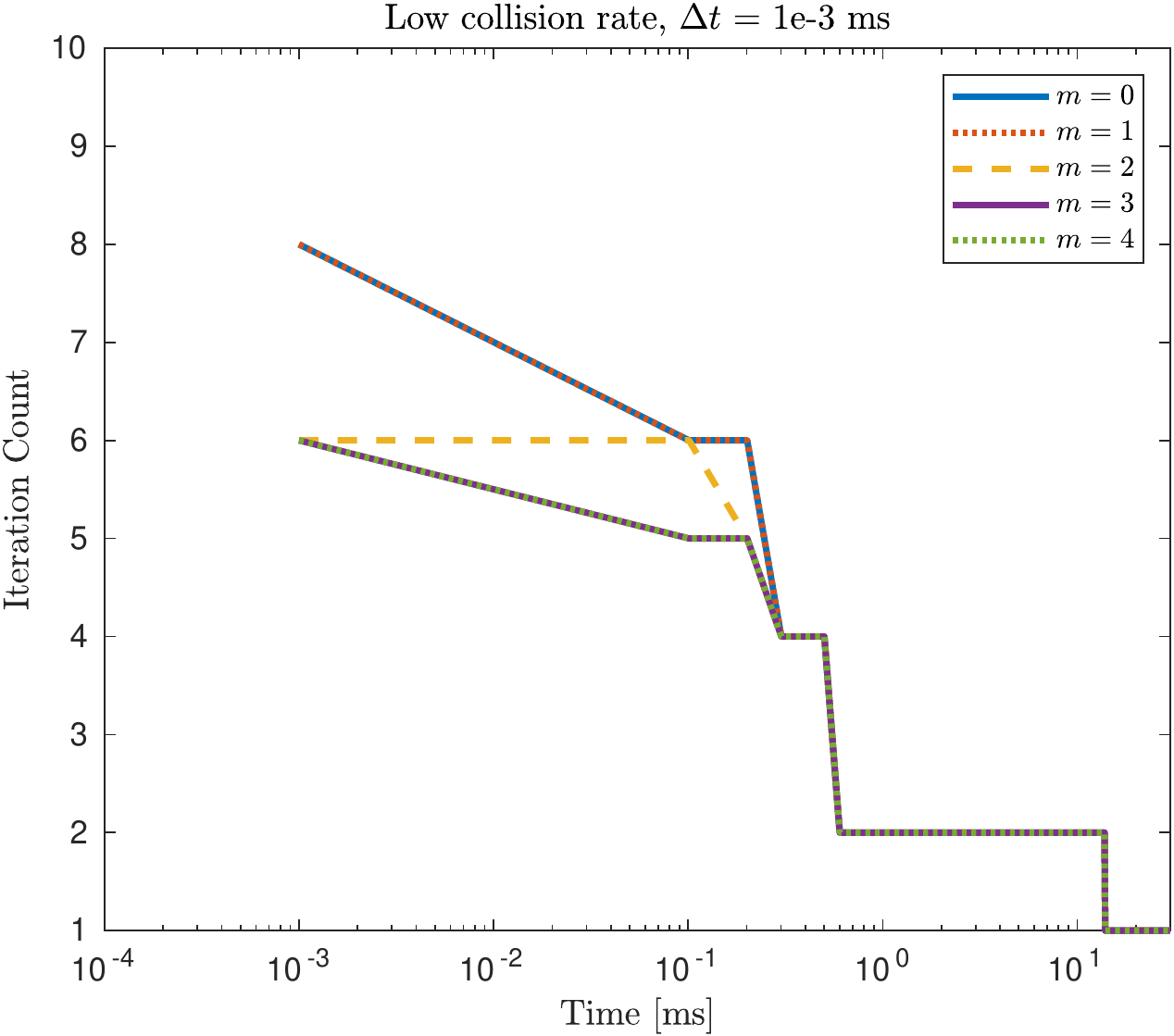}
	\label{fig:AA_2}}	
	
	\caption{Iteration counts of the Coupled AA solver with truncation parameter $m=0,\dots,4$ on relaxation problems.}
	\label{fig:AA}
	
\end{figure}

Next, we explore the effect of the truncation parameter $m$ on the convergence of Anderson acceleration by comparing the iteration count for the Coupled AA solver with different values of $m$, on both the high and low collision rate relaxation problems considered in the previous test.  
In this comparison, the time step size fixed to $\dt=10^{-3}$~ms, and $m$ varies from 0 to 4, where $m=0$ resembles the simple Picard iteration.
From the results reported in Figure~\ref{fig:AA}, we observe that, on these problems, Anderson acceleration does converge faster as the value of $m$ increases, while the marginal benefit becomes insignificant for $m\geq2$.
This result justifies our choice of $m=2$ in the numerical tests throughout the paper, since higher values of $m$ lead to larger memory footprints and more expensive least-squares solves in Anderson acceleration with minimal improvement in the iteration counts.

\begin{figure}[h]
	\captionsetup[subfigure]{justification=centering}
	\subfloat[Iteration Counts]
	{\includegraphics[width=0.48\linewidth]{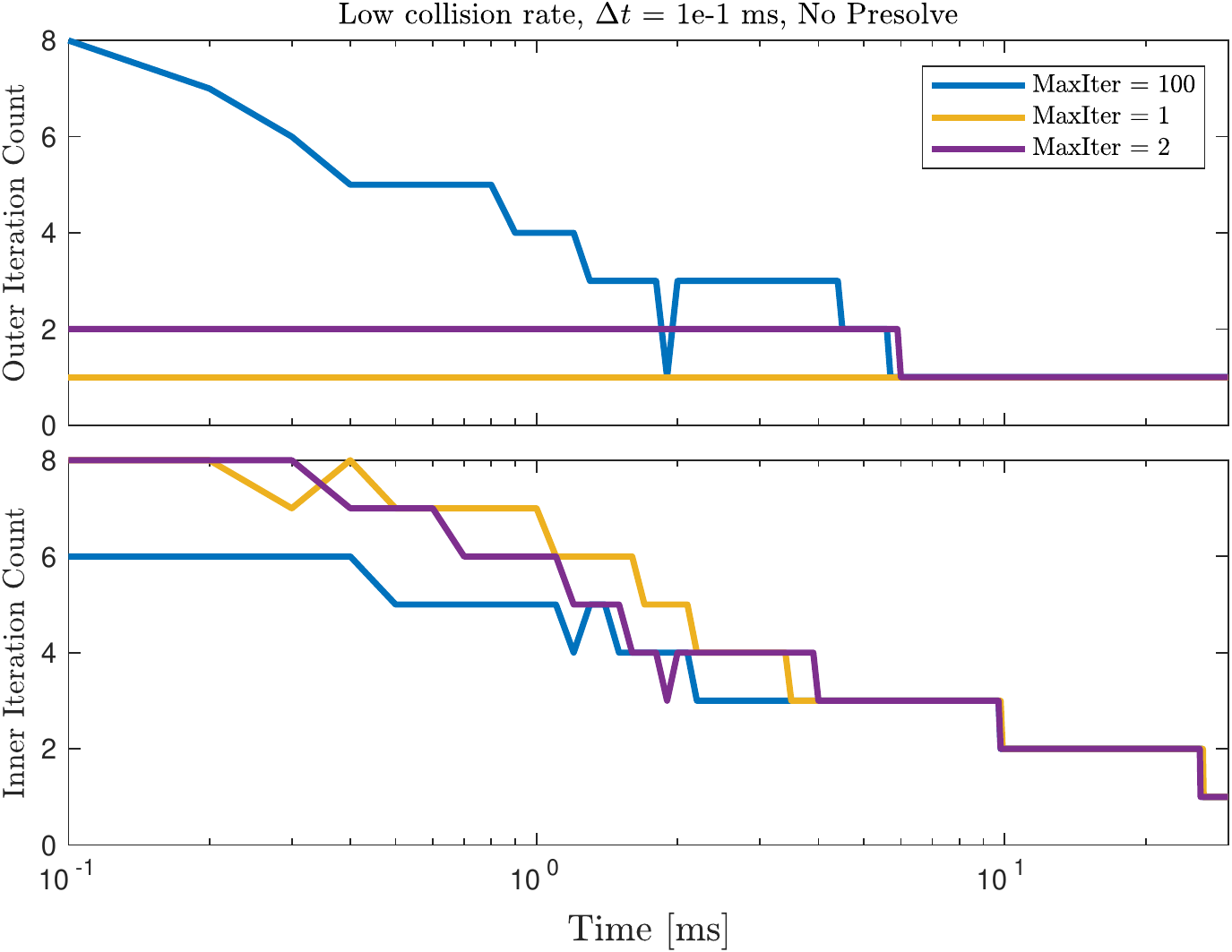}
		\label{fig:NoPre_Iter}}~~~
	\subfloat[Energy-integrated Number Densities]	
	{\includegraphics[width=0.487\linewidth]{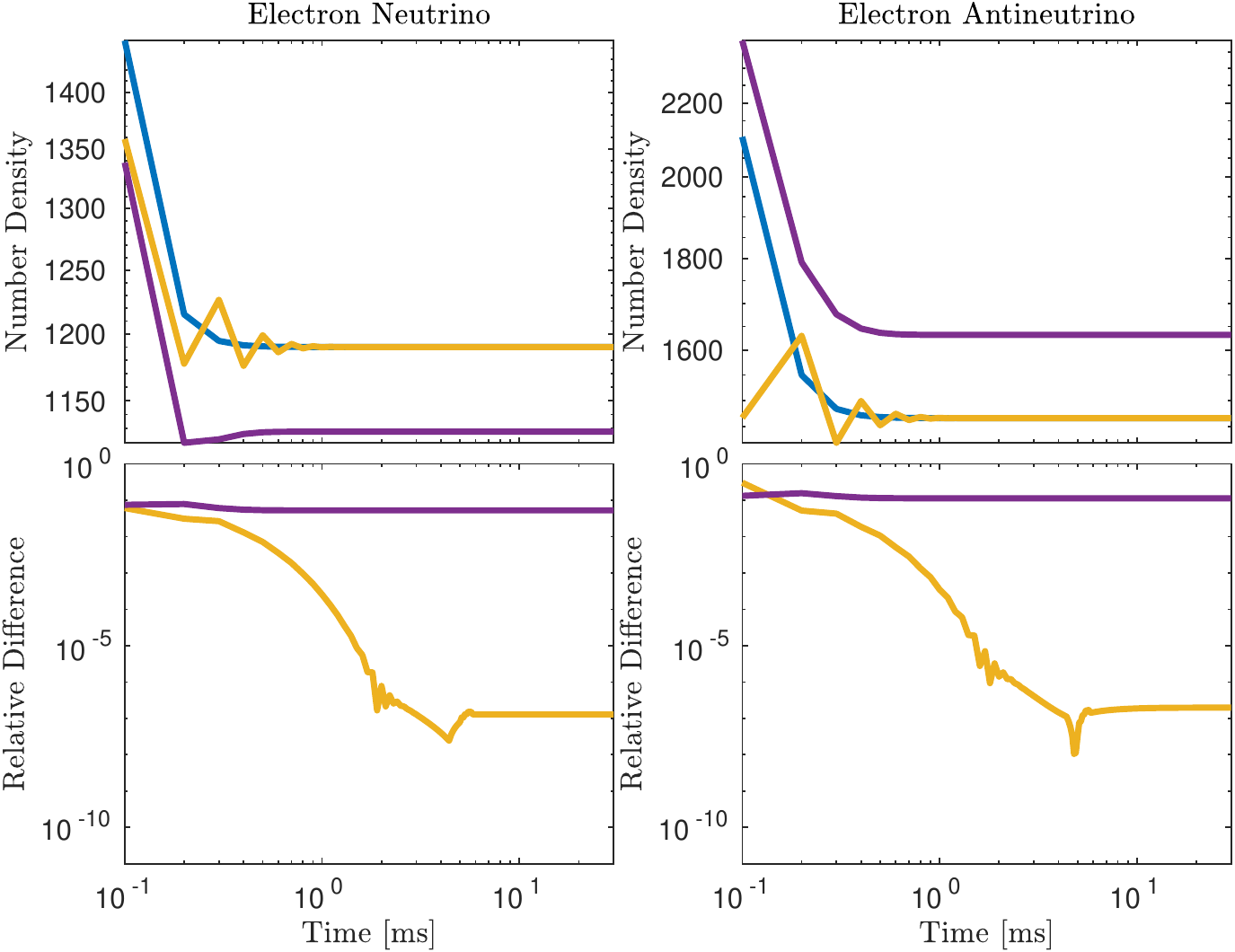}
		\label{fig:NoPre_Rad}}~~~	\\
	\subfloat[Temperature]
	{\includegraphics[width=0.345\linewidth]{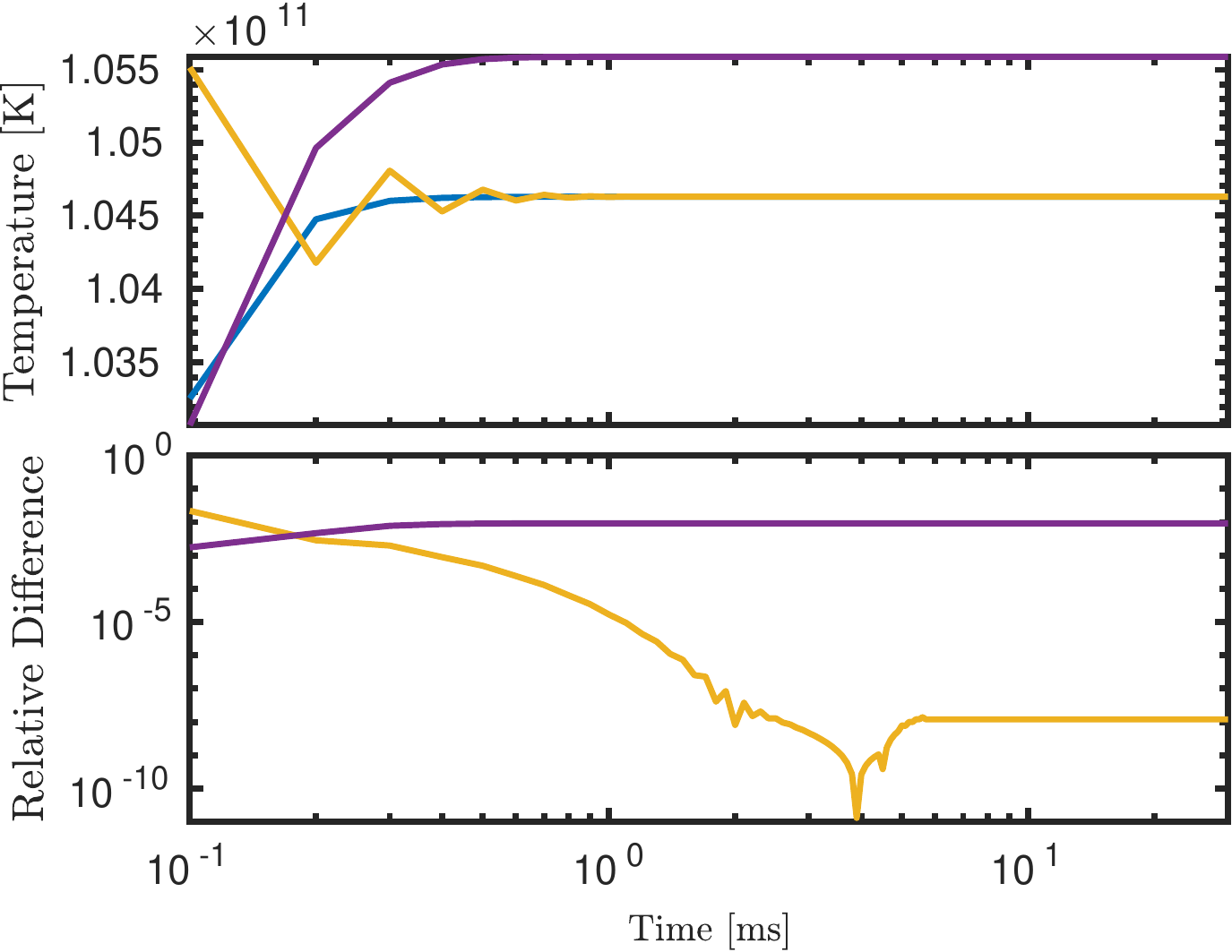}
		\label{fig:NoPre_T}}~~~
	\subfloat[Electron Fraction]	
	{\includegraphics[width=0.33\linewidth]{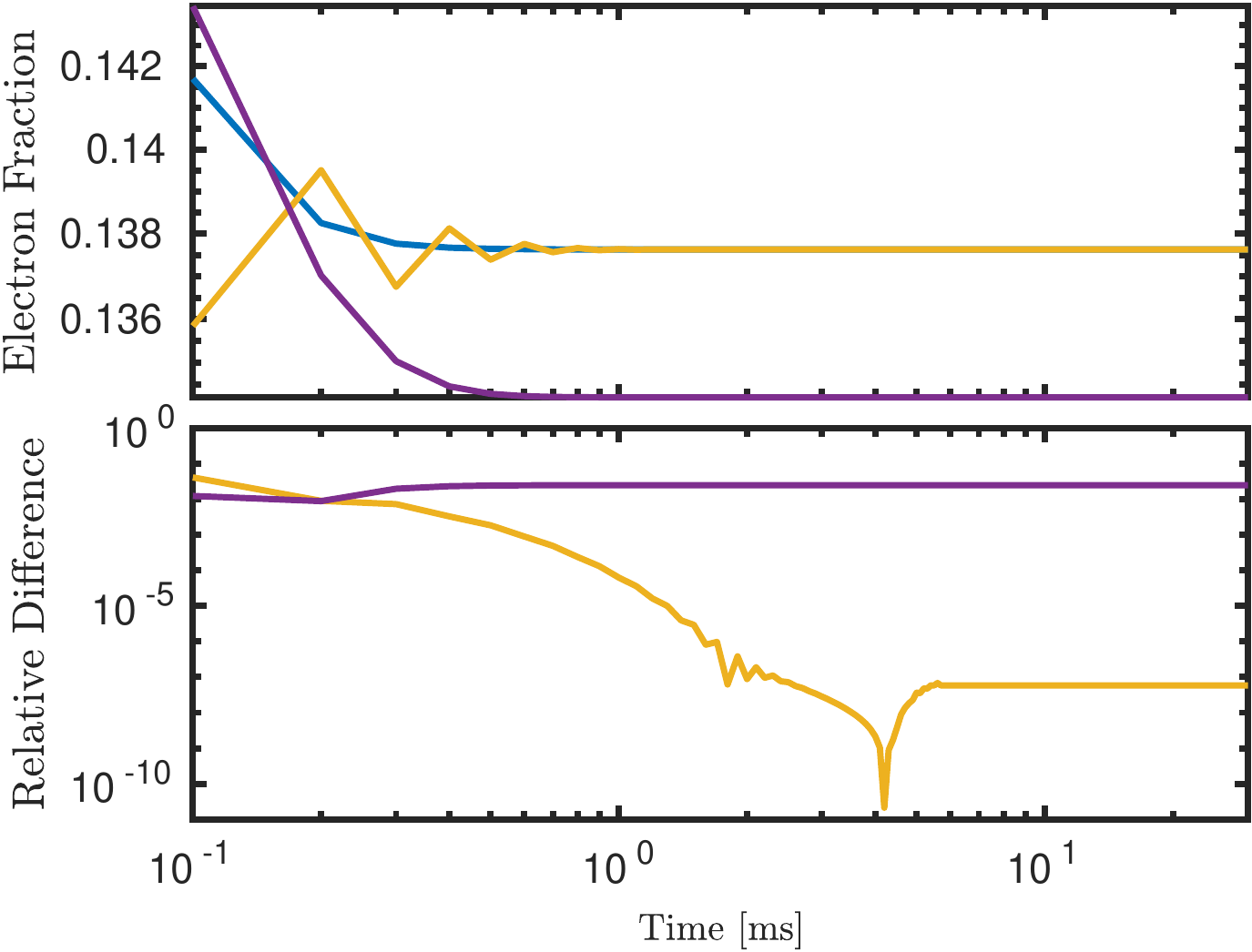}
		\label{fig:NoPre_Ye}}~~~
	\subfloat[Relative Change in Lepton Number and Energy]		{\includegraphics[width=0.315\linewidth]{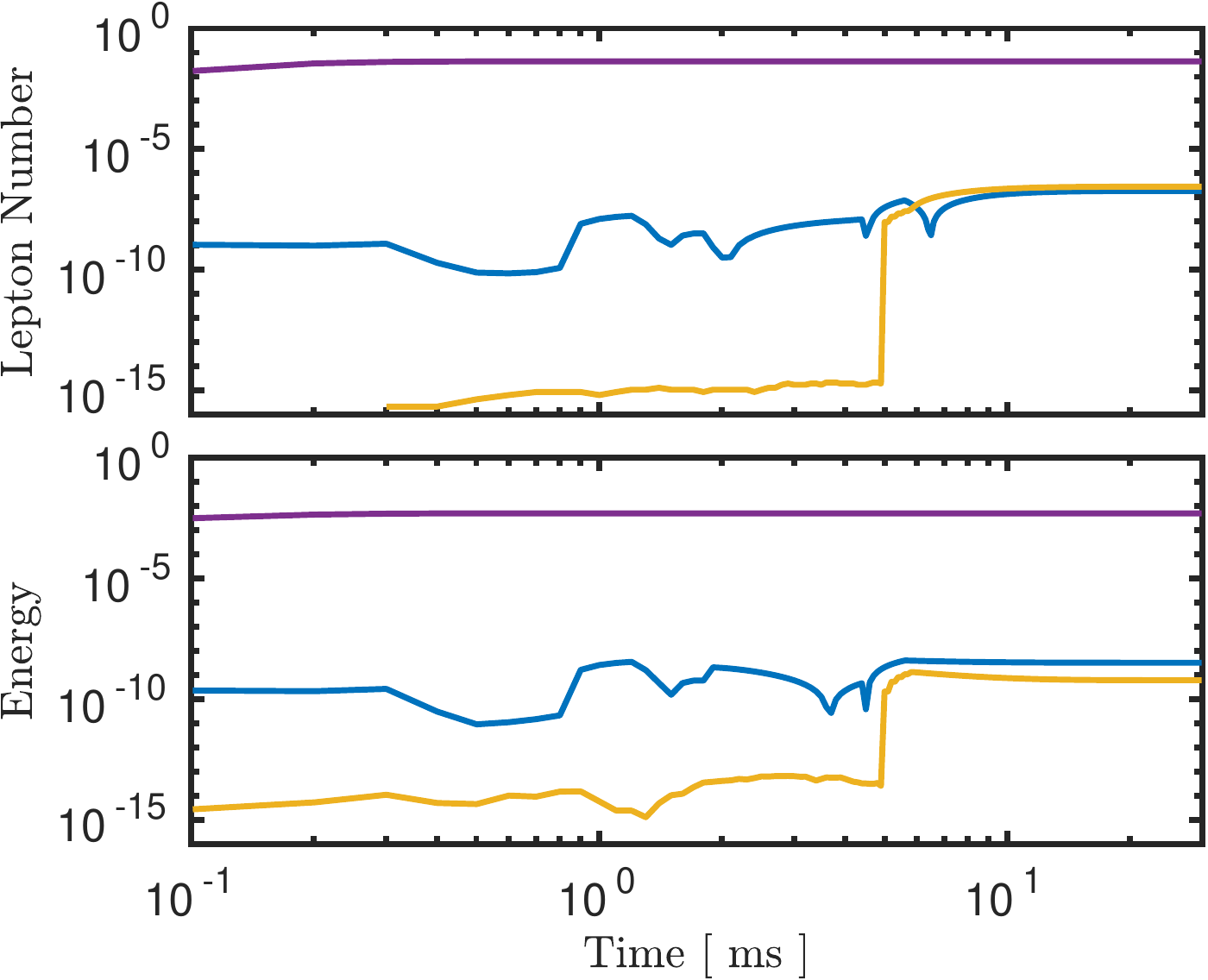}
		\label{fig:NoPre_Con}}~~~
	\caption{Iteration count, energy-integrated number densities, temperature, electron fraction, and lepton number and energy conservation results for the Nested AA solver on the relaxation problem with low collision rate $\rho = 1.032\times10^{12}$~g~cm$^{-3}$, $T = 8.806\times10^{10}$~K, $Y_e = 0.1347$. The solver is applied without the presolve step and run to various maximum outer iterations (MaxIter), with time step $\dt = 10^{-1}$~ms.}
	\label{fig:NoPre}
	\end{figure}

\begin{figure}[h]
	\captionsetup[subfigure]{justification=centering}
	\subfloat[Iteration Counts]
	{\includegraphics[width=0.48\linewidth]{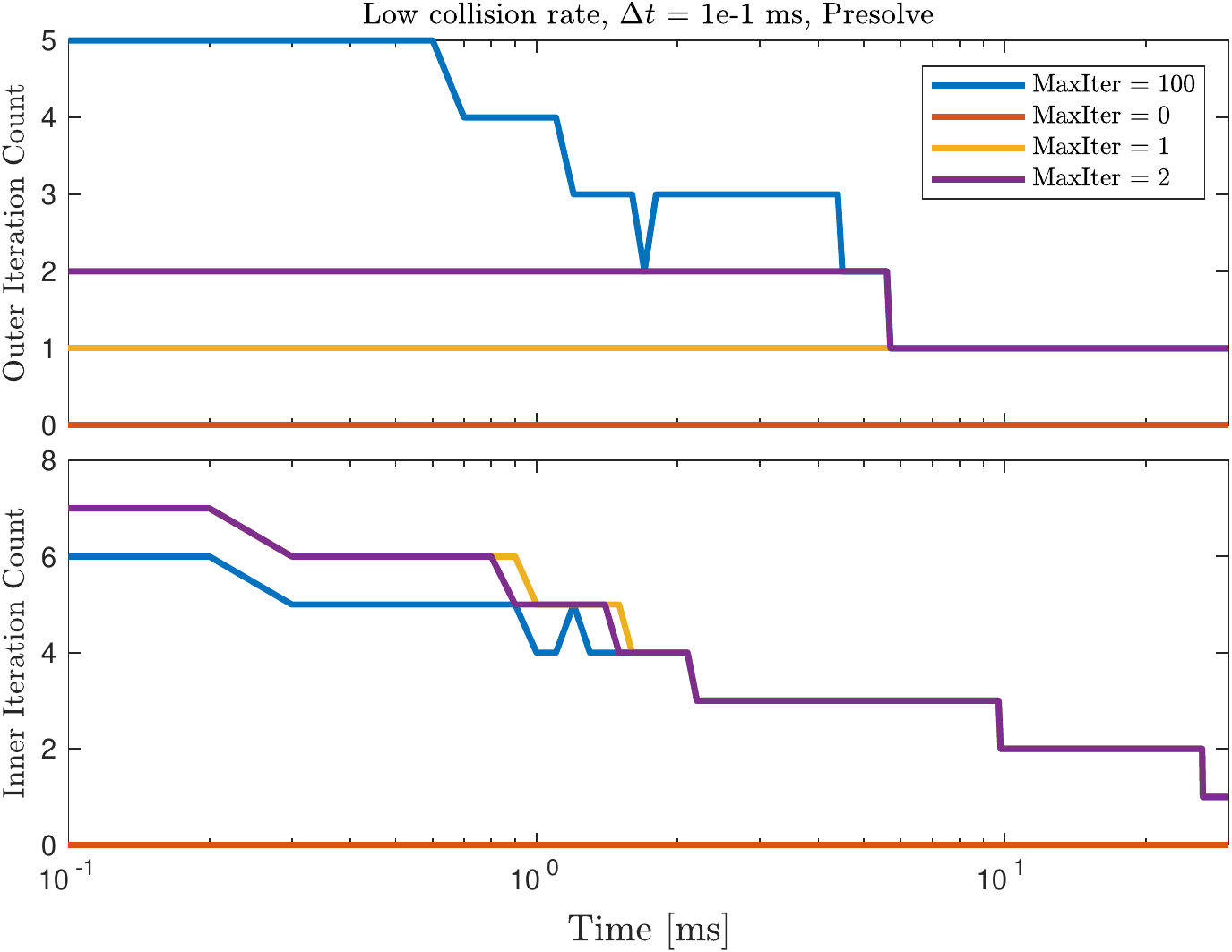}
		\label{fig:Pre_Iter}}~~~
	\subfloat[Energy-integrated Number Densities]	
	{\includegraphics[width=0.487\linewidth]{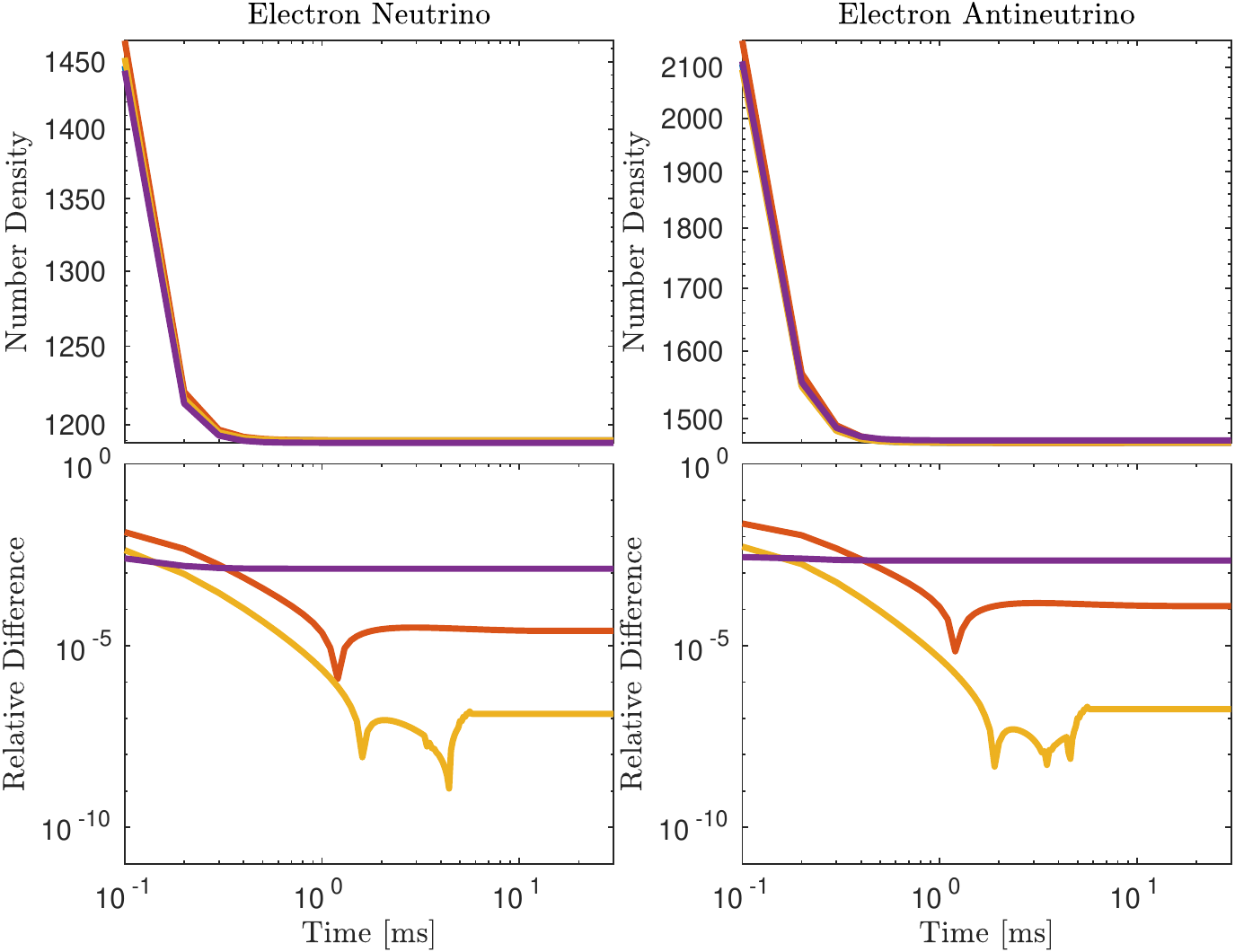}
		\label{fig:Pre_Rad}}~~~	\\
	\subfloat[Temperature]
	{\includegraphics[width=0.345\linewidth]{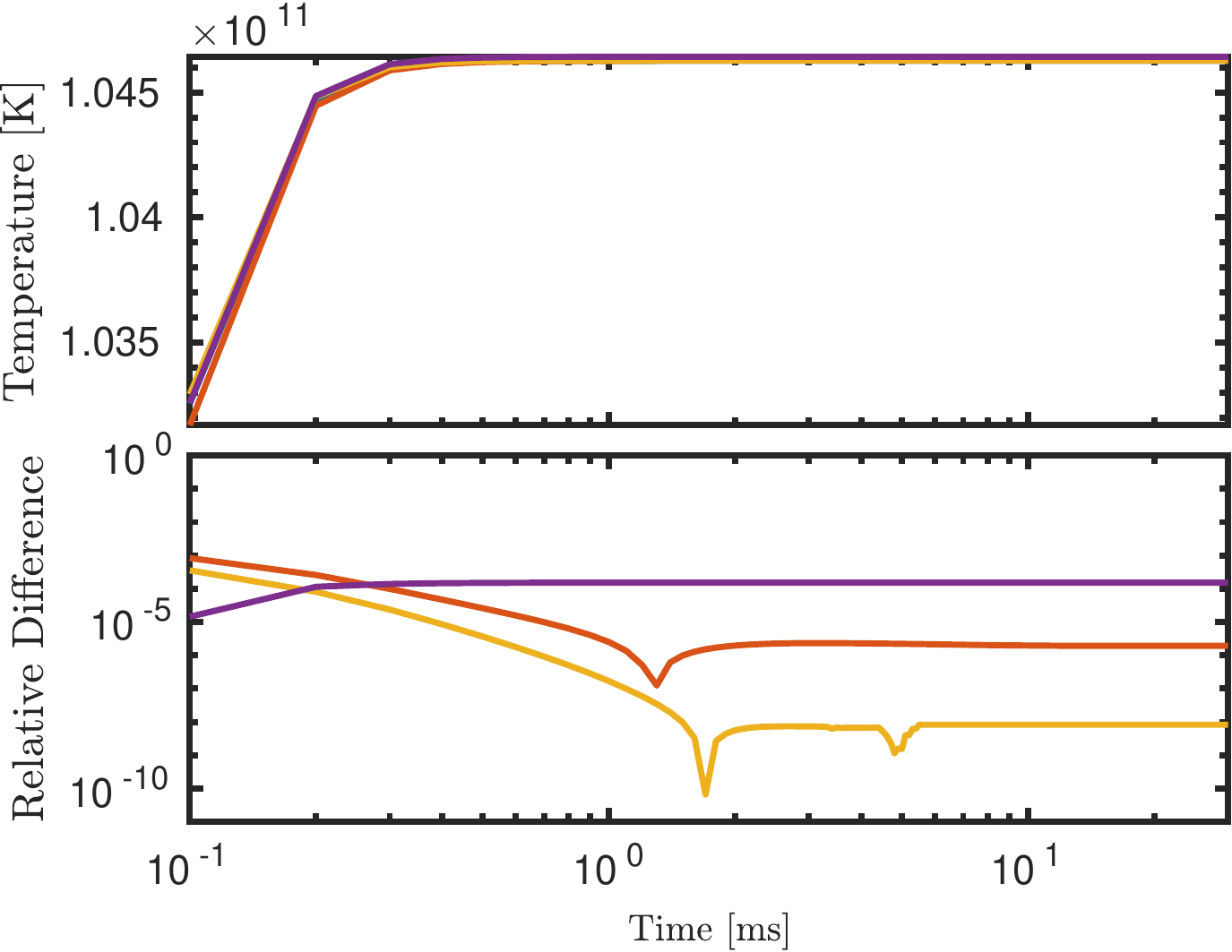}
		\label{fig:Pre_T}}~~~
	\subfloat[Electron Fraction]	
	{\includegraphics[width=0.33\linewidth]{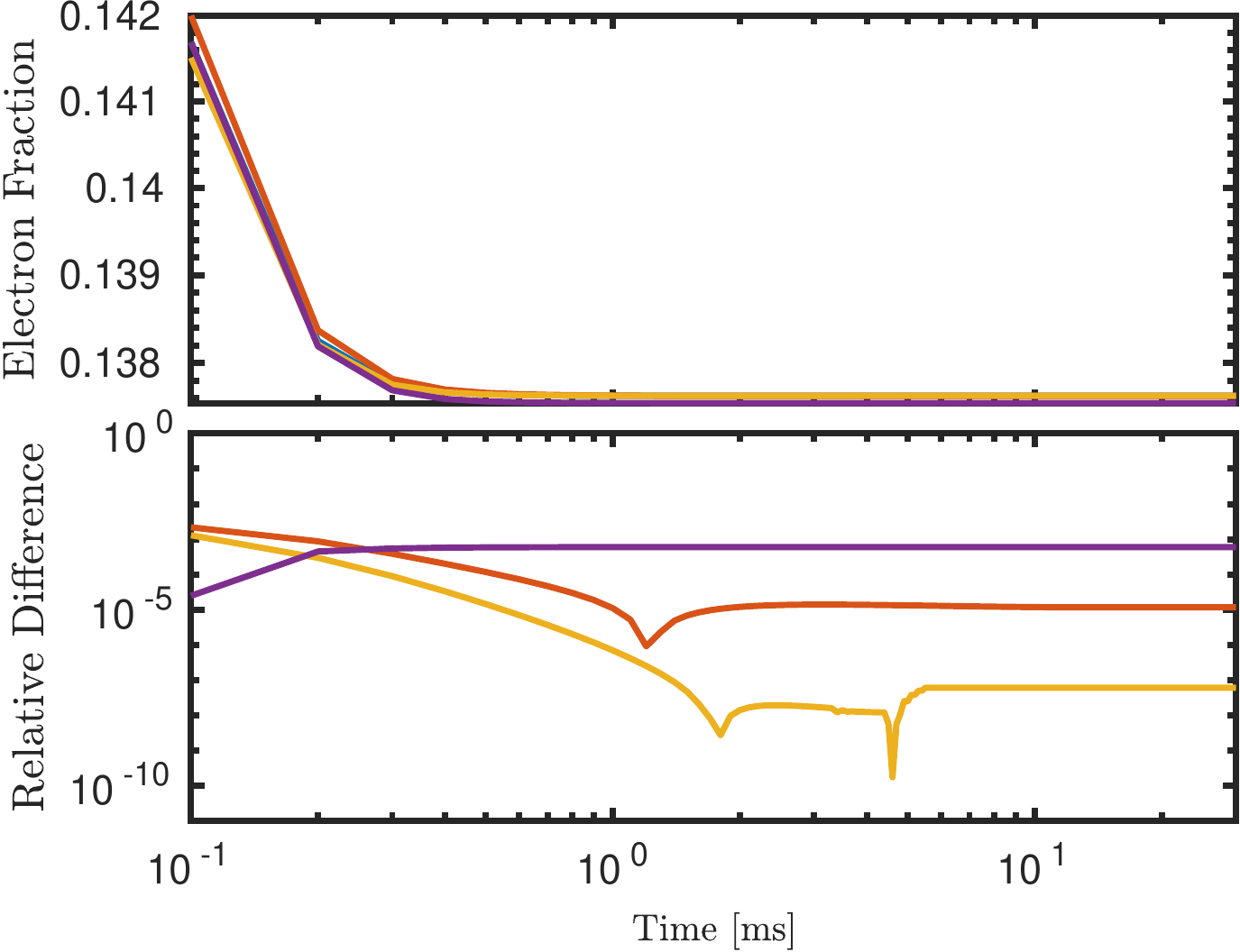}
		\label{fig:Pre_Ye}}~~~	
	\subfloat[Relative Change in Lepton Number and Energy]	
	{\includegraphics[width=0.315\linewidth]{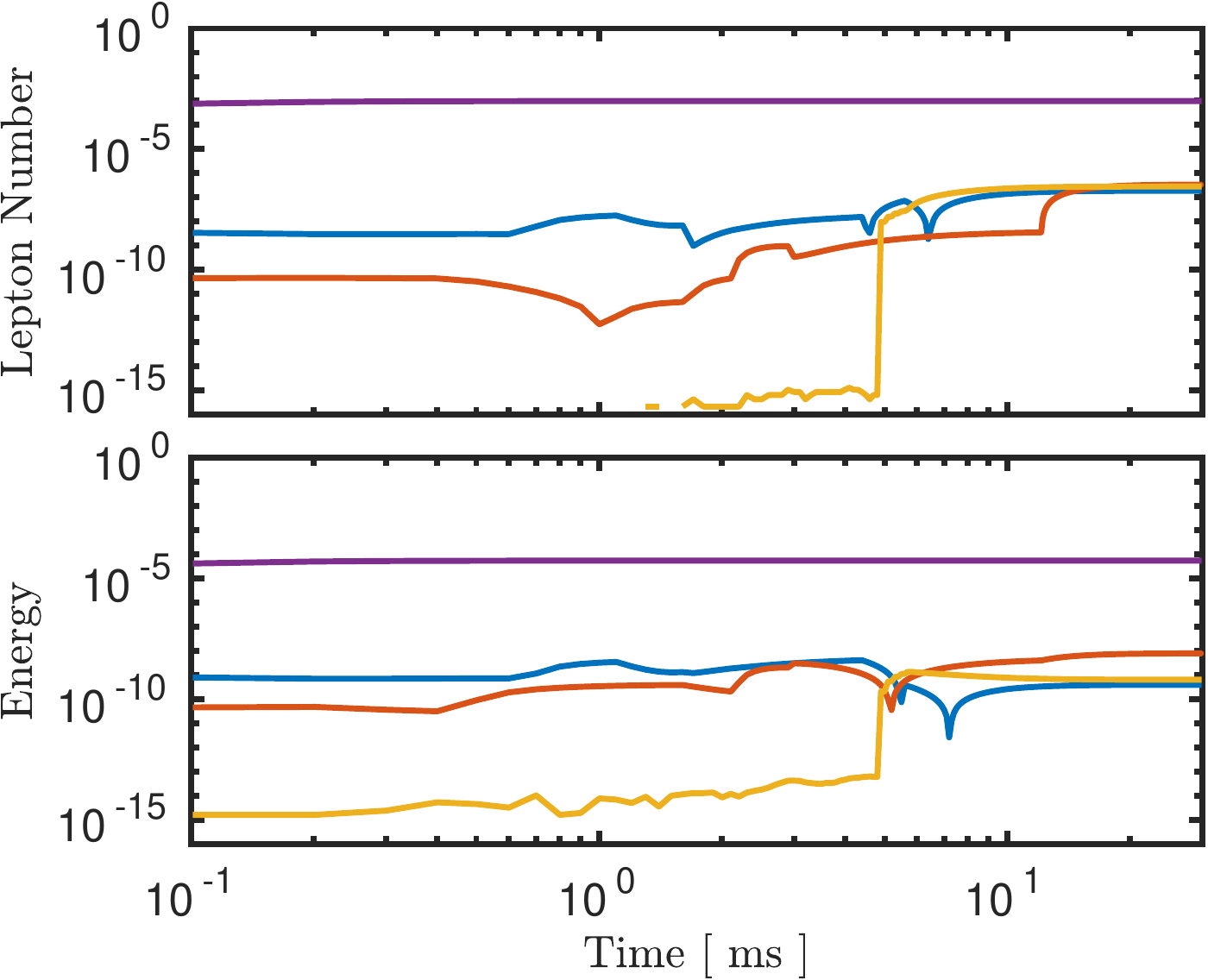}
		\label{fig:Pre_Con}}~~~		
	\caption{Iteration count, energy-integrated number densities, temperature, electron fraction, and lepton number and energy conservation results for the Nested AA solver on the relaxation problem with low collision rate $\rho = 1.032\times10^{12}$~g~cm$^{-3}$, $T = 8.806\times10^{10}$~K, $Y_e = 0.1347$. The solver is applied with the presolve step and run to various maximum outer iterations (MaxIter), with time step $\dt = 10^{-1}$~ms.}
	\label{fig:Pre}
	
\end{figure}

We next investigate how early termination of the iterative solvers affects solution accuracy.
We choose to test the Nested~AA solver with early termination of the outer loop, and compare the resulting solution to a converged reference solution at the final time.  
Here the solution process is terminated either when the convergence criteria \eqref{eq:C1} are satisfied or when the outer iteration count for solving Eq.~\eqref{eq:NestedOuter} reaches a preset maximum (MaxIter).  
We test the solver on the relaxation problem with low collision rate and time step $\dt = 10^{-1}$~ms, which is much larger than the usual time step for a stable explicit scheme.  
The choice of a large time step makes the effect of early termination more pronounced. 
Figure~\ref{fig:NoPre} reports the iteration counts for MaxIter = 1 and MaxIter = 2 on the test problem, along with the electron neutrino and antineutrino (energy-integrated) number densities, temperatures, and electron fractions at the final time. 
These results are compared to a fully converged reference solution (MaxIter = 100), where the nested fixed-point solver converges well before the nominal maximal outer iteration is reached, as shown in Figure~\ref{fig:NoPre_Iter}.
Here the ``presolve" step discussed in Section~\ref{subsec:implementation} is turned off. 
We note that when setting MaxIter = 1, the Nested AA solver (w/o presolve) resembles an approach used in earlier works such as \cite{just_etal_2015}, where the radiation quantities are updated using opacities computed from the lagged matter states, i.e., matter states at time $t^n$ are used to update the radiation quantities at $t^{n+1}$.  
The results in Figure~\ref{fig:NoPre} show that when MaxIter = 1, while the relative differences in the earlier time steps are rather significant (from 30\% to 0.5\%), the solution still converges to an identical (up to the solver tolerance) equilibrium at the final time.  
However, when MaxIter = 2, it is clear that the solution converges to a different equilibrium.
Indeed, it can be seen from Figure~\ref{fig:NoPre_Con} that the lepton number and energy are not conserved in the solution with MaxIter = 2, while they are conserved up to the solver tolerance in the fully converged solution (MaxIter = 100). When MaxIter = 2, the changes in lepton number and energy lead to a different equilibrium. 
As for the solution with MaxIter = 1, despite the fact that the solution process of the nonlinear system Eq.~\eqref{eq:CoupledSystem} is terminated early, the lepton number and energy are actually conserved in exact arithmetics. This is because, when MaxIter = 1, the early terminated nested iterates satisfy Eqs.~\eqref{eq:CoupledSystem_Ye}--\eqref{eq:CoupledSystem_eps} exactly, which enforces lepton number and energy convergences and leads to the correct equilibrium. However, the early terminated solutions generally do not satisfy Eqs.~\eqref{eq:CoupledSystem_JNe}--\eqref{eq:CoupledSystem_JNeb} and result in rather inaccurate solutions in the transient state.
In Figure~\ref{fig:Pre}, we repeat the test but with the presolve step turned on.  
Here we also include the option MaxIter = 0 into the comparison, in which the radiation and matter quantities are only updated in the presolve step, i.e., with the NES and pair processes ignored.  
From Figure~\ref{fig:Pre}, it can be observed that even with MaxIter = 0, the presolve step gives fairly accurate solution at the final time, as the lepton number and energy are conserved up to the solver tolerance in the presolve step.
When the maximum iteration is allowed to be higher, the presolve step improves the solution accuracy, at least in the earlier stage.  
We observe that while the solution is still inaccurate when MaxIter = 2, it is much closer to the reference solution when the presolve is turned on.

\subsection{Deleptonization problem}
\label{subsec:deleptonization}

We further investigate and compare the performance of the nonlinear solvers in a more realistic setting with a proto-neutron star deleptonization problem, using initial matter profiles from spherically symmetric CCSN simulations.  
In this test we solve the full moment equations presented in Section~\ref{subsec:moments}, using the DG phase-space discretization in Section~\ref{subsec:space_energy_DG} and the IMEX time integration scheme in Section~\ref{subsec:IMEX}.  

\begin{figure}[h]
	
	\captionsetup[subfigure]{justification=centering}
	\subfloat[Mass Density]
	{\begin{minipage}{0.49\textwidth}
		\includegraphics[width=\linewidth]{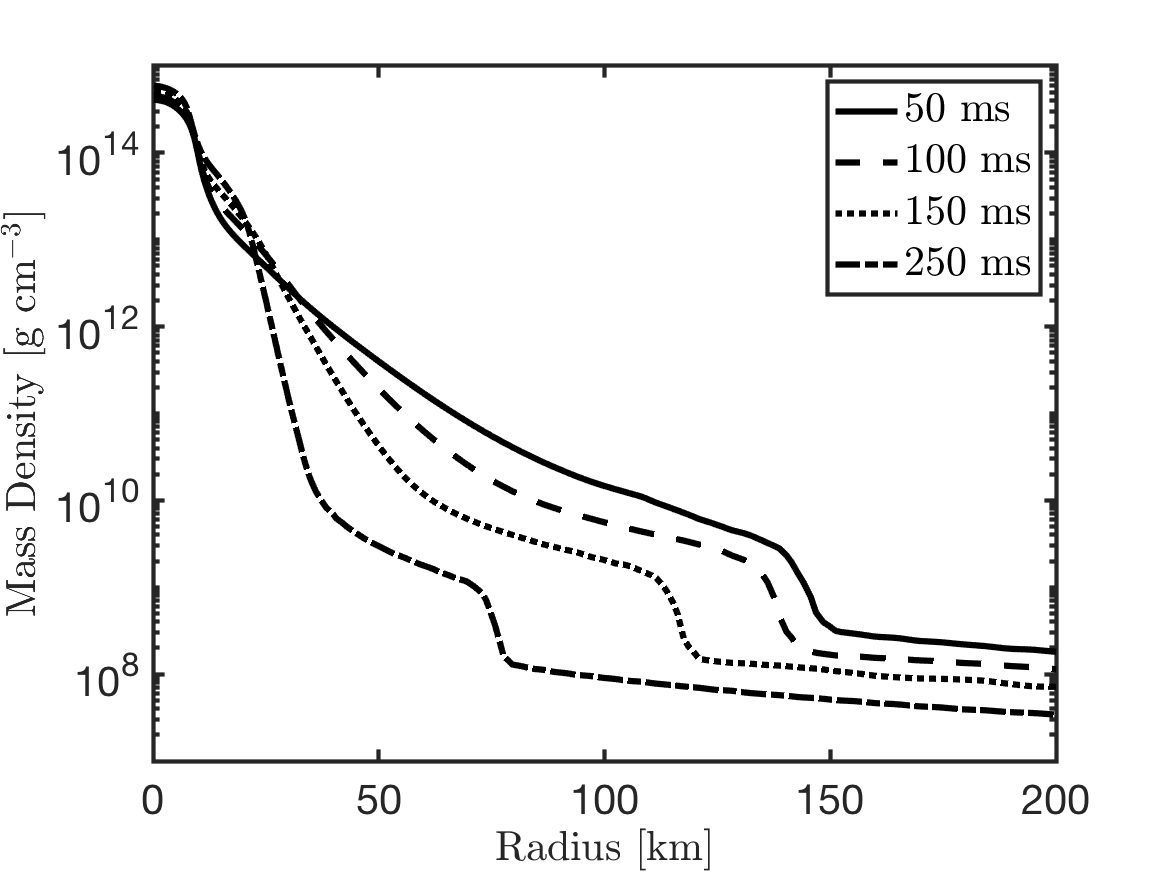}
		\label{fig:DeleptonizationWaveInitial_a}
	\end{minipage}
	}
	\subfloat[Temperature]
	{\begin{minipage}{0.49\textwidth}
			\includegraphics[width=\linewidth]{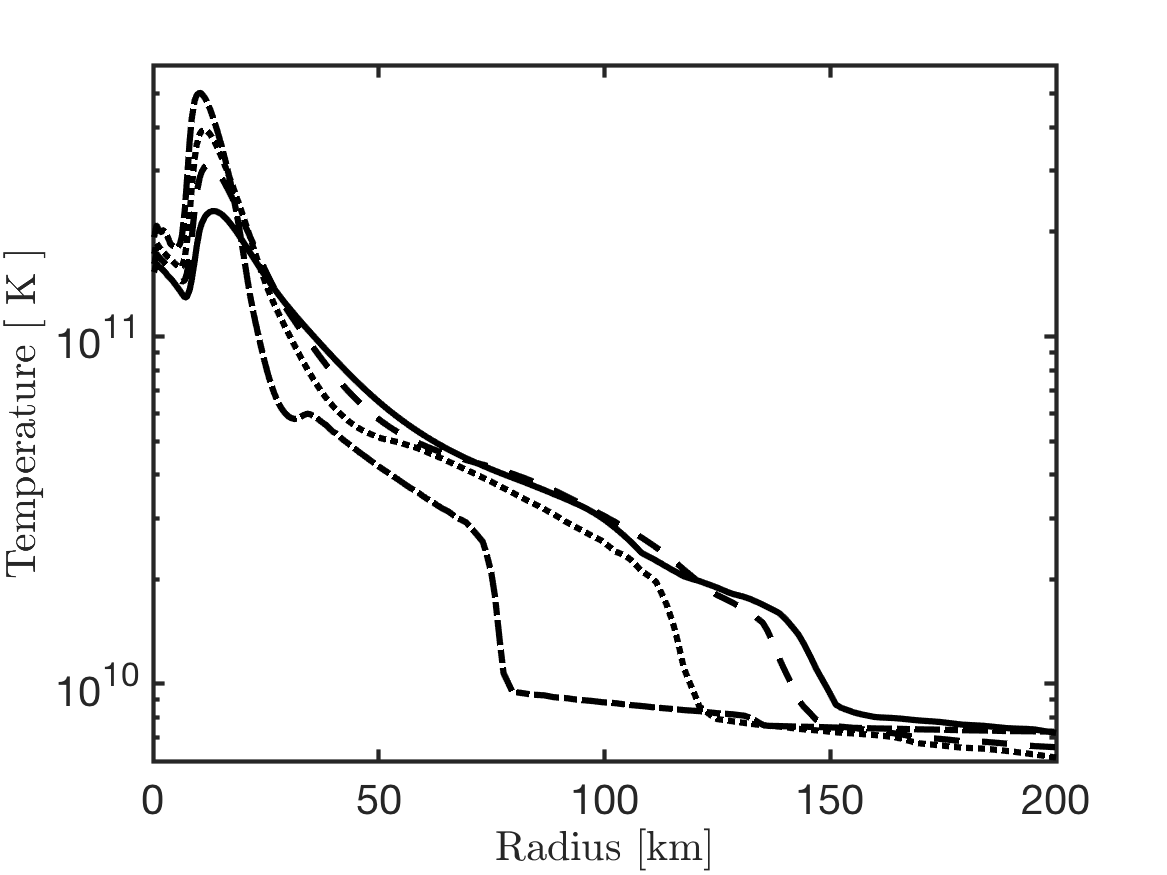}
			\label{fig:DeleptonizationWaveInitial_b}
		\end{minipage}
	} \\
	\subfloat[Electron Fraction]
	{\begin{minipage}{0.49\textwidth}
			\includegraphics[width=\linewidth]{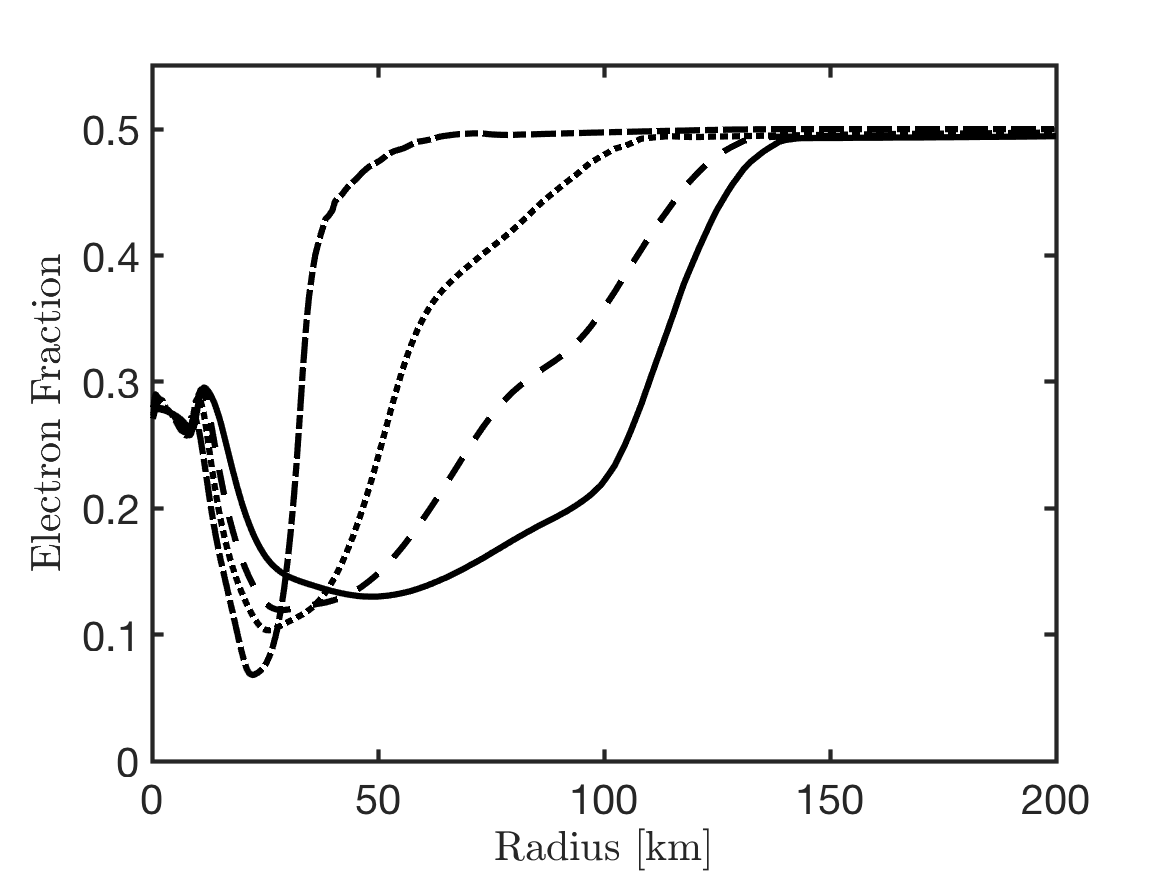}
			\label{fig:DeleptonizationWaveInitial_c}
		\end{minipage}
	}
	\subfloat[Neutrinosphere Radius]
	{\begin{minipage}{0.49\textwidth}
			\includegraphics[width=\linewidth]{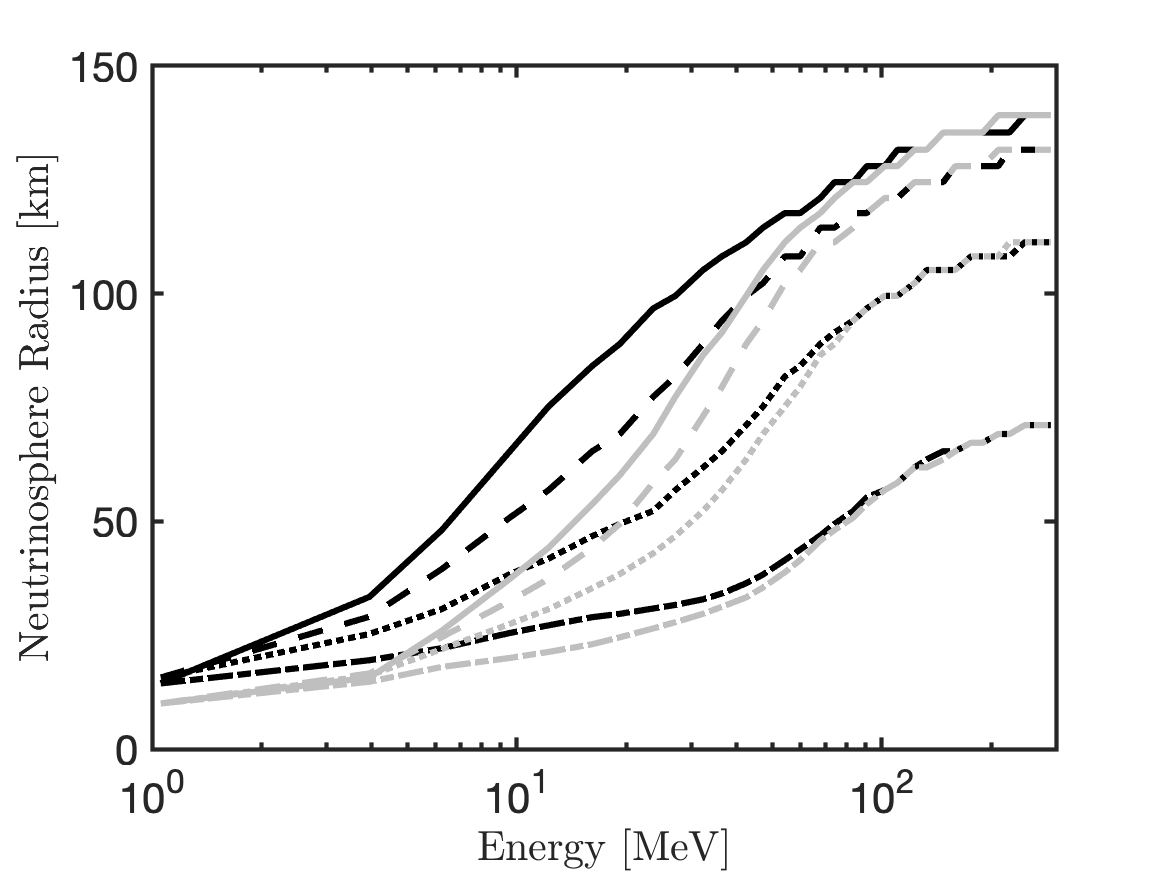}
			\label{fig:DeleptonizationWaveInitial_d}
		\end{minipage}
	}

	\caption{Initial matter profiles used in the deleptonization problem, taken from \cite{liebendorfer_etal_2005}.  Plotted versus radius are mass density (upper left panel), temperature (upper right panel), and electron fraction (lower left panel) for various post-bounce times: $50$~ms (solid), $100$~ms (dashed), $150$~ms (dotted), and $250$~ms (dash-dot).  The neutrinosphere radii, defined in Eq.~\eqref{eq:neutrinosphereRadius}, for the respective profiles --- for electron neutrinos (black) and electron antineutrinos (grey) --- are also plotted versus neutrino energy (lower right panel).}
	\label{fig:DeleptonizationWaveInitial}
	
\end{figure}

For this test, we adopt matter profiles from \cite{liebendorfer_etal_2005}.  
Specifically, we use profiles for mass density, temperature, and electron fraction obtained with the \textsc{vertex} code using a 15~$M_{\odot}$ progenitor from \cite{wooseleyWeaver_1995} (model G15 in \cite{liebendorfer_etal_2005}).  
Other thermodynamics quantities (e.g., internal energy and electron, proton, and neutron chemical potentials) are obtained from the tabulated SFHo EoS \citep{steiner_etal_2013}.
To investigate sensitivity to conditions encountered over an extended period covering the neutrino heating phase, we run the comparison on profiles taken at $t=50$, $100$, $150$, and $250$~{ms} after core bounce.  
The initial matter profiles are plotted in Figure~\ref{fig:DeleptonizationWaveInitial}.  

Since we do not have the radiation quantities from \cite{liebendorfer_etal_2005}, to initialize the radiation field, we adopt the analytical distribution function from the homogeneous sphere test, $f_{\mbox{\tiny \sc Hs}}$, (e.g., \cite{smit_etal_1997}), which is a solution to the steady state transport problem of radiation emanating from a sphere whose constant absorption opacity and emissivity inside a radius $R_{0}$ are $\chi_{0}$ and $\chi_{0}\,f_{0}$, respectively:
\begin{equation}
  f_{\mbox{\tiny \sc Hs}}[f_{0},\chi_{0},R_{0}](r,\mu)=f_{0}\,\big(1-e^{-\chi_{0}\,s[R_{0}](r,\mu)}\big),
  \label{distributionHS}
\end{equation}
where $f_{0}$ is an isotropic equilibrium distribution, 
\begin{equation}
  s[R_{0}](r,\mu)
  =\left\{
  \begin{array}{lll}
    r\,\mu+R_{0}\,g[R_{0}](r,\mu) & \mbox{if}\quad r<R_{0}, & \mu\in[-1,+1], \\
    2\,R_{0}\,g[R_{0}](r,\mu) & \mbox{if}\quad r \ge R_{0}, & \mu\in[(1-(R_{0}/r)^{2})^{1/2},+1], \\
    0 & \mbox{otherwise},
  \end{array}
  \right.
\end{equation}
and $g[R_{0}](r,\mu)=[1-(r/R_{0})^{2}(1-\mu^{2})]^{1/2}$.  

To adopt the homogeneous sphere distribution to the current setting, we first estimate the energy-dependent neutrinosphere radius $R_{\nu}(\varepsilon)$
\begin{equation}
  \int_{R_{\nu}(\varepsilon)}^{\infty}\chi(\varepsilon,r)\,dr = \f{2}{3},
  \label{eq:neutrinosphereRadius}
\end{equation}
where $\chi$ is the absorption opacity.  
(Neutrinosphere radii for the various initial profiles used here are plotted versus neutrino energy in the lower right panel of Figure~\ref{fig:DeleptonizationWaveInitial}.)  
The neutrino distribution function is then set to
\begin{equation}
  f(r,\varepsilon,\mu)=
  \left\{
  \begin{array}{ll}
    f_{\mbox{\tiny \sc Hs}}[f_{0}(r,\varepsilon),\chi(r,\varepsilon),R_{\nu}(\varepsilon)](r,\mu), & r < R_{\nu}(\varepsilon) \\
    f_{\mbox{\tiny \sc Hs}}[f_{0}(R_{\nu}(\varepsilon)),\chi(R_{\nu}(\varepsilon)),R_{\nu}(\varepsilon)](r,\mu), &  r \ge R_{\nu}(\varepsilon),
  \end{array}
  \right.
\end{equation}
where $f_{0}(r,\varepsilon)$ is taken to be the Fermi-Dirac distribution in Eq.~\eqref{eq:FermiDirac}.  
Finally, the initial moments are computed as
\begin{equation}
  \big\{\,\mathcal{J},\,\mathcal{H}\,\big\}(r,\varepsilon) = \f{1}{2}\int_{-1}^{1}f(r,\varepsilon,\mu)\,\mu^{\{0,1\}}\,d\mu.  
\end{equation}
(The same procedure is adopted to initialize the antineutrinos.)

\begin{figure}[h]
	\centering
	{\includegraphics[width=0.8\linewidth]{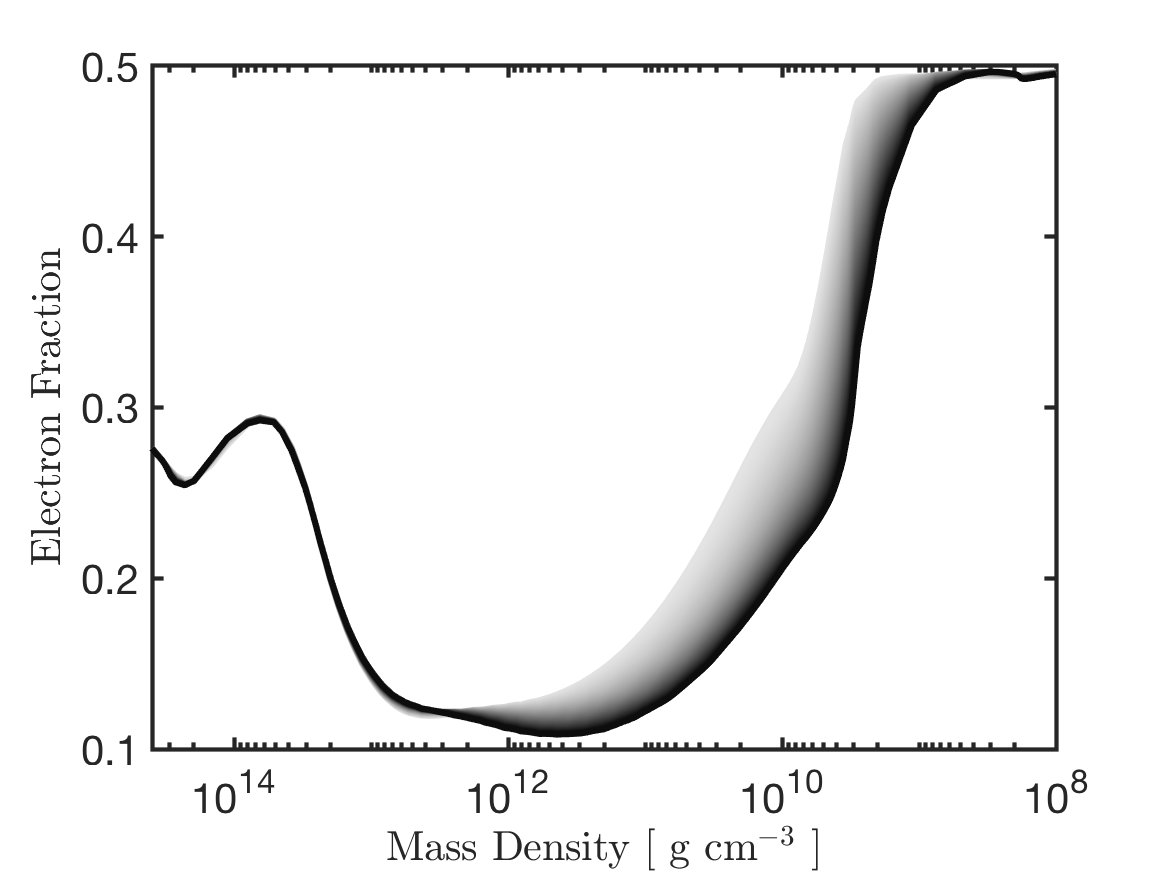}}
	\caption{Electron fraction versus mass density over $10$~ms of evolution from the initial state given by the $100$~ms post-bounce matter profile in Figure~\ref{fig:DeleptonizationWaveInitial}.  The time evolution is indicated by the grayscale, which goes from lightest (initial state) to darkest (final state).}
	\label{fig:DeleptonizationWaveEvolution}
\end{figure}

\begin{figure}[h]	
	\captionsetup[subfigure]{justification=centering}
	\subfloat[Lepton number and energy conservation]	
	{	\includegraphics[width=0.50\linewidth]{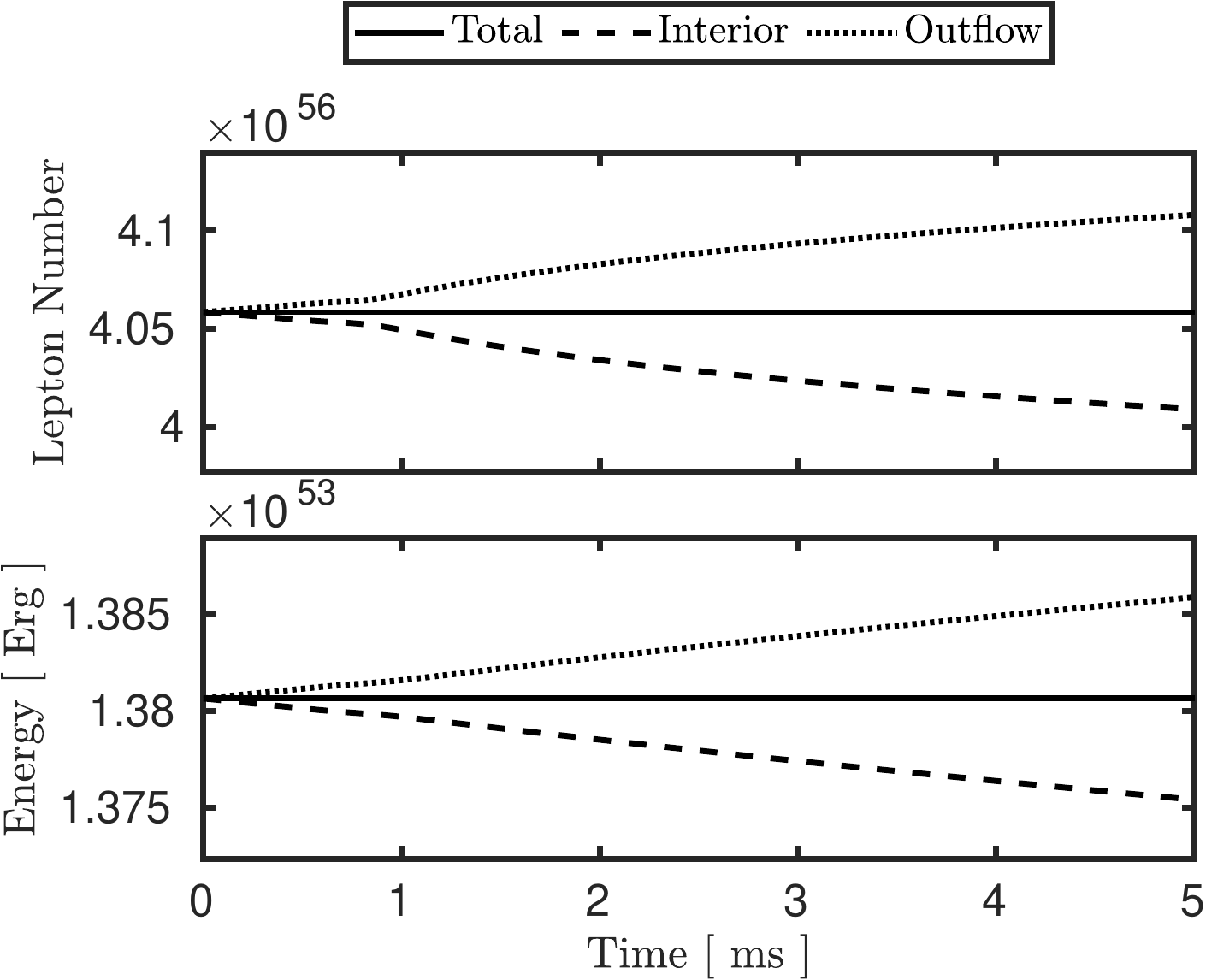}
		\label{fig:DLWave_Con}}~~~
	\subfloat[Evoluations of matter lepton number, neutrino lepton number, internal energy, and neutrino energy.]	
	{	\includegraphics[width=0.45\linewidth]{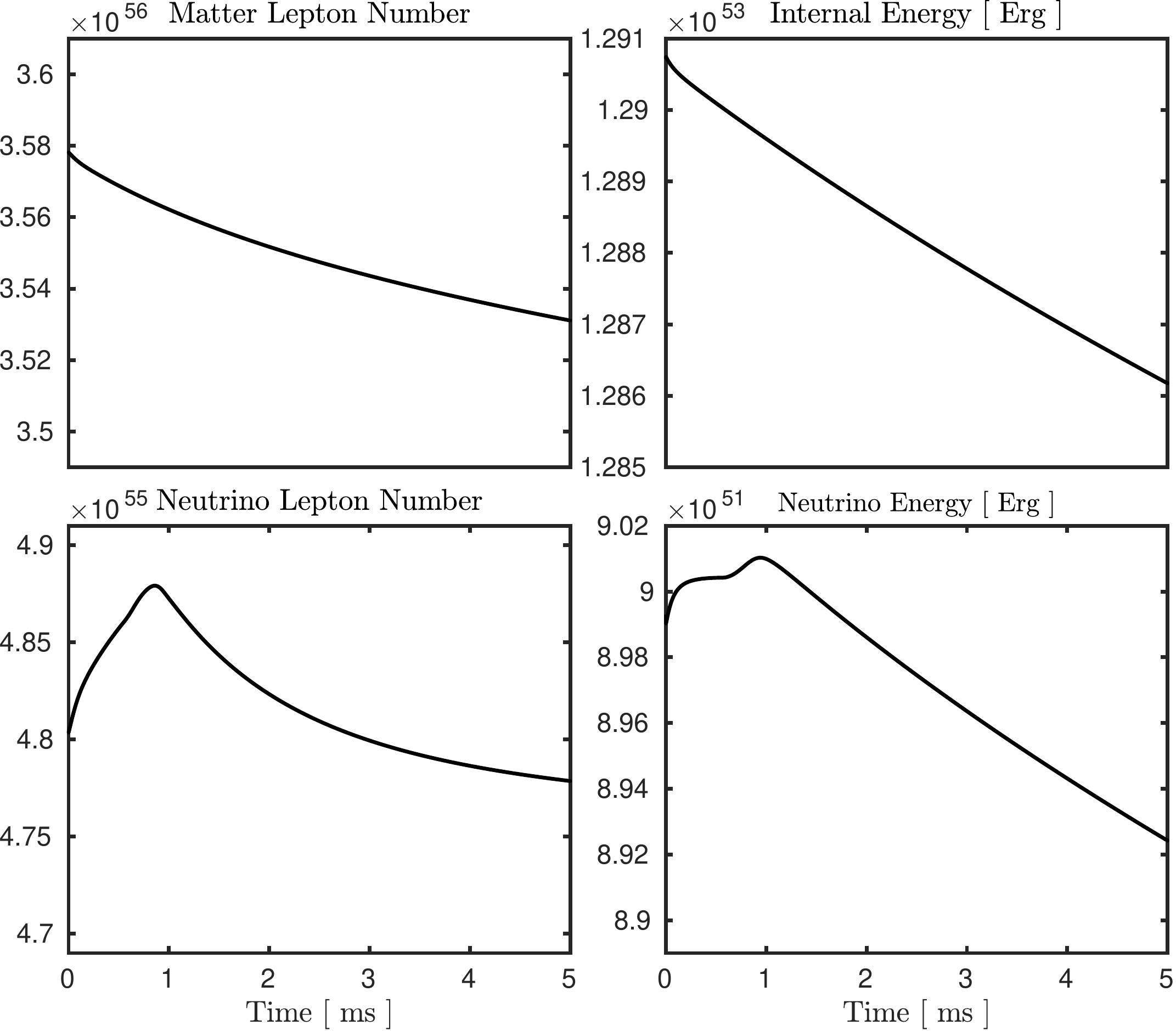}
		\label{fig:DLWave_ConParts}}
	\caption{Conservations in the deleptonization problem -- The left panel shows the lepton number and energy (see Eqs.~\eqref{eq:LeptonConservation} and \eqref{eq:EnergyConservation}) over $5$~ms of evolution starting from the $100$~ms post-bounce matter profile. Here the interior lepton number/energy in the computation domain are plotted in black dashed lines, and the accumulated outflow lepton number/energy are plotted in black dotted lines. Note that the accumulated outflows (black-dotted lines) are shifted up by the value of initial lepton number/energy for better illustration. The conserved lepton number and energy are plotted in black solid line. The right panel shows the evolution of components of the interior lepton number and interior energy, including matter lepton number (electron number), neutrino lepton number, internal energy, and neutrino energy.}
	\label{fig:DLWave_Conservation}
\end{figure}

The evolution of the electron fraction in the deleptonization problem is illustrated in Figure~\ref{fig:DeleptonizationWaveEvolution}, where we plot the electron fraction versus mass density over 10~ms of evolution, starting from the $100$~ms profile in Figure~\ref{fig:DeleptonizationWaveInitial} (dashed lines).  
Here the evolution of the electron fraction was generated from the DG-IMEX scheme with the proposed nonlinear solvers as described in Section~\ref{subsec:implementation}. We validated the result by comparing it against solutions from simulations on finer spatial-energy-temporal meshes, in which no noticeable differences were observed.
For higher densities ($\rho\gtrsim3\times10^{12}$~g~cm$^{-3}$), neutrinos are effectively trapped, and the electron fraction remains largely unchanged.  
For lower densities, neutrinos (created by electron capture on protons) are not trapped and escape the computational domain, which results in a lowering of the electron fraction (deleptonization).  
Figure~\ref{fig:DLWave_Conservation} shows conservations of lepton number and energy over $5$~ms of evolution in the deleptonization problem starting from the $100$~ms post-bounce matter profile. Here the lepton number and energy both comprise two parts: (i) the interior lepton number and energy in the computation domain, i.e., the time integrals of left-hand sides in Eqs.~\eqref{eq:LeptonConservation} and \eqref{eq:EnergyConservation}, respectively, and (ii) the accumulated outflow lepton number and energy at the boundary, which are respectively the negations of right-hand sides in Eqs.~\eqref{eq:LeptonConservation} and \eqref{eq:EnergyConservation}, integrated in time.
The conservation results of the lepton number/energy, as well as the evolutions of the interior lepton number/energy and the accumulated outflow lepton number/energy, are illustrated in Figure~\ref{fig:DLWave_Con}. The evolutions of individual components of the interior lepton number/energy, including matter lepton number (electron number), neutrino lepton number, internal energy, and neutrino energy, are shown in Figure~\ref{fig:DLWave_ConParts}.

For each profile, the deleptonization problems are simulated from the profile time ($50$, $100$, $150$, and $250$~ms after core bounce) to 5~ms after the profile time, which are referred to as the initial time $t_0 = 0$ and the final time $t_f = 5$~ms in the remainder of the paper. The IMEX time integration scheme discussed in Section~\ref{subsec:implementation} is used in the simulations, where the time step $\dt$ determined by the stability requirement for the explicit advection part.  
These problems are solved on the spatial domain $r\in[0,300]$~{km} and energy domain $[0,300]$~{MeV}, which are divided into $128$ and $16$ geometrically progressing elements, respectively.
Here the first spatial element has $\dx = 1$~{km}, the last spatial element has $\dx \approx 4.54$~{km}, the first energy element has $\Delta\varepsilon=4$~{MeV}, and the last energy element has $\Delta\varepsilon\approx50$~{MeV}.

\begin{figure}[h]
	\captionsetup[subfigure]{justification=centering}
	\subfloat[Profile: 50\,ms]	
	{	\includegraphics[width=0.45\linewidth]{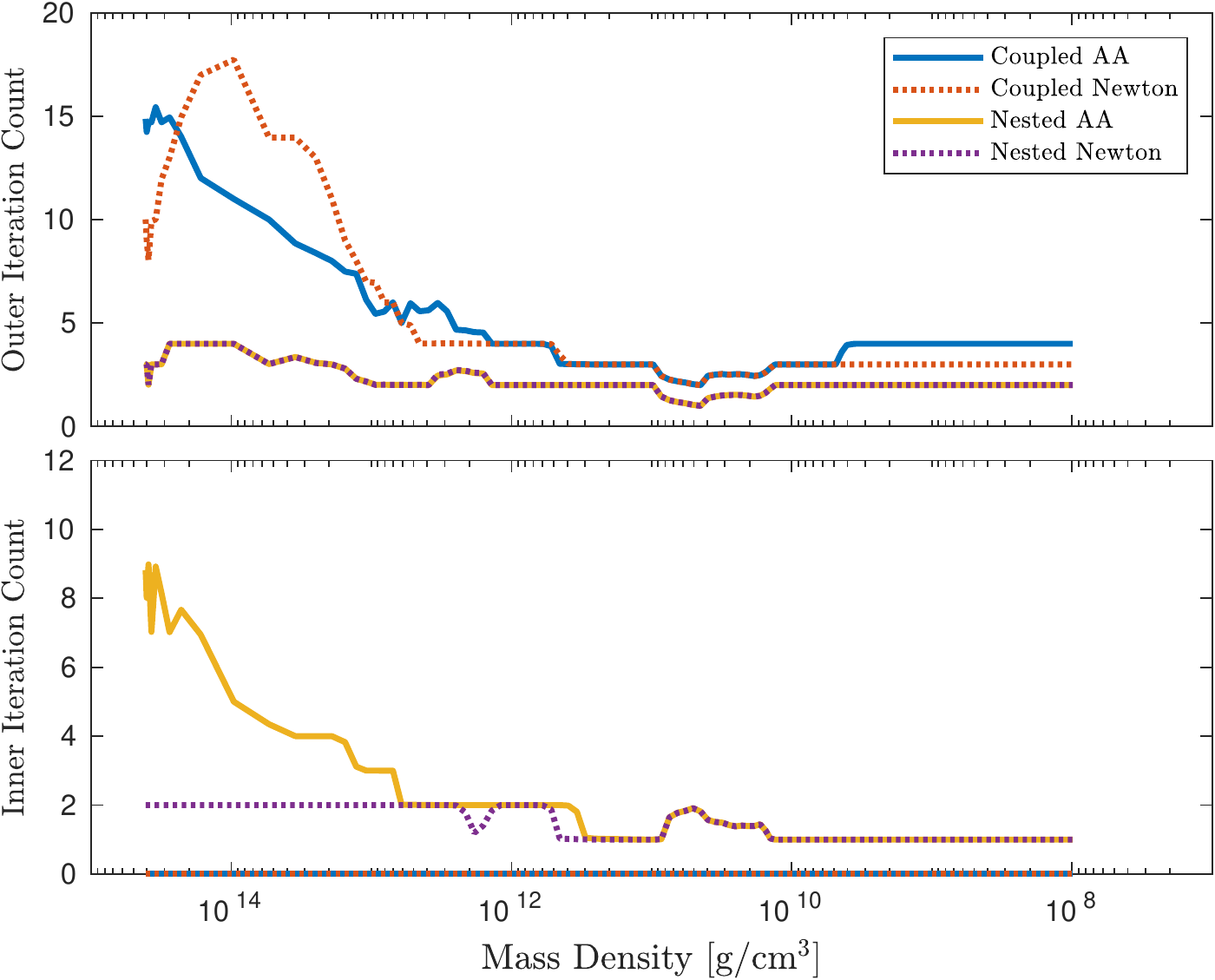}
			\label{fig:DL50ms}}~
	\subfloat[Profile: 100\,ms]	
	{	\includegraphics[width=0.45\linewidth]{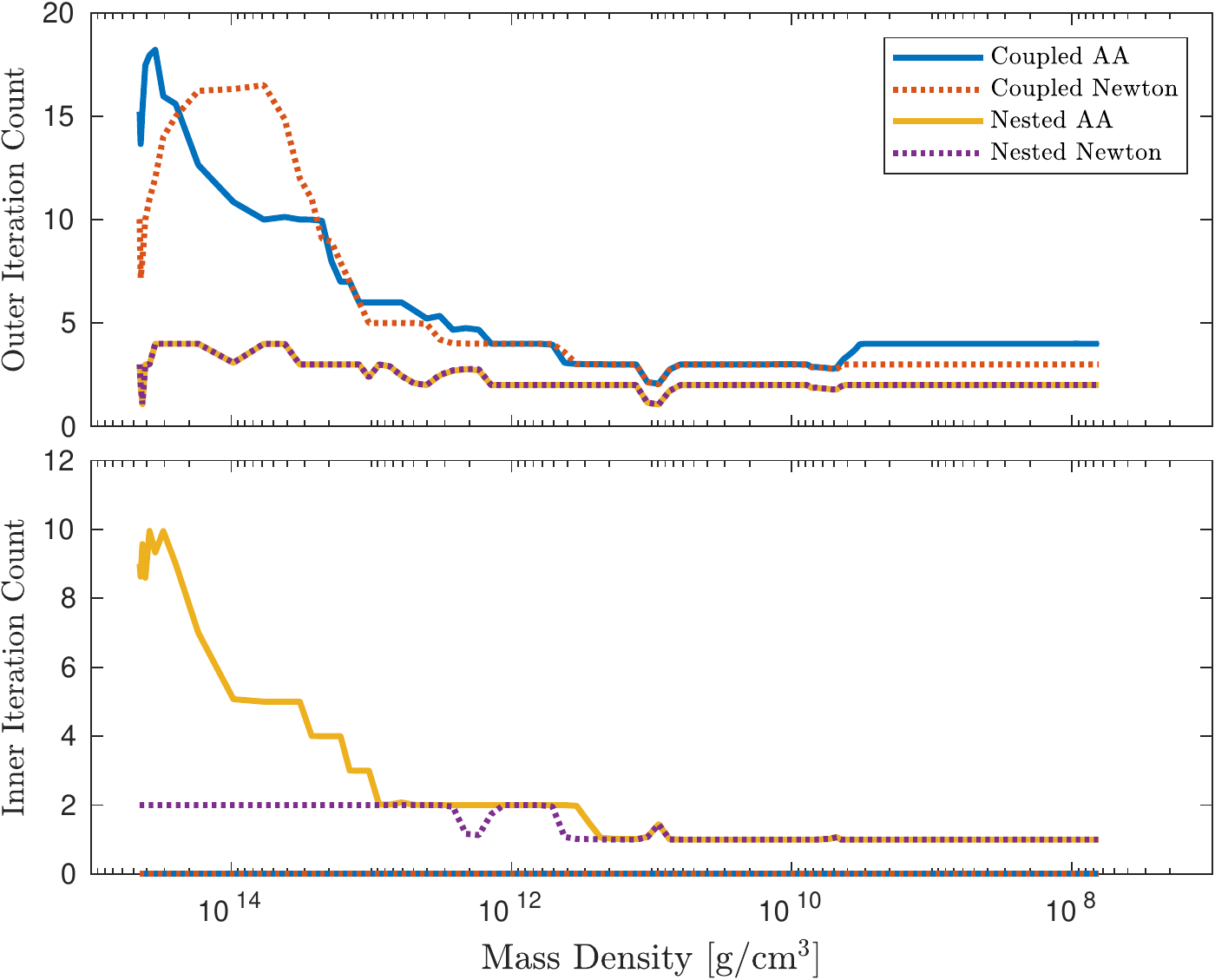}
		\label{fig:DL100ms}}\\
	\subfloat[Profile: 150\,ms]	
	{	\includegraphics[width=0.45\linewidth]{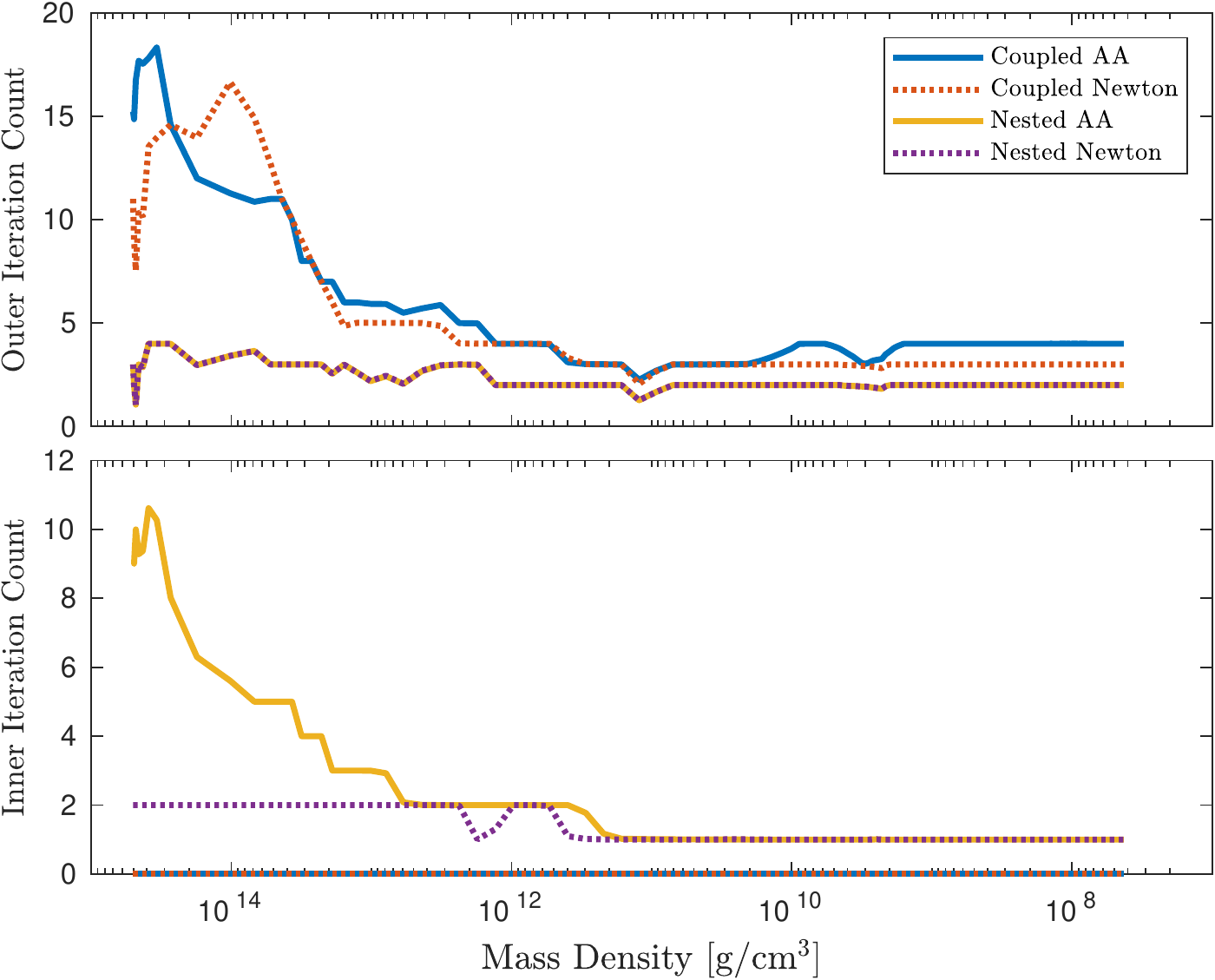}
		\label{fig:DL150ms}}~
	\subfloat[Profile: 250\,ms]	
	{	\includegraphics[width=0.45\linewidth]{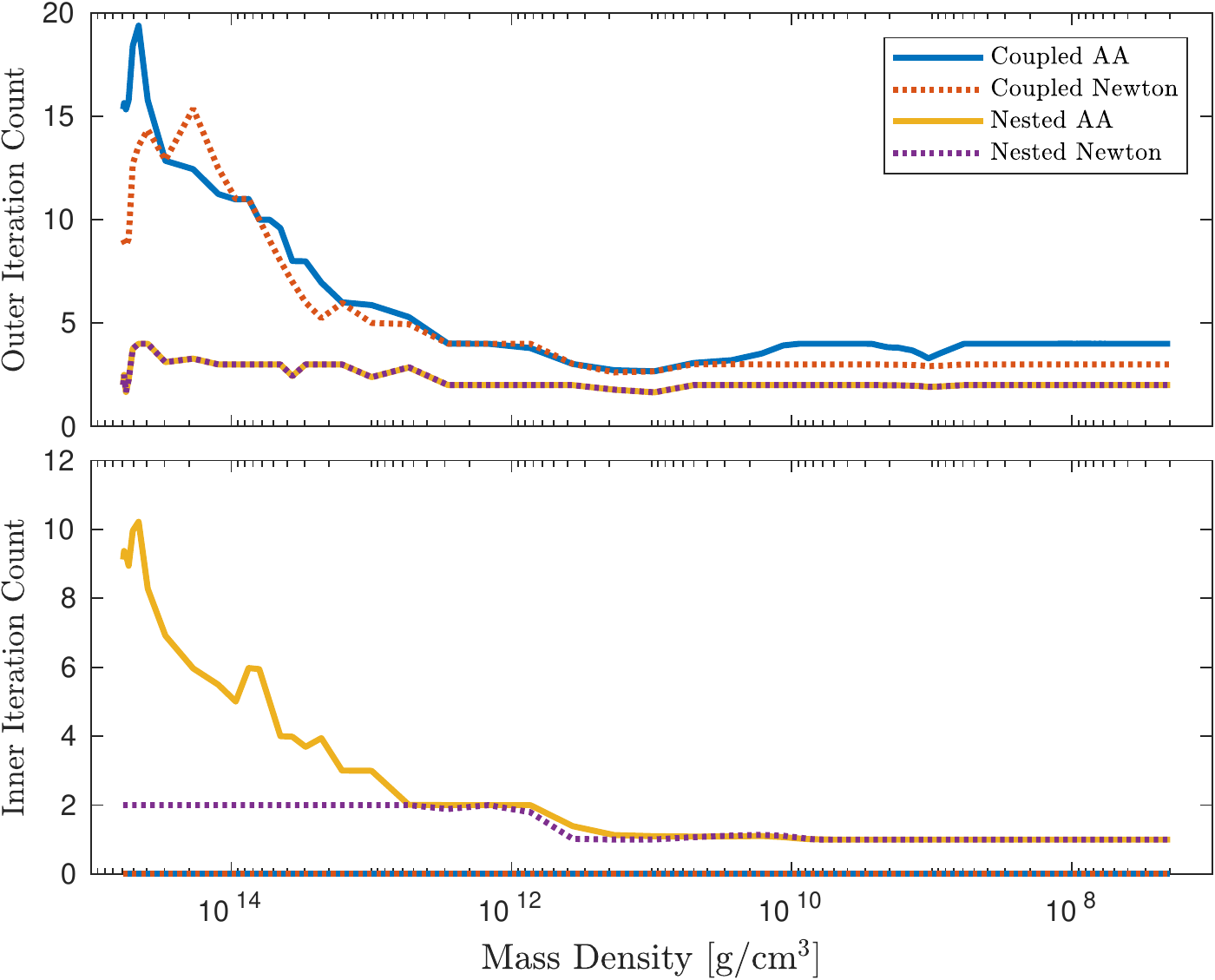}
		\label{fig:DL250ms}}~

	\caption{Time-averaged iteration counts of the nonlinear solvers on deleptonization problems -- The time-averaged iteration counts of the Coupled AA, Coupled Newton, Nested AA, and Nested Newton solvers on deleptonization problems with various profiles.}
	\label{fig:DLWave}
\end{figure}

Figure~\ref{fig:DLWave} shows the iteration counts of the nonlinear solvers on the deleptonization problem for various profiles.  
The results for profiles taken at $50$, $100$, $150$, and $250$~ms after core bounce are illustrated in Figures~\ref{fig:DL50ms}, \ref{fig:DL100ms}, \ref{fig:DL150ms}, and \ref{fig:DL250ms}, respectively.  
As in Figure~\ref{fig:Relaxation}, the top plot in these figures shows the ``outer" iteration counts of each solver, while the bottom plot shows the averaged ``inner" iteration counts of the nested solvers, where the inner iteration counts are averaged over the number of times that the inner equations~\eqref{eq:NestedInner} or \eqref{eq:NestedNewtonInner} were solved.
In addition, here both the outer and inner iteration counts are averaged over time (from $t_0$ to $t_f$), and the averaged iteration counts are plotted against the mass density, which corresponds to spatial locations for each profile.  
(We have found that the number of iterations at a given location varies little from $t_0$ to $t_f$.)

\begin{figure}[h]
	\centering
	\includegraphics[width=0.75\linewidth]{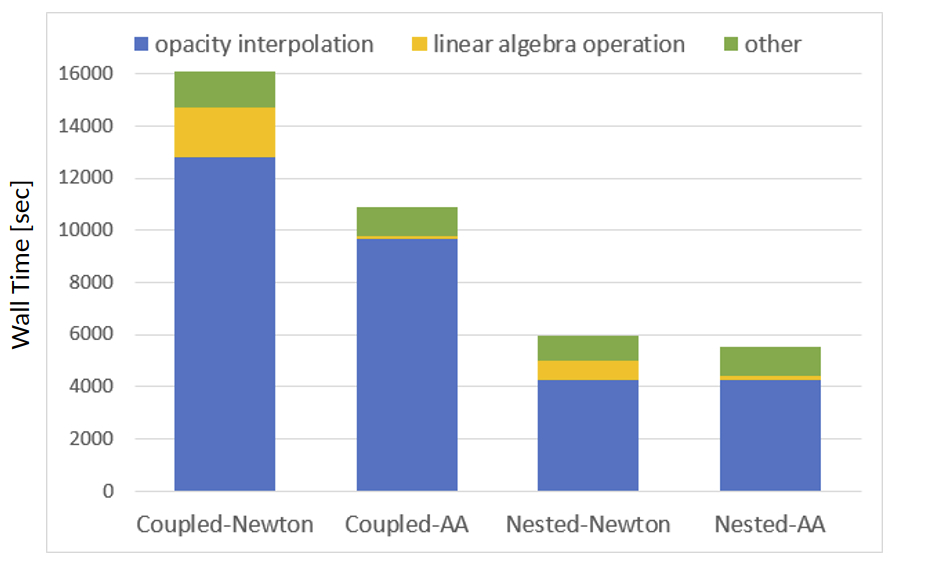}
	\caption{Computation time breakdown for the nonlinear solvers on the deleptonization problem with 100~ms profile -- The opacity interpolation time ($t_{\rm{Op}}$ in Table~\ref{table:DLTimingRatio}) and the dense linear algebra operation time ($t_{\rm{LA}}$ in Table~\ref{table:DLTimingRatio}) are the blue and yellow sections of the columns, respectively. The green sections represent the time spent on all other computations.}
	\label{fig:column_chart}
\end{figure}

From Figure~\ref{fig:DLWave}, we observe that the simulations with all four profiles give consistent results ---  the nested solvers require fewer outer iterations than the coupled ones do, and the Nested~AA solver requires more inner iteration to converge than the Nested Newton solver does, especially for harder problems (at higher mass density).  
This observation also agrees with the results reported in Section~\ref{subsec:relaxation} on the relaxation problems.
Computational times for these simulations are reported in the top panel of Table~\ref{table:DLTimingRatio}, where the tests~\#1, \#5, \#9, \#13 correspond to simulations shown in Figure~\ref{fig:DL50ms}, tests~\#2, \#6, \#10, \#14 correspond to simulations shown in Figure~\ref{fig:DL100ms}, and so on. 
In Table~\ref{table:DLTimingRatio}, we report the total computation time $t_{\rm{Tot}}$ as well as the detailed timing measurements in each simulation, such as the computational time spent on (i) solving the nonlinear system in Eq.~\eqref{eq:CoupledSystem} in the implicit step ($t_{\rm{Im}}$), (ii) the opacity evaluations/interpolations when solving Eq.~\eqref{eq:CoupledSystem} ($t_{\rm{Op}}$), (iii) linear algebra operations, such as the least-squares solve in Anderson acceleration or the assembly and inversion of Jacobian matrices in Newton's method ($t_{\rm{LA}}$), (iv) the presolve step ($t_{\rm{Ps}}$), (v) the explicit update of the advection term ($t_{\rm{Ex}}$), and (vi) the positivity limiter ($t_{\rm{PL}}$)%
\footnote{Here the positivity limiters are applied after each explicit and implicit update to enforce realizability of the moments. The computation time for the positivity limiter ($t_{\rm{PL}}$) reported in Table~\ref{table:DLTimingRatio} is relatively large (compared to $t_{\rm{Ex}}$), and we expect that $t_{\rm{PL}}$ could be further reduced by a more sophisticated implementation. In any case, it does not affect the observations we made in this paper.}.  
The reported timing results in the top panel of Table~\ref{table:DLTimingRatio} are all linearly scaled such that the highest reported computational time (16,085~seconds from test~\#2) is scaled to $100$, which corresponds to the total computation time when the Coupled Newton solver is used to solve the deleptonization problem with the 100~ms profile.  
For each simulation, we also record the solver configurations, such as the value of the truncation parameter $m$ in Anderson acceleration, the maximum allowed iteration (MaxIter), and whether the presolve step is performed or not.  
We also note that these reported data are not mutually exclusive, e.g., $t_{\rm{Tot}}\approx t_{\rm{Im}}+t_{\rm{Ex}}+t_{\rm{PL}}$, and $t_{\rm{Im}}\approx t_{\rm{Op}}+t_{\rm{LA}}+t_{\rm{Ps}}$.
Figure~\ref{fig:column_chart} provides a column chart that visualizes the results with the 100~ms profile reported in the top panel of Table~\ref{table:DLTimingRatio}. There is no qualitative difference between the results from problems with different profiles.
The column chart in Figure~\ref{fig:column_chart} confirms that the majority of the computational time in these simulations is spent on opacity evaluation/interpolation, which is proportional to the outer iteration counts.  
It also shows that the nested solvers indeed speed up the computations by taking inner iterations to reduce the number of outer iterations.  
Further, we observe that the lower computational cost on the linear algebra operations ($t_{\rm{LA}}$) makes the Nested AA solver outperform the Nested Newton solver in terms of the total computation time by about 8\%, despite the higher inner iteration counts.  
These results indicate that the Nested AA solver leads to the least computation time for the deleptonization problems among all the tested solvers.  

\begin{figure}[h]
	\centering
	\captionsetup[subfigure]{justification=centering}
	\includegraphics[width=0.7\linewidth]{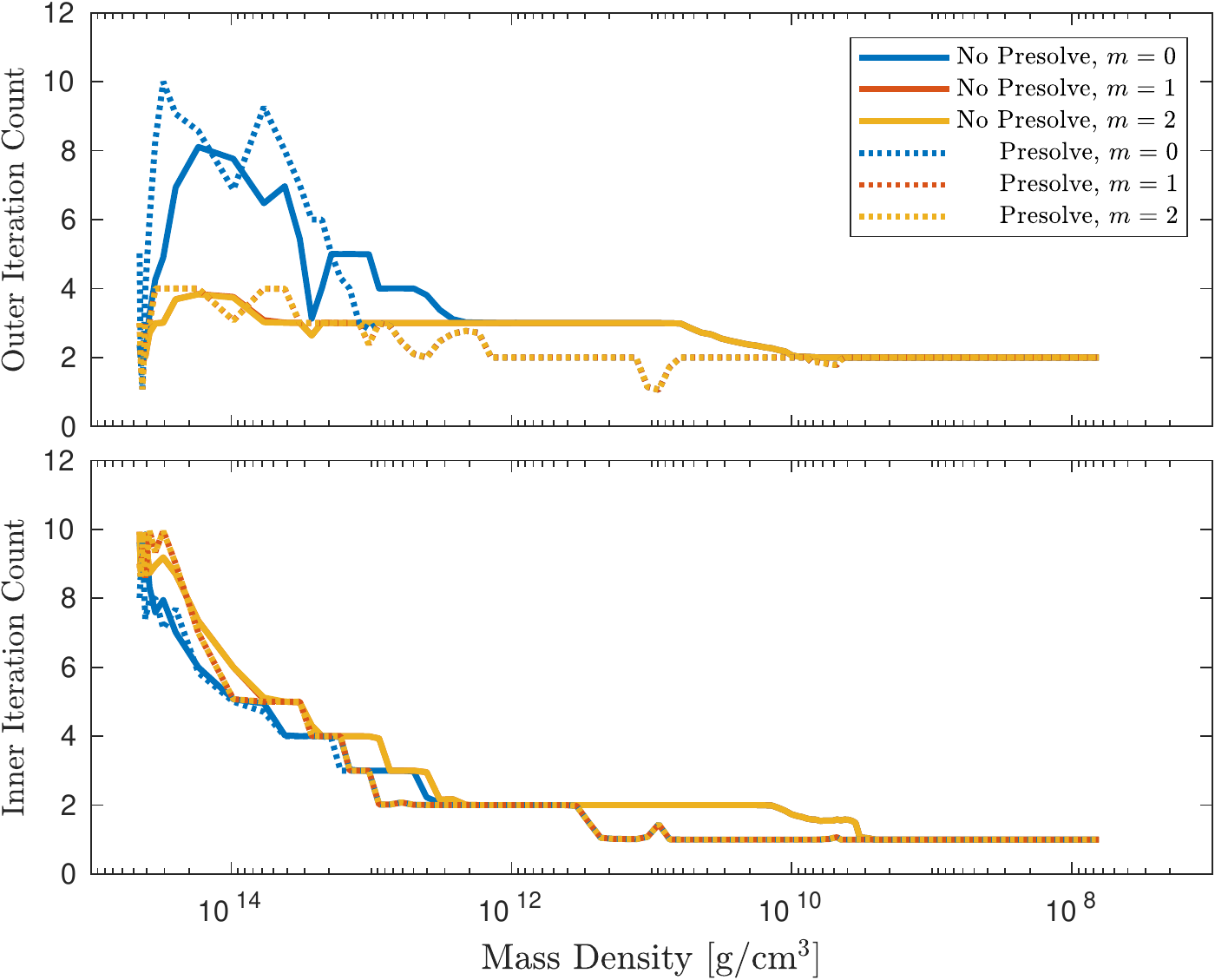}
	\caption{Time-averaged iteration counts for the Nested AA algorithm with various solver configurations on the deleptonization problem with $100$~ms profile. Results for the Nested AA algorithm without/with presolve and with truncation parameter value $m = 0,1,2$ in the outer loop are reported. Here the iteration counts for $m=2$ overlap with the corresponding reuslts for $m=1$, which implies that there is no benefit moving from $m=1$ to $m=2$ when solving the outer loop problem.}
	\label{fig:NestedAAiter}
\end{figure}

To further analyze the performance of the Nested~AA solver, we experiment with different solver parameters on the deleptonization problem using the $100$~ms profile.  
Specifically, we tested the Nested~AA solver with the Anderson acceleration truncation parameter set to $m = 0$, $1$, and $2$ for solving the outer system in Eq.~\eqref{eq:NestedOuter}.  
This experiment is to verify the benefit of Anderson acceleration on solving the smaller outer system in Eq.~\eqref{eq:NestedOuter}, which has only two unknowns.  
In addition, for each choice of $m$, we initialize the solver with and without the ``presolve" step introduced in Section~\ref{subsec:implementation}, which helps us assess the effect of the ``presolve" step in a more realistic setting.  
The resulting iteration counts are shown in Figure~\ref{fig:NestedAAiter}, which are averaged over the time from $t_0$ to $t_f$ as in Figure~\ref{fig:DLWave}. 
The computation times are reported in the bottom panel of Table~\ref{table:DLTimingRatio} (tests~\#17$\sim$\#21).  
From these results, we observe that moving from Picard iteration $(m=0)$ to Anderson acceleration $(m>0)$ does reduce the outer iteration count, and thus improves the computation time by around 18\%, especially for harder problems (at higher mass density).  
However, unlike the results for Coupled AA solver reported in Figure~\ref{fig:AA}, increasing the truncation parameter from $m=1$ to $m=2$ does not lead to any observable reduction in the either iteration counts or computation time.  
This result is not unexpected, since Anderson acceleration is applied here to solve the outer system in Eq.~\eqref{eq:NestedOuter} with only two unknowns, while the results reported in Figure~\ref{fig:AA} are from solving the fully coupled system in Eq.~\eqref{eq:FP}.  
Another observation from Figure~\ref{fig:NestedAAiter} is that there is no clear benefit in applying the presolve step on this problem.  
The presolve step does slightly reduce the iteration counts, however, the additional cost of the presolve step wipes out the gains from fewer iterations.  
We suspect that the diminished benefit of the presolve step is due to the fact that in the deleptonization problem, the explicit time step restriction from the advection term limits the stiffness of the implicit problem and forces the implicit update to be small, which makes the effect of presolve insignificant.  
For deleptonization problems, larger implicit time steps could potentially be achieved by techniques such as subcycling the explicit steps, or by adopting the general multirate framework proposed by \cite{Sandu_2019}. 
We expect to see a greater impact of the presolve step on these problems with larger time steps, as is observed in the relaxation tests.

\begin{figure}[h]
	\centering
	\captionsetup[subfigure]{justification=centering}
	\subfloat[Time-averaged Iteration Counts]	
	{	\includegraphics[width=0.49\linewidth]{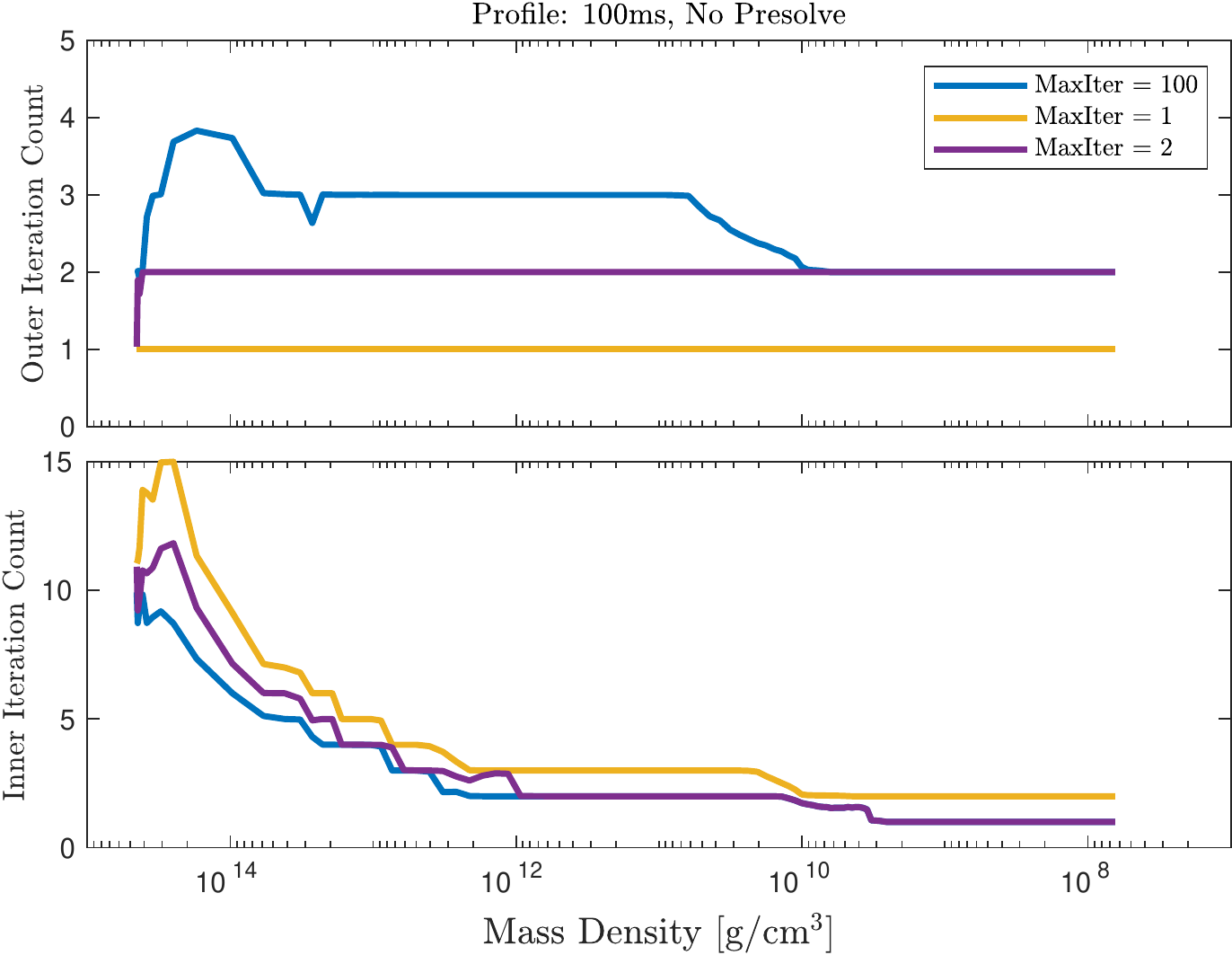}
		\label{fig:MaxIter_NoPre}}~~~~
	\subfloat[Energy-integrated Number Densities at $t_f = 5$~{ms}]	
	{	\includegraphics[width=0.49\linewidth]{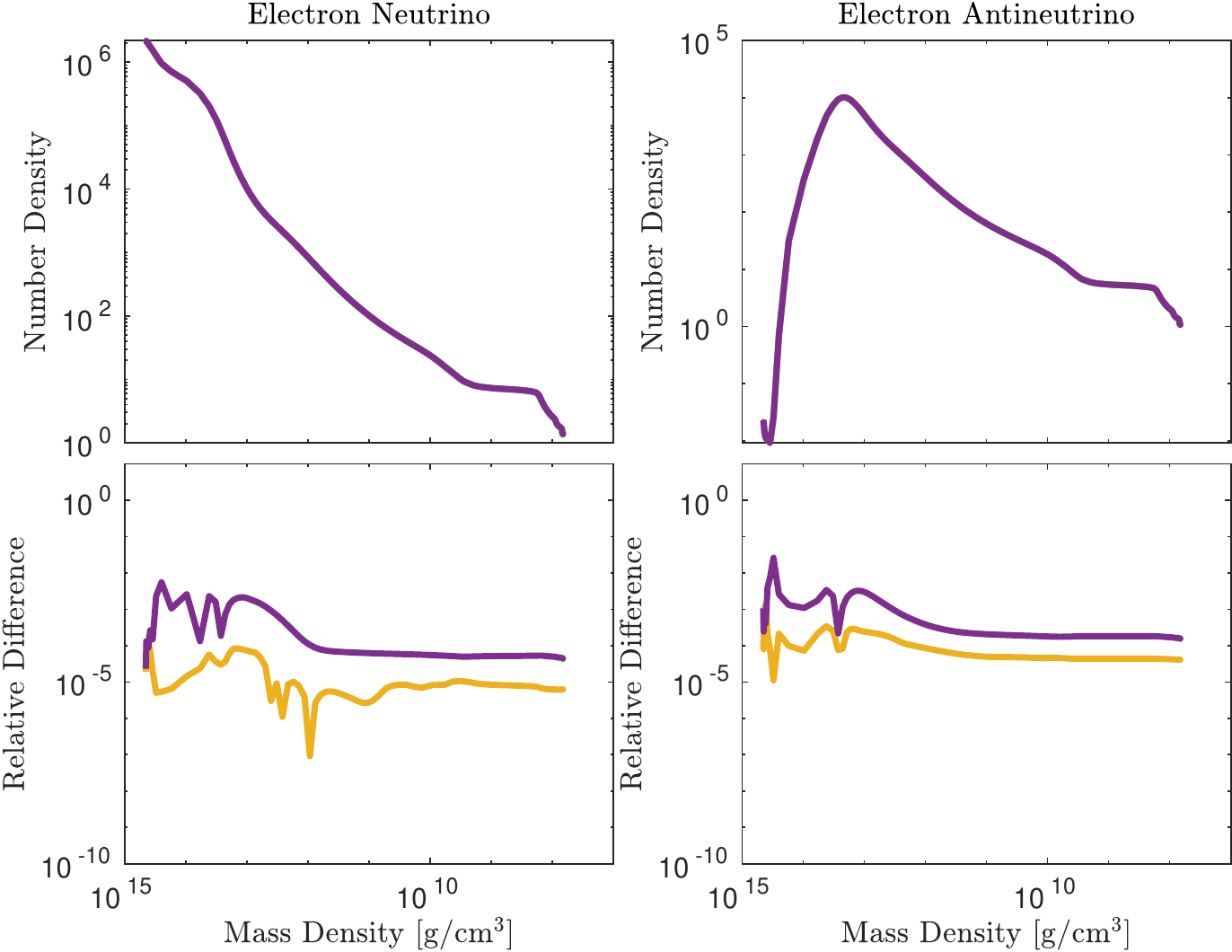}
		\label{fig:MaxIter_NoPre_Rad}}\\
	\subfloat[Temperature at $t_f = 5$~{ms}]	
	{	\includegraphics[width=0.51\linewidth]{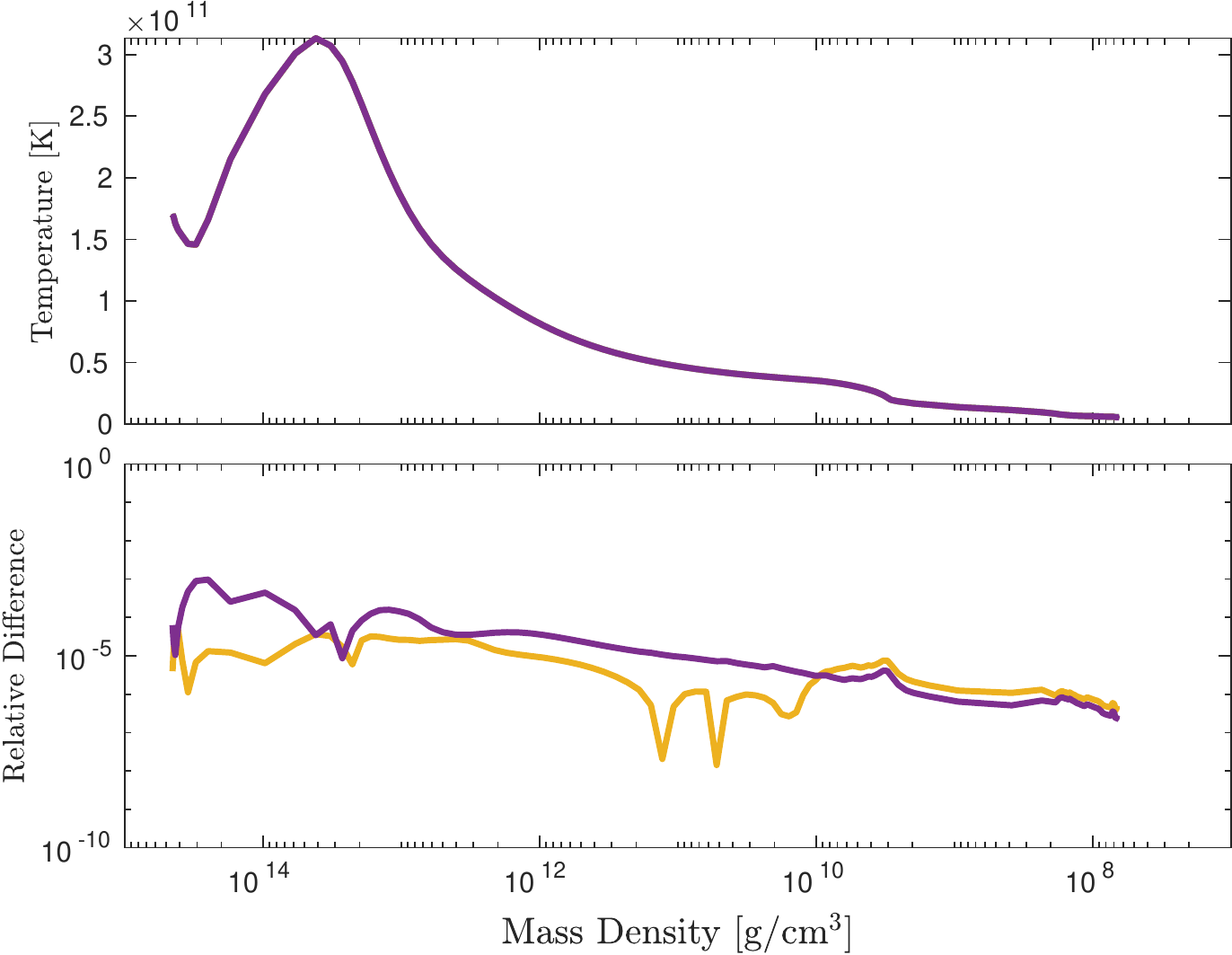}
		\label{fig:MaxIter_NoPre_T}}~
	\subfloat[Electron Fraction at $t_f = 5$~{ms}]	
	{	\includegraphics[width=0.5\linewidth]{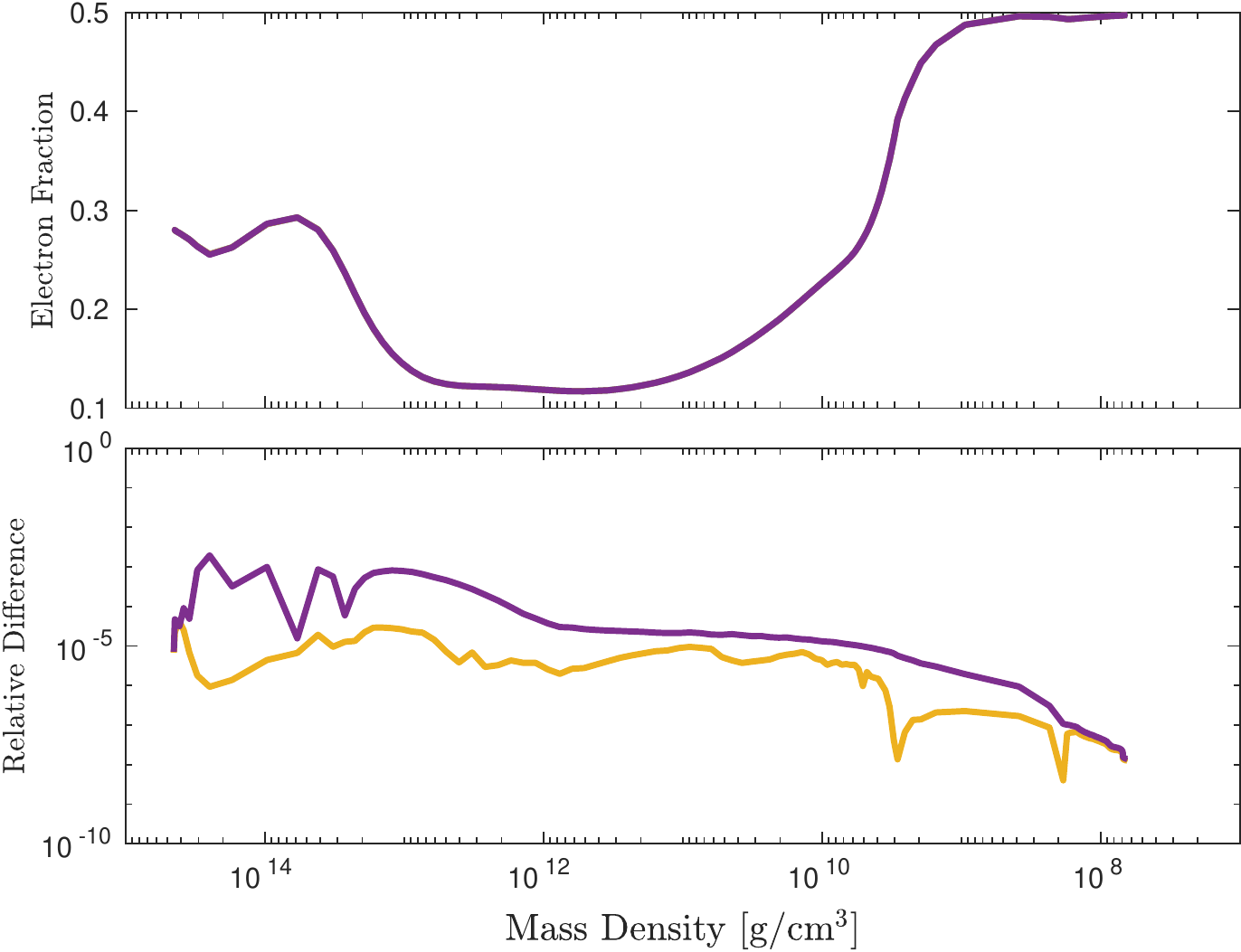}
		\label{fig:MaxIter_NoPre_Ye}}~
	
	\caption{Time-averaged iteration counts and final-time ($t_f=5$~{ms}) solutions for the Nested AA solver without the presolve step, running to various maximum outer iterations (MaxIter). The blue line (MaxIter = $100$) is considered as a reference.}
	\label{fig:NestedAAMaxiterNoPre}
\end{figure}

\begin{figure}[h]
	\centering
	\captionsetup[subfigure]{justification=centering}
	\subfloat[Time-averaged Iteration Counts]	
	{	\includegraphics[width=0.49\linewidth]{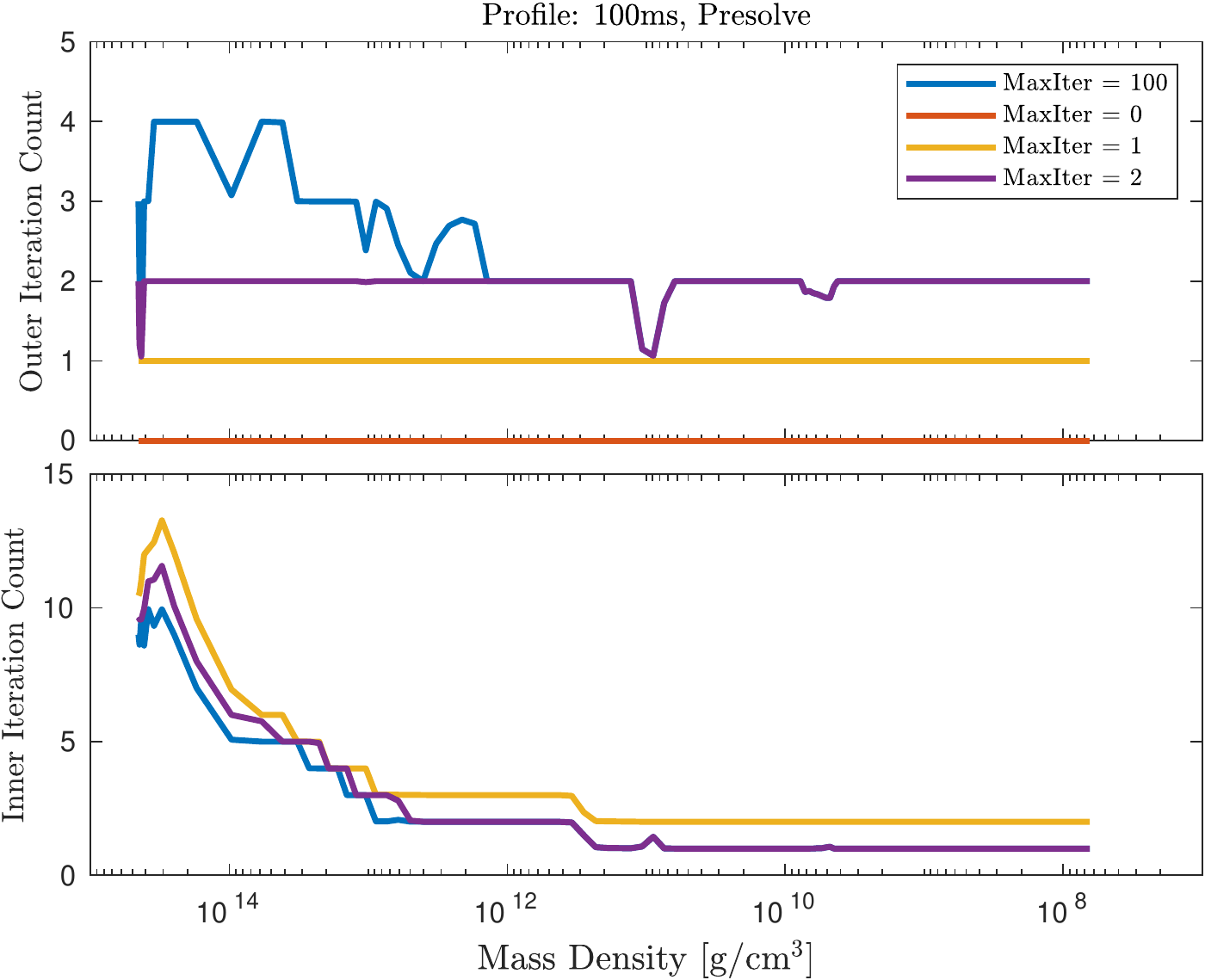}
		\label{fig:MaxIter_Pre}}~~~~
	\subfloat[Energy-integrated Number Densities at $t_f = 5$~{ms}]	
	{	\includegraphics[width=0.515\linewidth]{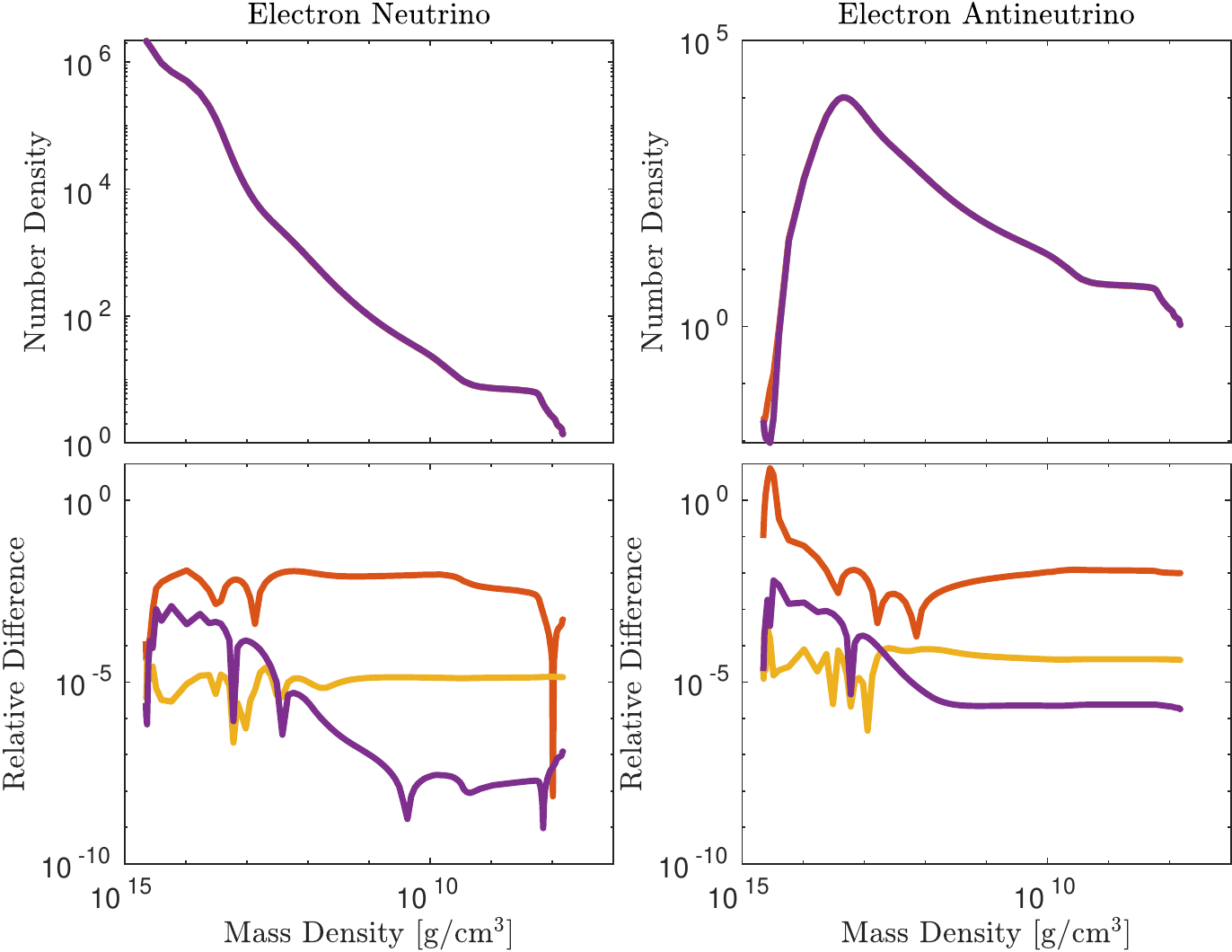}
		\label{fig:MaxIter_Pre_Rad}}\\
	\subfloat[Temperature at $t_f = 5$~{ms}]	
	{	\includegraphics[width=0.51\linewidth]{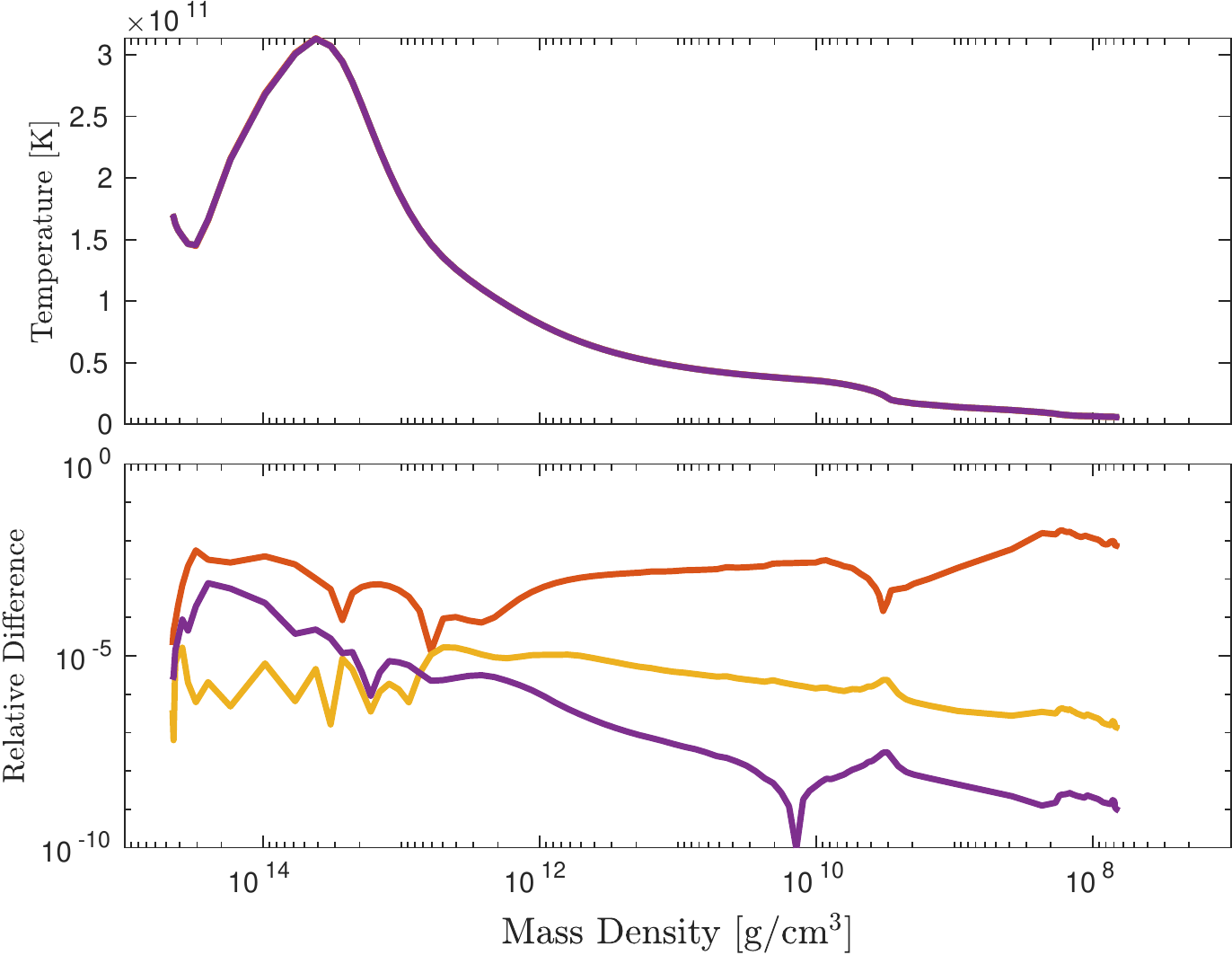}
		\label{fig:MaxIter_Pre_T}}~
	\subfloat[Electron Fraction at $t_f = 5$~{ms}]	
	{	\includegraphics[width=0.5\linewidth]{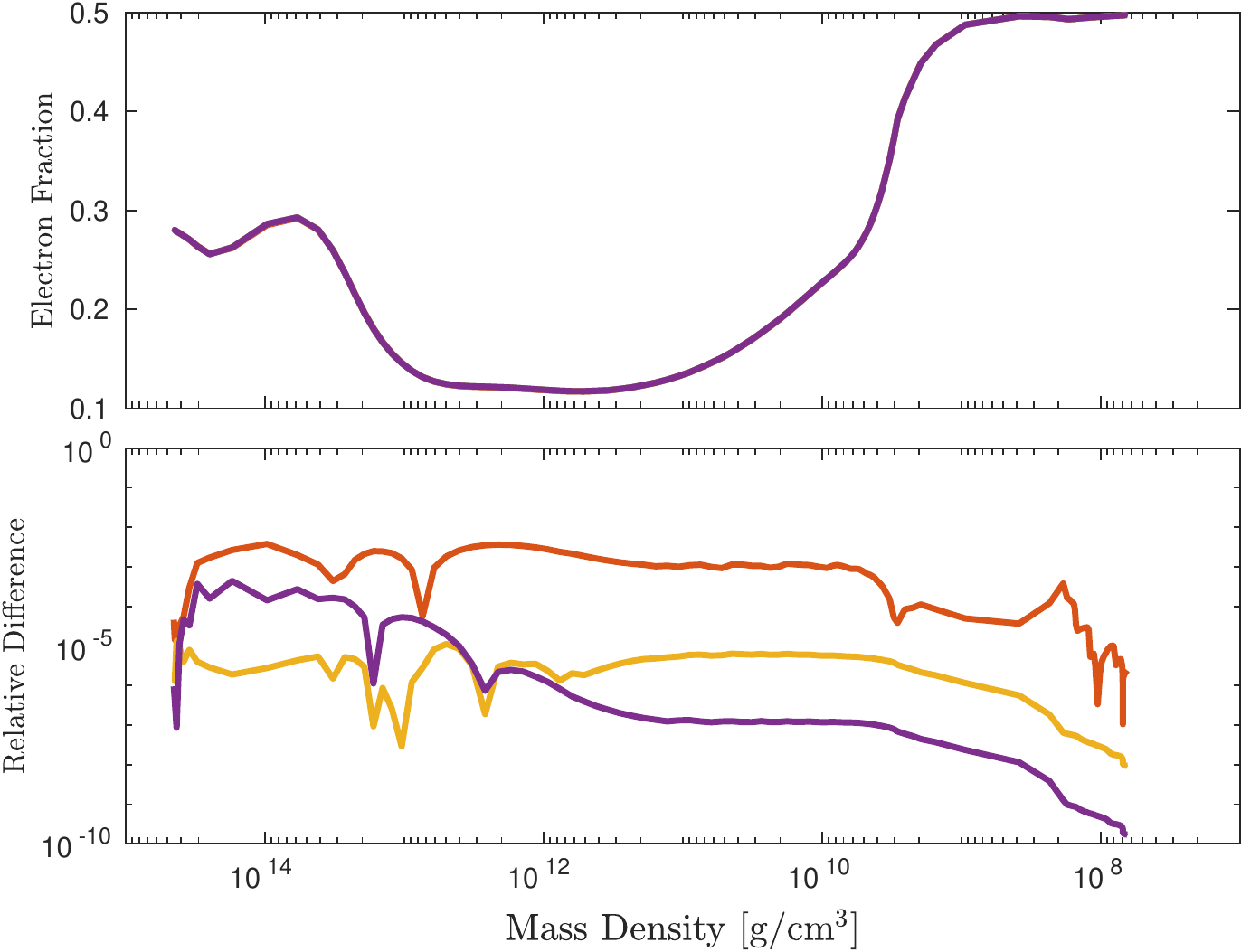}
		\label{fig:MaxIter_Pre_Ye}}~~~~~~~~~~~~~~~~
	
	\caption{Time-averaged iteration counts and final-time ($t_f=5$~{ms}) solutions for the Nested AA solver with the presolve step, running to various maximum outer iterations (MaxIter). The blue line (MaxIter = $100$) is considered as a reference.}
	\label{fig:NestedAAMaxiterPre}
\end{figure}

Finally, we investigate the effect of early termination for the Nested~AA solver on the deleptonization problems.  
In these tests, the solver configurations are identical to the ones reported in Figures~\ref{fig:NoPre} and \ref{fig:Pre} for the relaxation problems, e.g., the outer loop of the Nested AA solver is terminated early by restricting the maximum number of outer iterations (MaxIter).  
Here we test the solver on the deleptonization problem with the 100~ms initial matter profile for MaxIter = 1, 2, and 100, and the presolve step discussed in Section~\ref{subsec:implementation} turned off. We then repeat the test for MaxIter = 0, 1, 2, and 100, with the presolve step turned on.  
The results with and without the presolve step are shown in Figures~\ref{fig:NestedAAMaxiterNoPre} and \ref{fig:NestedAAMaxiterPre}, respectively, where the iteration counts are reported, along with the electron neutrino and antineutrino (energy-integrated) number densities, temperatures, and electron fractions at the final time $t=5$~ms.  
Here the fully converged solutions (MaxIter = 100) are considered as reference solutions, where the nested solver converges well before the nominal maximal outer iteration is reached, as shown in Figures~\ref{fig:MaxIter_NoPre} and \ref{fig:MaxIter_Pre}.
As mentioned in the earlier subsection, the Nested~AA solver with MaxIter = 1 and the presolve step turned off is effectively lagging the opacities by computing them from the matter states at the previous time step, while updating the radiation quantities at the current time step (see, e.g., \cite{just_etal_2015}).  
From Figure~\ref{fig:NestedAAMaxiterNoPre}, we observe that the early terminated solutions are in good agreement with the reference solution. 
However, allowing two outer iterations does not give a better solution than the one with only a single outer iteration, which is possibly due to the issue on lepton number and energy conservation for early terminated solutions, as discussed in Section~\ref{subsec:relaxation}. 
Another potential reason is that Anderson acceleration does not guarantee monotone decreasing of the residual (see \cite{pollock2019anderson, evans2020proof,kelley2018numerical,Toth-Kelley-2015} for convergence analysis of AA).
The behavior of residuals from Anderson acceleration alternating increasing and decreasing has been observed in \cite{pollock2019anderson}, and a potential explanation is given from \cite[Theorem~4.5]{pollock2019anderson}.
Similar results can be observed in Figure~\ref{fig:NestedAAMaxiterPre}, where the presolve step is in effect.  
Here the comparison includes the case that MaxIter = 0, where both the radiation and matter quantities are solely updated in the presolve step, with the NES and pair processes omitted.  
The relative difference in the solution with MaxIter = 0 is mostly around $10^{-2}$ to $10^{-3}$, however, the difference could go up to $10^1$ for the energy-integrated antineutrino number density in the high mass density region.  
We also note that the presolve step reduces the difference in the solution with MaxIter = 2 by a few orders of magnitude in the low mass density region.  
The reason is that, in the low density region, the nonlinear solve usually converges within two iterations with the presolve step.  
The results in Figures~\ref{fig:NestedAAMaxiterNoPre} and \ref{fig:NestedAAMaxiterPre} suggest that, for problems requiring few iterations, limiting MaxIter to one can give sufficiently accurate solutions, while reducing the computation time by roughly a factor of two from the fully converged case, as shown in the results for tests~\#22$\sim$\#27 in Table~\ref{table:DLTimingRatio}.

\begin{table}[h]
	\caption{Overview of nonlinear solver timing results for the deleptonization problem.}
	\label{table:DLTimingRatio}
	\small
	Top panel: The detailed computational time is reported for the results shown in Figure~\ref{fig:DLWave}. Here the four iterative solvers are compared on four initial matter profiles with identical solver parameters.
	The results are linearly scaled so that the highest measurement (boxed, 16,085 seconds) is scaled to 100.
	\begin{center}
		\begin{tabular}{rlrrrrrrrrccc}
			\toprule
			\# & {Solver} & {Profile} & $t_{\rm{Tot}}$ & $t_{\rm{Im}}$ & $t_{\rm{Op}}$ & $t_{\rm{LA}}$ & $t_{\rm{Ps}}$ & $t_{\rm{Ex}}$ & $t_{\rm{PL}}$ & $m$ & {Presolve} & {MaxIter }\\
			\midrule
			\midrule
			1 & Coupled-Newton &  50~ms & 98.8 & 96.8 & 78.6 & 11.7 & 2.7 & 0.8 & 1.0 & -- & yes & 100 \\
			2 & Coupled-Newton & 100~ms & \boxed{100} & 97.9 & 79.5 & 11.9 & 2.7 & 0.8 & 1.0 & -- & yes & 100 \\
			3 & Coupled-Newton & 150~ms & 97.5 & 95.4 & 77.5 & 11.5 & 2.7 & 0.8 & 1.0 & -- & yes & 100 \\
			4 & Coupled-Newton & 250~ms & 92.9 & 90.8 & 73.8 & 10.8 & 2.7 & 0.8 & 1.0 & -- & yes & 100 \\
			\midrule
			5 & Coupled-AA &  50~ms & 65.2 & 63.2 & 57.6 &  0.9 & 2.7 & 0.8 & 1.0 & 2 & yes & 100 \\
			6 & Coupled-AA & 100~ms & 67.8 & 65.8 & 60.0 &  1.0 & 2.7 & 0.8 & 1.0 & 2 & yes & 100 \\
			7 & Coupled-AA & 150~ms & 69.0 & 67.0 & 61.1 &  1.0 & 2.7 & 0.8 & 1.0 & 2 & yes & 100 \\
			8 & Coupled-AA & 250~ms & 67.5 & 65.4 & 59.7 &  1.0 & 2.6 & 0.8 & 1.0 & 2 & yes & 100 \\
			\midrule
			9 & Nested-Newton  &  50~ms & 36.6 & 34.5 & 26.0 &  4.8 & 2.7 & 0.8 & 1.0 & 2 & yes & 100  \\
			10 & Nested-Newton & 100~ms & 36.9 & 34.9 & 26.4 &  4.7 & 2.7 & 0.8 & 1.0 & 2 & yes & 100  \\
			11 & Nested-Newton & 150~ms & 36.4 & 34.3 & 26.1 &  4.5 & 2.7 & 0.8 & 1.0 & 2 & yes & 100  \\
			12 & Nested-Newton & 250~ms & 35.7 & 33.7 & 25.8 &  4.3 & 2.6 & 0.8 & 1.0 & 2 & yes & 100  \\
			\midrule
			13 & Nested-AA &  50~ms & 33.7 & 31.6 & 26.0 &  0.9 & 2.7 & 0.8 & 1.0 & 2 & yes & 100  \\
			14 & Nested-AA & 100~ms & 34.3 & 32.2 & 26.5 &  0.9 & 2.7 & 0.8 & 1.0 & 2 & yes & 100  \\
			15 & Nested-AA & 150~ms & 33.8 & 31.7 & 26.0 &  0.9 & 2.7 & 0.8 & 1.0 & 2 & yes & 100  \\
			16 & Nested-AA & 250~ms & 33.1 & 31.0 & 25.5 &  0.8 & 2.6 & 0.8 & 1.0 & 2 & yes & 100  \\
			\bottomrule
		\end{tabular}
	\end{center}

		~\\
	Bottom panel: The detailed computational time is reported for the results shown in Figures~\ref{fig:NestedAAiter}--\ref{fig:NestedAAMaxiterPre}. Here the Nested-AA solver is tested on the 100~ms profile with various solver parameters.
	The results are linearly scaled so that the total time in Test~\#14 (boxed, 5,518 seconds) is scaled to 100.		
	\begin{center}
		\begin{tabular}{rlrrrrrrrrccc}
			\toprule
			\# & {Solver} & {Profile} & $t_{\rm{Tot}}$ & $t_{\rm{Im}}$ & $t_{\rm{Op}}$ & $t_{\rm{LA}}$ & $t_{\rm{Ps}}$ & $t_{\rm{Ex}}$ & $t_{\rm{PL}}$ & $m$ & {Presolve} & {MaxIter }\\
			\midrule
			\midrule
			14 & Nested-AA & 100~ms & \boxed{100} & 94.0 & 77.1 & 2.6 & 7.8 & 2.4 & 2.9 & 2 & yes & 100  \\
			\midrule
			17 & Nested-AA & 100~ms & 117.8 & 111.9 & 96.1 & 2.0 & ----- & 2.4 & 2.9 & 0 &  no & 100 \\
			18 & Nested-AA & 100~ms &  99.9 &  94.0 & 81.5 & 1.5 & ----- & 2.4 & 2.9 & 1 &  no & 100 \\
			19 & Nested-AA & 100~ms &  99.7 &  93.8 & 81.3 & 1.5 & ----- & 2.4 & 2.9 & 2 &  no & 100 \\
			\midrule
			20 & Nested-AA & 100~ms & 118.2 & 112.3 & 91.4 & 3.3 & 7.7 & 2.4 & 2.9 & 0 & yes & 100 \\
			21 & Nested-AA & 100~ms &  99.3 &  93.4 & 76.7 & 2.6 & 7.7 & 2.4 & 2.9 & 1 & yes & 100 \\
			\midrule
			22 & Nested-AA & 100~ms &  47.2 &  41.2 & 34.0 & 0.8 & ----- & 2.4 & 2.9 & 2 & no & 1 \\
			23 & Nested-AA & 100~ms &  83.9 &  77.9 & 67.2 & 1.2 & ----- & 2.5 & 2.9 & 2 & no & 2 \\
			24 & Nested-AA & 100~ms & 100.9 &  94.9 & 82.3 & 1.5 & ----- & 2.4 & 2.9 & 2 & no & 100 \\
			\midrule
			25 & Nested-AA & 100~ms &  16.9 &  10.8 &  3.7 & 1.2 & 7.6 & 2.5 & 2.9 & 2 & yes & 0 \\
			26 & Nested-AA & 100~ms &  54.2 &  48.2 & 36.2 & 1.9 & 7.7 & 2.4 & 2.9 & 2 & yes & 1 \\
			27 & Nested-AA & 100~ms &  89.1 &  83.0 & 68.2 & 2.2 & 7.8 & 2.4 & 2.9 & 2 & yes & 2 \\
			\bottomrule
		\end{tabular}
	\end{center}

	~\\
	{
	Brief descriptions of data reported in this table:
	\begin{center}
\boxed{
	\begin{tabular}{lll}
$t_{\rm{Tot}}$: total simulation time;  & $t_{\rm{Im}}$: implicit solution time;  & $t_{\rm{Op}}$: opacity interpolation time;  \\ 
$t_{\rm{LA}}$: dense linear algebra time;  & $t_{\rm{Ps}}$: initial presolve time;  & $t_{\rm{Ex}}$: explicit update time;  \\
$t_{\rm{PL}}$: positivity limiter time;  & \multicolumn{2}{l}{$m:$ Anderson acceleration truncation parameter;} \\
	\multicolumn{3}{l}{Presolve: whether the presolve step is performed in solver initialization;}\\
	\multicolumn{3}{l}{MaxIter: maximum allowed (outer) iteration for (nested) iterative solvers;}\\
	\multicolumn{3}{c}{Relations: $\quad t_{\rm{Tot}}\approx t_{\rm{Im}}+t_{\rm{Ex}}+t_{\rm{PL}}\:,\qquad t_{\rm{Im}}\approx t_{\rm{Op}}+t_{\rm{LA}}+t_{\rm{Ps}}\:.$}
		\end{tabular}
	}
	\end{center}
	}
\end{table}

\clearpage

\section{Summary and discussion}
\label{sec:conclusion}

We have investigated several iterative solvers for nonlinear systems arising from the discretization of a non-relativistic two-moment model for neutrino transport with opacities from \cite{bruenn_1985}, coupled with static matter configurations.  
Specifically, we have incorporated the nonlinear solvers in a DG-IMEX scheme, as implemented in the toolkit for high-order neutrino-radiation hydrodynamics (\thornado).  
Within the IMEX time integration scheme currently adopted in \thornado, updating the neutrino transport and matter equations requires solving a coupled nonlinear system on the radiation moments and matter states (internal energy and electron fraction).  
We have considered two approaches to solve the nonlinear system --- a coupled approach that directly solves the fully coupled system, and a nested approach that formulates the nonlinear system as a nested system with the outer system governing the matter states and the inner system governing the neutrino number densities.  
The nested approach is introduced to reduce the number of opacity evaluations/interpolations required in the solution procedure, and thus becomes more efficient than the coupled approach.  
Two iterative solvers --- the Anderson accelerated fixed point solver and Newton's method --- are implemented for both the coupled and nested approaches.  
We have tested the four solvers on relaxation problems with various collision rates and time steps, as well as on proto-neutron star deleptonization problems with post-bounce matter profiles from spherically symmetric CCSN simulations.  

Numerical results confirm that both nested solvers indeed require fewer iterations to converge (and thus less computational time) than the coupled solvers, due to the fewer number of opacity interpolations performed in the solution procedure.  
The nested Anderson acceleration solver requires more inner iterations to converge, but, due to the low cost per iteration, less computation time than the nested Newton's method, which is a consequence of the heavier dense linear algebra operations in Newton's method.  
In addition to the advantage in computation time, another benefit for using solvers based on Anderson acceleration over Newton's method is the simplicity of implementation, particularly for solving problems in which the derivatives are not readily available, such as the coupled nonlinear systems in CCSN simulations considered in this paper.
For the test problems considered in this paper, we also observe that forcing the nested Anderson acceleration solver to terminate after the first outer iteration could lead to reasonably accurate results.  
This observation confirms that, on these problems, solving the nonlinear coupled system in the implicit step using lagged opacity kernels from the previous time step could give a sufficiently accurate solution, which has also be observed by others (e.g., \cite{just_etal_2015}).  

Moving forward, we will continue to expand on the capabilities in \thornado, and incorporate the more comprehensive physics needed for realistic CCSN models.  
Next steps toward this goal include (i) incorporate special and general relativistic effects into the neutrino transport model, (ii) include muon and tau neutrinos, (iii) update the opacity set to include, e.g., modern electron capture rates, bremsstrahlung, and inelastic scattering on nucleons, (iv) couple the neutrino transport equations with fluid equations to self-consistently model neutrino-radiation hydrodynamics, (v) port the nonlinear solvers to modern hardware architectures (e.g., GPUs), and further analyze implementation performance, and (vi)  compare {\thornado} to other well-established CCSN simulation codes, such as \textsc{agile-boltztran} \citep{liebendorfer_etal_2004}, following the approaches in \cite{just_etal_2015} and \cite{OConnor_etal_2018}. 
We are making progress in these directions, and plan to report on the results in future publications.

\acknowledgments
This manuscript has been authored, in part, by UT-Battelle, LLC, under contract DE-AC05-00OR22725 with the US Department of Energy (DOE). The US government retains and the publisher, by accepting the article for publication, acknowledges that the US government retains a nonexclusive, paid-up, irrevocable, worldwide license to publish or reproduce the published form of this manuscript, or allow others to do so, for US government purposes. DOE will provide public access to these results of federally sponsored research in accordance with the DOE Public Access Plan (\texttt{http://energy.gov/downloads/doe-public-access-plan}).

This research was supported by the Exascale Computing Project (17-SC-20-SC), a collaborative effort of the U.S. Department of Energy Office of Science and the National Nuclear Security Administration.
This research used resources of the Oak Ridge Leadership Computing Facility at the Oak Ridge National Laboratory, which is
supported by the Office of Science of the U.S. Department of Energy under Contract No. DE-AC05-00OR22725.

Support for DOI 10.13139/OLCF/1735948 dataset is provided by the U.S. Department of Energy, project AST137 under Contract DE-AC05-00OR22725. Project AST137 used resources of the Oak Ridge Leadership Computing Facility at Oak Ridge National Laboratory, which is supported by the Office of Science of the U.S. Department of Energy under Contract No. DE-AC05-00OR22725
\clearpage

\appendix

\section{Kernel Derivatives}
\label{appendix:kernelDerivatives}

The NES and pair kernels in Eqs.~\eqref{eq:NESemissitivity}, \eqref{eq:NESOpacity}, \eqref{eq:pairEmissivity}, and \eqref{eq:pairOpacity} are tabulated in terms of temperature $T$ and degeneracy parameter $\eta=\mu_{e}/kT$; i.e., for given $\varepsilon$ and $\varepsilon'$ they are functions of the form
\begin{equation}
  \Phi=\Phi(T,\eta).  
  \label{eq:kernel_T_Eta}
\end{equation}
However, for the coupled Newton's method in Section~\ref{subsec:coupled_Newton}, we need kernel derivatives with respect to internal energy $\epsilon$ and electron fraction $Y_{e}$.  
The electron chemical potential, along with all the other quantities given by the EoS, is tabulated in terms of $\rho$, $T$, and $Y_{e}$.  
On the one hand, the variation of the kernel in Eq.~\eqref{eq:kernel_T_Eta} is
\begin{align}
d\Phi 
&= \Big(\pderiv{\Phi}{T}\Big)_{\eta}\,dT+\Big(\pderiv{\Phi}{\eta}\Big)_{T}\,d\eta \nonumber \\
&=\f{1}{kT}\Big(\pderiv{\Phi}{\eta}\Big)_{T}\,\Big(\pderiv{\mu_{e}}{\rho}\Big)_{T,Y_{e}}\,d\rho
+\Big\{\,\Big(\pderiv{\Phi}{T}\Big)_{\eta} + \Big(\pderiv{\Phi}{\eta}\Big)_{T}\,\Big[\,\f{1}{kT}\Big(\pderiv{\mu_{e}}{T}\Big)_{\rho,Y_{e}}-\f{\eta}{T}\,\Big]\,\Big\}\,dT \nonumber \\
&\hspace{12pt}
+\f{1}{kT}\Big(\pderiv{\Phi}{\eta}\Big)_{T}\,\Big(\pderiv{\mu_{e}}{Y_{e}}\Big)_{\rho,T}\,dY_{e},
\label{eq:variation_T_Eta}
\end{align}
where we used
\begin{align}
d\eta &= \Big(\pderiv{\eta}{\rho}\Big)_{T,Y_{e}}\,d\rho+\Big(\pderiv{\eta}{T}\Big)_{\rho,Y_{e}}\,dT+\Big(\pderiv{\eta}{Y_{e}}\Big)_{\rho,T}\,dY_{e} \nonumber \\
&=\f{1}{kT}\Big(\pderiv{\mu_{e}}{\rho}\Big)_{T,Y_{e}}\,d\rho 
+ \Big[\,\f{1}{kT}\Big(\pderiv{\mu_{e}}{T}\Big)_{\rho,Y_{e}}-\f{\eta}{T}\,\Big]\,dT + \f{1}{kT}\Big(\pderiv{\mu_{e}}{Y_{e}}\Big)_{\rho,T}\,dY_{e}.  
\end{align}
On the other hand, by considering the kernel as a function of $\epsilon$ and $Y_{e}$, the variation is
\begin{align}
d\Phi
&=\Big(\pderiv{\Phi}{\epsilon}\Big)_{Y_{e}}\,d\epsilon+\Big(\pderiv{\Phi}{Y_{e}}\Big)_{\epsilon}\,dY_{e} \nonumber \\
&=\Big(\pderiv{\Phi}{\epsilon}\Big)_{Y_{e}}\,\Big(\pderiv{\epsilon}{\rho}\Big)_{T,Y_{e}}\,d\rho
+\Big(\pderiv{\Phi}{\epsilon}\Big)_{Y_{e}}\,\Big(\pderiv{\epsilon}{T}\Big)_{\rho,Y_{e}}\,dT \nonumber \\
&\hspace{12pt}
+\Big[\,\Big(\pderiv{\Phi}{\epsilon}\Big)_{Y_{e}}\,\Big(\pderiv{\epsilon}{Y_{e}}\Big)_{\rho,T}+\Big(\pderiv{\Phi}{Y_{e}}\Big)_{\epsilon}\,\Big]\,dY_{e},
\label{eq:variation_Epsilon_Ye}
\end{align}
where we used
\begin{equation}
d\epsilon
=\Big(\pderiv{\epsilon}{\rho}\Big)_{T,Y_{e}}\,d\rho+\Big(\pderiv{\epsilon}{T}\Big)_{\rho,Y_{e}}\,dT+\Big(\pderiv{\epsilon}{Y_{e}}\Big)_{\rho,T}\,dY_{e}.  
\end{equation}
Comparing Eqs.~\eqref{eq:variation_T_Eta} and \eqref{eq:variation_Epsilon_Ye}, we have
\begin{align}
\Big(\pderiv{\Phi}{T}\Big)_{\eta} + \Big[\,\f{1}{kT}\Big(\pderiv{\mu_{e}}{T}\Big)_{\rho,Y_{e}}-\f{\eta}{T}\,\Big]\,\Big(\pderiv{\Phi}{\eta}\Big)_{T}
&=\Big(\pderiv{\epsilon}{T}\Big)_{\rho,Y_{e}}\,\Big(\pderiv{\Phi}{\epsilon}\Big)_{Y_{e}}, \\
\f{1}{kT}\Big(\pderiv{\mu_{e}}{Y_{e}}\Big)_{\rho,T}\,\Big(\pderiv{\Phi}{\eta}\Big)_{T}
&=\Big(\pderiv{\epsilon}{Y_{e}}\Big)_{\rho,T}\,\Big(\pderiv{\Phi}{\epsilon}\Big)_{Y_{e}}+\Big(\pderiv{\Phi}{Y_{e}}\Big)_{\epsilon}.  
\end{align}
Solving for $(\partial\Phi/\partial\epsilon)_{Y_{e}}$ and $(\partial\Phi/\partial Y_{e})_{\epsilon}$ we obtain the derivatives we need for Newton's method in terms of derivatives that can be computed directly from the opacity and EoS tables
\begin{align}
\Big(\pderiv{\Phi}{\epsilon}\Big)_{Y_{e}}
&=\Big[\,\f{1}{kT}\Big(\pderiv{\mu_{e}}{T}\Big)_{\rho,Y_{e}}-\f{\eta}{T}\,\Big]/\Big(\pderiv{\epsilon}{T}\Big)_{\rho,Y_{e}}\,\Big(\pderiv{\Phi}{\eta}\Big)_{T} + \Big(\pderiv{\epsilon}{T}\Big)_{\rho,Y_{e}}^{-1}\,\Big(\pderiv{\Phi}{T}\Big)_{\eta}, \\
\Big(\pderiv{\Phi}{Y_{e}}\Big)_{\epsilon}
&=\Big\{\,\f{1}{kT}\Big(\pderiv{\mu_{e}}{Y_{e}}\Big)_{\rho,T} - \Big(\pderiv{\epsilon}{Y_{e}}\Big)_{\rho,T}\,\Big[\,\f{1}{kT}\Big(\pderiv{\mu_{e}}{T}\Big)_{\rho,Y_{e}}-\f{\eta}{T}\,\Big]/\Big(\pderiv{\epsilon}{T}\Big)_{\rho,Y_{e}}\,\Big\}\,\Big(\pderiv{\Phi}{\eta}\Big)_{T} \nonumber \\
&\hspace{12pt}
- \Big\{\,\Big(\pderiv{\epsilon}{Y_{e}}\Big)_{\rho,T}/\Big(\pderiv{\epsilon}{T}\Big)_{\rho,Y_{e}}\,\Big\}\,\Big(\pderiv{\Phi}{T}\Big)_{\eta}.
\end{align}
As discussed in Section~\ref{subsec:implementation} (and following, e.g., \cite{mezzacappaMesser_1999}), tabulated quantities are evaluated using bilinear or trilinear interpolation, while derivatives are estimated by direct differentiation of the respective interpolation formulae.

\bibliographystyle{aasjournal}
\bibliography{reference}

\end{document}